\documentclass[11pt]{article}
\usepackage{pifont}
\usepackage{amsfonts}
\usepackage{subfigure}

\usepackage{color}
\usepackage{amssymb}
\usepackage{amsmath}
\usepackage{epsfig}

\begin{document}
\newtheorem{theorem}{Theorem}
\newtheorem{corollary}{Corollary}
\newtheorem{conjecture}{Conjecture}
\newtheorem{definition}{Definition}
\newtheorem{lemma}{Lemma}
\newtheorem{algorithm}{Algorithm}
\newtheorem{remark}{Remark}
\newtheorem{idea}{Idea}
\newtheorem{observation}{Observation}

\newcommand{\define}{\stackrel{\triangle}{=}}

\pagestyle{empty}

\def\sDoF{\overline{\mbox{\normalfont \scriptsize DoF}}}

\def\QED{\mbox{\rule[0pt]{1.5ex}{1.5ex}}}
\def\proof{\noindent{\it Proof: }}

\date{}

\title{Genie Chains: Exploring Outer Bounds on the Degrees of Freedom of MIMO  Interference Networks}
\author{\normalsize  Chenwei Wang\thanks{This paper was presented in part at IEEE International Symposium on
Information Theory (ISIT), Cambridge, Massachusetts, July 2012. This work was primarily done when Chenwei Wang was with the Center for Pervasive Communications and Computing at the University of California, Irvine.}, Hua Sun$^{\dag}$ and Syed A. Jafar$^{\dag}$\\
        $^*${\small Mobile Network Technology Group,}\\
        {\small DOCOMO Innovations Inc., Palo Alto, CA 94304,}\\
        $^{\dag}${\small Department of Electrical Engineering and Computer Science,}\\
        {\small University of California, Irvine, Irvine, CA 92697}\\
      {\small \it E-mail~:~\{chenweiw, huas2, syed\}@uci.edu}\\
       }

%% Notes
\maketitle

\thispagestyle{empty}
\begin{abstract}

In this paper, we propose a novel genie chains approach to obtain information theoretic degrees of freedom (DoF) outer bounds for MIMO wireless interference networks.
This new approach creates a chain of mappings from genie signals provided to a receiver to the exposed
signal spaces at that receiver, which then serve as the genie
signals for the next receiver in the chain subject to certain linear independence requirements,
essentially converting an information theoretic DoF outer bound problem into a
linear algebra problem. Several applications of the genie chains approach are presented.

%To show the applications of the new genie chains approach, we study the DoF characterizations of several MIMO wireless interference networks, including the  MIMO interference channel, the many-to-one MIMO interference channel, and the MIMO $X$ channel. In particular, we partially characterize the DoF of the $K=4$ user $M_T\times M_R$ MIMO Gaussian interference channel where each TX is equipped with $M_T$ and each RX is equipped with $M_R$ antennas. Expressing the DoF characterization as a function of the ratio $\gamma=M/N$, where $M=\min(M_T,M_R)$ and  $N=\max(M_T,M_R)$, we show that when $\gamma\leq \frac{K-1}{K(K-2)}=\gamma_0$, the DoF value per user is piecewise linear depending on $M$ and $N$ alternately, similar to DoF characterization of the $K=3$ setting. In contrast, when $\gamma >\gamma_o$, we show that the achievable $\frac{MN}{M+N}$ DoF value per user is also information theoretic DoF outer bound for many regimes. Notably, the $\gamma>\gamma_o$ regime, which is the dominant regime for $K>3$ users and is main focus of this paper, is not encountered in the recently solved $K=3$ user setting,  highlighting the fundamentally different character of the DoF for $K>3$ versus the $K=3$ setting.
\end{abstract}

\newpage

\allowdisplaybreaks

\section{Introduction}\label{sec:introduction}

Recently, Wang \emph{et. al.} characterized the spatially normalized degrees of freedom (DoF) for the
$K=3$ user $M_T\times M_R$ interference channel in \cite{Wang_Gou_Jafar_3userMxN} where each transmitter is equipped with $M_T$ antennas,  each receiver with $M_R$ antennas,  and $M_T$, $M_R$ can take
{\em arbitrary} positive integer values.\footnote{A strictly weaker set of DoF results for the 3 user $M_T\times M_R$  wireless interference channel, restricted to linear precoding schemes without symbol extensions, is obtained independently by Bresler \emph{et. al.} in \cite{Bresler_Cartwright_Tse}  in parallel work. The information theoretic outer bounds of Wang \emph{et. al.} in \cite{Wang_Gou_Jafar_3userMxN} match the linear outer bounds of Bresler \emph{et. al.} in \cite{Bresler_Cartwright_Tse}, and the achievability in both \cite{Wang_Gou_Jafar_3userMxN} and \cite{Bresler_Cartwright_Tse} is based on linear schemes. Since information theoretic outer bounds imply linear outer bounds (but not vice versa), the results of Bresler \emph{et. al.} are strictly contained in the results of Wang \emph{et. al.}.} The DoF characterization is comprised of a piece-wise linear mapping with infinitely many linear intervals over the range of the parameter $\gamma=M/N$ where $M=\min(M_T,M_R),N=\max(M_T,M_R)$, shedding light on several interesting elements such as redundant antenna dimensions, decomposability, subspace alignment chains and the feasibility of linear interference alignment. However, existing insights do not suffice beyond the 3-user $M_T\times M_R$ interference channel. In particular, finding good DoF outer bounds for $K$-user MIMO wireless interference networks continues to be a challenge. It is this challenge of finding good DoF outer bounds that motivates this work.

%Intuitively, the  distinction between $K>3$ and $K=3$ settings has  to do with the distinction of over-constrained (proper) versus under-constrained (improper) settings, which in turn require structured (block diagonal, achieved by symbol extensions) and unstructured (e.g., by spatial scaling) interference alignment solutions.

%Specifically, for $K=3$, the interference alignment problem is never over-constrained. This is evident from the observation that the information theoretic DoF outer bound for $K=3$ is achievable with only regular beamforming schemes, based on only \emph{spatial} extensions (i.e., scaling the number of antennas at each node by the same factor) and no time/frequency channel extensions are required. The distinction is important because spatial extensions create unstructured (completely generic) channels, whereas time/frequency extensions create structured (block diagonal) channels and the structure of these channels can create the opportunities for more sophisticated alignment approaches that would not be possible with only unstructured (spatial) extensions. However, unlike $K=3$, the $K>3$ setting is often over constrained.

%One obvious reason is that in contrast with the $K=3$ case where it
%suffices to achieve the optimal DoF using {\em non-asymptotic
%linear} interference alignment schemes, in more than 3 users setting
%the {\em asymptotic} alignment for the achievability comes into
%play.

\begin{figure}[!ht]
\centering
\includegraphics[width=5.0in]{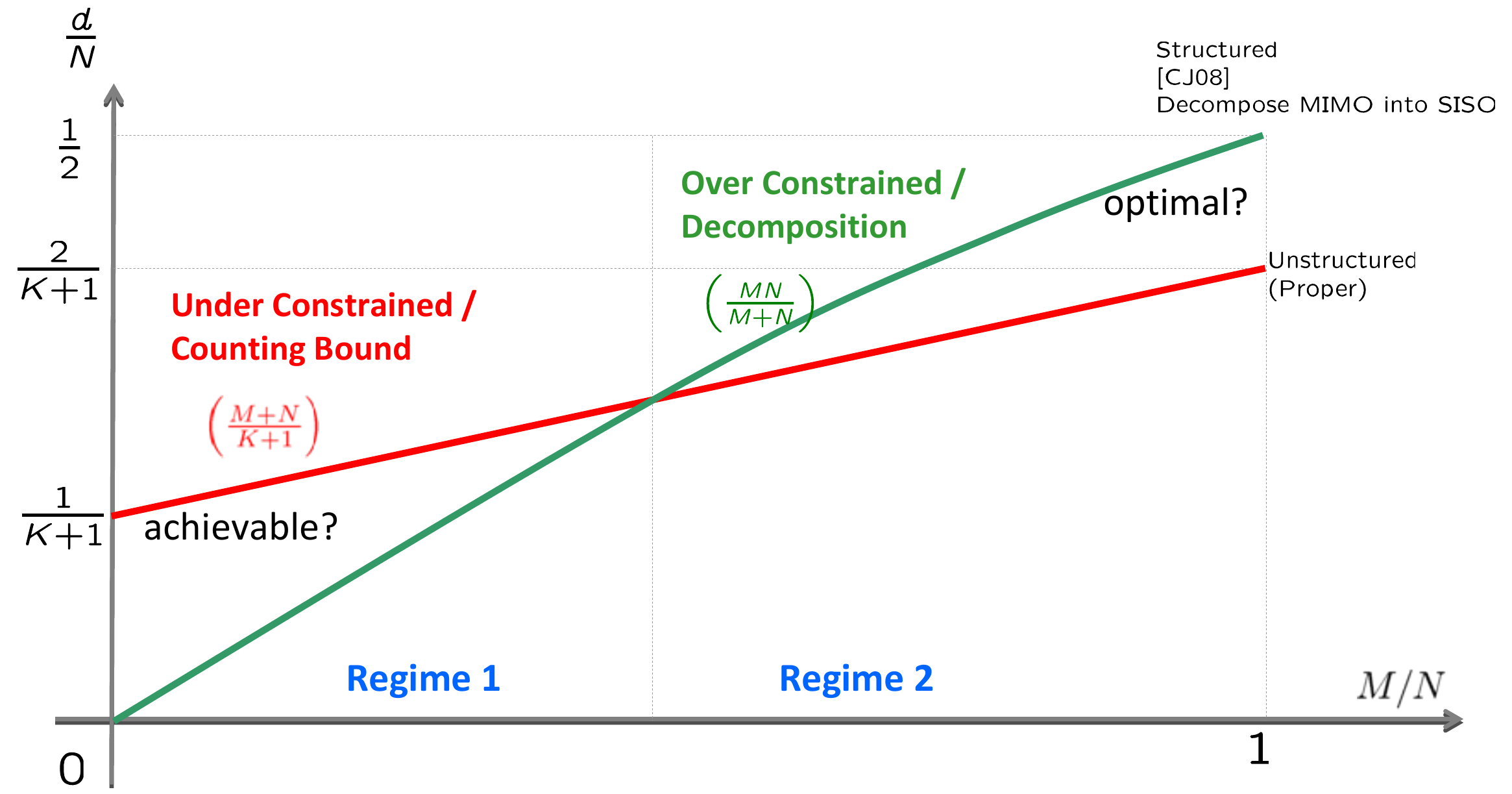}
\caption{The DoF counting bound and the decomposition bound of the $K$-user $M\times N$ MIMO interference channel} %\vspace{-0.1in}
\label{fig:two_regimes}
\end{figure}

In order to clarify what we expect from good DoF outer bounds, it is worthwhile to summarize our expectation of the DoF results of MIMO wireless interference networks. This is simply our projection based on all previously known results,  re-affirmed by our results in this work, and may be seen as a weak conjecture for the general results that so far remain elusive. We will focus on the $K$-user $M_T\times M_R$ wireless interference network and use the Figure \ref{fig:two_regimes} as an illustration. In this figure, the horizonal axis denotes the ratio $\gamma=M/N$, and the vertical axis denotes the DoF per user normalized by $N$. As in the 3-user setting, we  use the notation $M=\min(M_T,M_R), N=\max(M_T,M_R)$. There are two curves in the figure. The red straight line, which we label as the ``counting" \emph{outer} bound, plots the value $d=\frac{M+N}{K+1}$, and the green curve, which we label as the ``decomposition" \emph{inner} bound, plots the value $d=\frac{MN}{M+N}$. An understanding of these two curves is essential to the understanding the DoF of the $K$-user $M_T\times M_R$ interference channel.

A dichotomy is evident in the existing DoF results for $K$-user MIMO wireless interference networks. On the one hand, we have the question of linear DoF, i.e., the DoF achievable  (almost surely) by linear precoding  \emph{without} symbol extensions in time/frequency. Spatial extension, i.e., scaling of antennas  at every node by the same factor, is allowed in this setting. The key distinction between spatial extensions and time/frequency extensions is that the former can only produce generic (structureless) channels whereas the latter give rise to structured (block-diagonal) channel matrices. The linear schemes studied along this research avenue are designed mainly for unstructured generic channels, so they do not benefit from the channel structure, but they \emph{may} be hurt by it if the channel structure causes an overlap of desired and interfering signals. The key to the linear DoF question is the distinction of proper versus improper systems, introduced by Yetis \emph{et. al.} in \cite{Cenk_Gou_Jafar} through the counting bound. A system is proper if $d\leq\frac{M+N}{K+1}$ and improper otherwise. The counting bound is obtained simply by counting the number of alignment constraints and comparing it to the number of design variables. If the number of constraints exceeds the number of variables the system is labeled improper. It is labeled proper otherwise. Yetis \emph{et. al.} conjecture that improper systems are infeasible (when restricted to linear schemes over unstructured channels), whereas proper systems are feasible (through linear schemes over unstructured channels) \emph{provided} they are information theoretically feasible, i.e., that they satisfy the information theoretic DoF bounds.  The first conjecture of Yetis \emph{et. al.} is proved by Bresler \emph{et. al.} in \cite{Bresler_Cavendish_Tse} and by Razaviyayn \emph{et. al.} in \cite{Razaviyayn_MIMO}. The second conjecture of Yetis \emph{et. al.} is consistent with all  DoF results known so far, including the 3-user case, but has not been proved in general.

On the other hand, we have the question of information theoretic DoF, i.e., DoF achievable (almost surely) by linear and non-linear schemes, with no constraints on symbol extensions. It has been observed, and indeed it has been conjectured by Jafar in \cite{Jafar_ITA} that linear schemes over arbitrarily long symbol extensions are still sufficient to achieve the optimal DoF, if generic time-variations are allowed. In the absence of time-variations, more sophisticated schemes, e.g., those based on rational alignments, may be involved. As far as spatial extensions are concerned, there is the spatial scale invariance conjecture by Jafar in \cite{Jafar_ITA, Wang_Gou_Jafar_3userMxN} that claims that if the number of antennas at every node is scaled by a certain factor, then the information theoretic DoF will scale by the same factor. The spatial scale invariance conjecture is consistent with all known results but has not been proved in general. This is in part because few good information theoretic outer bounds are  known.  However, the most important aspect of this discussion is the achievability result by \cite{Ghasemi_Motahari_Khandani_MIMO}, that shows that in a $K$-user $M_T\times M_R$ wireless interference channel, each user is able to achieve $\frac{MN}{M+N}$ DoF by first decomposing multiple antenna nodes into multiple single antenna nodes, and then using the asymptotic alignment scheme of Cadambe and Jafar \cite{Cadambe_Jafar_int} (the CJ scheme) over the resulting SISO network,  precoding over linear vector space dimensions if channels are time-varying, and over rational scalar dimensions if the channels are constant.

The counting bound is an outer bound on the linear DoF, thus restricted to linear precoding schemes with no symbol extensions. The decomposition bound is an inner bound on information theoretic DoF, thus with no restrictions on the type of coding scheme or the use of symbol extensions. At first sight, the two have little to do with each other. And yet, the two seem to play an important joint role as we explain next. First, note that there are two distinct regimes, labeled Regime 1 and Regime 2 in Figure \ref{fig:two_regimes}, where the counting bound dominates the decomposition bound and the decomposition bound dominates the counting bound, respectively.  Regime 1 is relatively well  understood, especially because of the recent insights from the DoF characterization of the 3-user $M_T\times M_R$ MIMO interference channel by Wang \emph{et. al.} \cite{Wang_Gou_Jafar_3userMxN}. Note that the 3-user setting contains only Regime 1. This is easily seen because when $K=3$, the counting bound $\frac{M+N}{K+1}=\frac{M+N}{4}$ is always greater than or equal to the decomposition bound $\frac{MN}{M+N}$. That is,
\begin{align}
\frac{M+N}{4}-\frac{MN}{M+N}&=\frac{(N-M)^2}{4(M+N)}\geq 0.
\end{align}
As we will see in this work, the insights from the 3-user case generalize in a relatively straightforward manner to \emph{most} of Regime 1 of the $K$ user setting: in both cases the optimal DoF curve (for both information theoretic DoF and linear DoF) is piecewise linear, with the linear segments bouncing between the counting bound and the decomposition bound, as they do in the 3-user interference channel.

For this work, it is Regime 2 that is most intriguing. Some interesting observations can be made here. First, note that because the decomposition bound dominates the counting bound, the second conjecture of Yetis \emph{et. al.} would suggest that proper systems in this regime should be feasible with linear precoding and no symbol extensions. Because improper systems are already known to be infeasible, if the conjecture holds, it would settle the linear feasibility question for all systems in Regime 2, i.e., the counting bound would be optimal for linear DoF. This is indeed an interesting observation. However, the main question that interests us in this work has to do with the information theoretic DoF, and the information theoretic optimality of the decomposition bound in Regime 2. To test such a hypothesis, we need better information theoretic DoF outer bounds. So we will develop a novel ``genie-chains" approach that will give us an information theoretic outer bound in terms of a linear algebra problem, specifically requiring the computation of the ranks of certain matrices. The downside is that these matrices  become large  as the MIMO dimensions $M_T, M_R$ increase, so that we face computational bottlenecks. The upside, however, is that for most practically reasonable values of $M_T, M_R$, as well as for certain sub-regimes of Regime 2, we are able to compute the outer bound, and indeed verify that it matches the decomposition inner bound. We summarize these observations  in a loosely stated conjecture, that the decomposition bound is DoF optimal in \emph{most} of Regime 2.

\section{System Model}\label{sec:model}

%In this exposition, since we introduce the notion of genie chains through several examples of $K$ user MIMO interference channel, we will first introduce the system model of MIMO interference channel in this section. However, notice that the tool of genie chains can be applied to other wireless networks as well, as shown in Section \ref{section:app}.

Consider a fully connected $K$-user MIMO interference channel where
there are $M_T$ and $M_R$ antennas at each transmitter (TX) and
receiver (RX), respectively, and each TX has one independent message,
intended for its corresponding RX. Denote by ${\bf H}^{[ji]}$ the $M_R\times M_T$
channel matrix from TX $i$ to RX $j$ where $i,j\!\in
\!\mathcal{K}\triangleq \{1,\cdots,K\}$. For simplicity, we
assume that the channel coefficients are independently drawn from a
continuous distribution. %, and will stay constant during the entire
%transmission once they are drawn (our results do not change if channels are assumed time-varying).
While we will assume that the channels are constant for simplicity,
we note that it is straightforward to extend our DoF outer bounds to the setting where the channel
coefficients are varying in time/frequency. Global channel
knowledge is assumed to be available at all nodes. For codebooks spanning $n$ channel uses, at time index
$t\in \{1,2,\cdots,n\}$, TX $i$ sends a complex-valued $M_T\times 1$
signal vector ${\bf X}^{[i]}(t)$, which satisfies an average power
constraint $\frac{1}{n}\sum_{t=1}^n\mathbb{E}[\|{\bf
X}^{[i]}(t)\|^2]\leq \rho$. At the RX
side, RX $j$ observes an $M_R\times 1$ signal vector ${\bf
\bar{Y}}^{[j]}(t)$ at time index $t$, which
is given by:%\vspace{-0.05in}
\begin{eqnarray}
{\bf \bar{Y}}^{[j]}(t)=\underbrace{\sum_{i=1}^K {\bf H}^{[ji]}{\bf X}^{[i]}(t)}_{\triangleq {\bf Y}^{[j]}(t) } +{\bf
Z}^{[j]}(t)
\end{eqnarray}
\noindent where ${\bf Z}^{[j]}(t)$ is an $M_R\times 1$ column vector
representing the i.i.d. circularly symmetric complex additive white
Gaussian noise (AWGN) at RX $j$, each entry of which is an i.i.d. Gaussian random variable with zero-mean and unit-variance.

As a function of the signal-to-noise ratio (SNR) parameter $\rho$, let $R_k({\rho})=R(\rho)$ denote the  symmetric capacity, i.e., the highest rate simultaneously achievable by
each user. We define  $d(K,M_T,M_R)\triangleq \lim_{\rho\rightarrow
\infty}R(\rho)/\log\rho$ as the symmetric DoF per user. Here, the
user index $k$ is interpreted {\em modulo} $K$ so that, e.g., User 1
is the same as User $K\!+\!1$, etc. %Furthermore, we define the DoF
%per user normalized by the spatial dimension $\bar{d}(K,M_T,M_R)$ as
%\vspace{-0.05in}
%\begin{eqnarray*}
%\bar{d}(K,M_T,M_R)= \max_{n\in\mathbb{Z}^+}d(K,nM_T,nM_R)/n.
%\end{eqnarray*}
The dependence on $K, M_T, M_R$ may be dropped for compact notation
when no ambiguity would be caused. Moreover, we use $o(x)$ to
represent any function $f(x)$ such that $\lim_{x\rightarrow
\infty}f(x)/x = 0$. Furthermore, we define $M=\min(M_T, M_R),
N=\max(M_T, M_R)$.

\section{A Vector Space Perspective}

In this section, we introduce a vector space perspective, and its associated notation, terminology and basic properties, that we will later use for information theoretic DoF outer bounds.

Consider a TX with $M$ antennas, which transmits the $M\times 1$ vector $ {\bf X}(i)$ over the $i^{th}$ channel use, and satisfies an average transmit power constraint $\frac{1}{n}\sum_{i=1}^n\mathbb{E}[\| {\bf X}(i)\|^2]\leq \rho$ across $n$ channel uses. We will denote by $ {\bf X}^n=\{ {\bf X}(1),  {\bf X}(2), \cdots,  {\bf X}(n)\}$, the $n$ vectors sent over the  $n$ channel uses. When referring to the vector transmitted over a single channel use, we will suppress the channel use index for brevity (whenever the particular channel use index is not significant) and simply refer to it as the $M\times 1$ vector $ {\bf X}=[X_1, X_2,\cdots, X_M]^T$.

The vector $ {\bf X}$ lies in the $M$-dimensional vector space spanned by the columns of the $M\times M$ identity matrix.  We are interested  in
\begin{enumerate}
\item Projections of ${\bf X}$ into vector subspaces,
\item Additive Gaussian noise.
\end{enumerate}
The notation and the underlying concepts are best explained through examples. Suppose $M=3$, i.e., we are operating in a $3$-dimensional vector space, and let us consider the following 2-dimensional vector subspace:
\begin{eqnarray}
{\bf L}=\mbox{column span}\left(\left[\begin{array}{rr}1&2\\1&0\\0&3\end{array}\right]\right)\label{eq:basis}.
\end{eqnarray}
Choosing a basis for this subspace, such as the one shown in (\ref{eq:basis}), let us project $ {\bf X}$ into this basis, say $B_1({\bf L})$, giving us:
\begin{eqnarray}
B_1({\bf L})^T{\bf X}=\left[\begin{array}{rr}1&2\\1&0\\0&3\end{array}\right]^T\left[\begin{array}{c}  X_1\\  X_2\\  X_3\end{array}\right]=\left[\begin{array}{c}  X_1+  X_2\\2  X_1+3  X_3\end{array}\right].
\end{eqnarray}
Note that a different choice of basis for the same subspace,  say $B_2({\bf L})^T = A_{2\times 2}B_1({\bf L})^T$, where $A_{2\times 2}$ is an arbitrary $2\times 2$ full rank matrix, will give us a different projected vector, such as:
\begin{eqnarray}
A_{2\times 2}B_1({\bf L})^T{\bf X}=\left[\begin{array}{rr}1&1\\-1&2\end{array}\right]\left[\begin{array}{rr}1&2\\1&0\\0&3\end{array}\right]^T\left[\begin{array}{c}  X_1\\  X_2\\  X_3\end{array}\right]=\left[\begin{array}{rr}3&3\\1&-1\\3&6\end{array}\right]^T\left[\begin{array}{c}  X_1\\  X_2\\  X_3\end{array}\right]=\left[\begin{array}{c}3  X_1+  X_2+3  X_3\\  3X_1 -  X_2+ 6  X_3\end{array}\right].
\end{eqnarray}
However, as we will soon establish, since we are interested only in DoF, the choice of basis is not important for our purpose. Only the span of the space itself is significant.

Next, let us also bring in additive noise into the picture.  Given any vector of random variables ${\bf U}=[U_1, U_2, \cdots, U_k]^T$, let us define the differential entropy of its noisy version as
\begin{eqnarray}
\hbar({\bf U})\triangleq h({\bf U}+{\bf Z})= h(U_1+Z_1, U_2+Z_2, \cdots, U_k+Z_k)
\end{eqnarray}
where $h(\cdot)$ is the standard differential entropy function, ${\bf Z}=[Z_1, Z_2, \cdots, Z_k]^T$ is a circularly symmetrically additive white Gaussian noise vector that is independent with ${\bf U}$ and ${\bf Z}\sim\mathcal{CN}({\bf 0}, {\bf I})$. Similar definitions are used for joint and conditional differential entropies, i.e.,
\begin{eqnarray}
\hbar({\bf U}, {\bf V})&\!\!\!\!\triangleq \!\!\!\!& h({\bf U}+{\bf Z}, {\bf V}+{\bf Z}')\\
&\!\!\!\!=\!\!\!\!& h(U_1+Z_1, U_2+Z_2, \cdots, U_k+Z_k,V_1+Z'_1, V_2+Z'_2, \cdots, V_k+Z'_k),\\
\hbar({\bf U}|{\bf V})&\!\!\!\!\triangleq \!\!\!\!& \hbar({\bf U},{\bf V}) - \hbar({\bf V})\\
&\!\!\!\!=\!\!\!\!& h({\bf U}+{\bf Z}, {\bf V}+{\bf Z}') - {h}({\bf V}+{\bf Z}') = {h}({\bf U}+{\bf Z} | {\bf V}+{\bf Z}')\\
&\!\!\!\!=\!\!\!\!& h(U_1+Z_1, U_2+Z_2, \cdots, U_k+Z_k|V_1+Z'_1, V_2+Z'_2, \cdots, V_k+Z'_k),
\end{eqnarray}
where ${\bf V} = [V_1,V_2,\cdots,V_k]^T$ is another vector of random variables and ${\bf Z'} = [Z'_1,Z'_2,\cdots,Z'_k]^T \sim\mathcal{CN}({\bf 0},{\bf I})$, which is independent with ${\bf U}, {\bf V}$ and ${\bf Z}$.

\begin{lemma}\label{lemma:noise_distortion}
Consider an arbitrary subspace ${\bf L}$ of the $M$-dimensional vector space $\mathbb{C}^M$ and let $B_i({\bf L})$, $B_j({\bf L})$ be two arbitrary choices for the basis of ${\bf L}$. We have
\begin{eqnarray}
\hbar(B_i({\bf L})^T {\bf X})=h(B_j({\bf L})^T {\bf X}+\tilde{\bf Z})+o(\log\rho),
\end{eqnarray}
where $\tilde{\bf Z}\sim\mathcal{CN}({\bf 0}, \tilde{\bf K})$, and $\tilde {\bf K}$ is a non-singular covariance matrix. We require that ${\bf L}, B_i({\bf L}), B_j({\bf L}),\tilde{\bf K}$ are held fixed as $\rho\rightarrow\infty$.
\end{lemma}
\proof We defer the proof to Appendix \ref{app:lemma_noise_distortion}. \hfill\QED
%\textcolor{red}{A corollary for the conditional version is also needed.}

According to Lemma \ref{lemma:noise_distortion},  as long as the subspace ${\bf L}$, its basis representation $B({\bf L})$ and the additive noise terms $\tilde {\bf Z}$ do not depend on the SNR,  $\rho$, and the noise in the projected subspace is non-singular, then all that matters is the subspace ${\bf L}$ within which ${\bf X}$ is projected. Neither the particular choice of basis representation, nor the specific form of the noise covariance matrix is relevant.

In light of this observation, we will henceforth simplify our notation by referring to $\hbar(B({\bf L})^T{\bf X})$ as $\hbar({\bf L}\!\circ\!{\bf X})$ instead, where the symbol ``$\circ$" denotes the projection operation, with the understanding that the given representation of ${\bf L}$ is equivalent to any other basis representation of the same space for our purpose.

Lemma 1 extends easily to joint and conditional differential entropies as well, for which still only the space matters, not the specific basis representation chosen. For two subspaces ${\bf L}^{[1]}, {\bf L}^{[2]}$ of $\mathbb{C}^M$, we define $\hbar({\bf L}^{[1]} \!\circ\! {\bf X}, {\bf L}^{[2]} \!\circ\! {\bf X})$ and $\hbar({\bf L}^{[1]} \!\circ\! {\bf X}| {\bf L}^{[2]} \!\circ\! {\bf X})$ in a similar way to refer to the joint and conditional differential entropies of $\bf{X}$ projected in corresponding spaces, respectively.

%Denote by ${\bf L}\subset{\bf X}$ as an $|{\bf L}|$ dimensional subspace of ${\bf
%X}$ where $|{\bf L}|$ represents the dimensions of the subspace
%${\bf L}$. If ${\bf L}$ is a minimal basis representation, then $|{\bf L}|$ is simply the number
%of column vectors in ${\bf L}$. Writing ${\bf
%L}\!=\!\{L_1,\cdots,L_{|{\bf L}|}\}$ we recognize ${\bf L}$ as a collection of $|{\bf L}|$
%linear combinations  of the
%$M$ variables $X_m,~m=1,\cdots,M$ comprising ${\bf X}$. The
%linear combination corresponding to $L_i$ is given by ${\bf L}_i^T{\bf X}$
%where the column vector ${\bf L}_i$, the basis of ${\bf L}$,
%contains linear combination coefficients for the $i^{th}$ equation.
%If ${\bf L}_i$ are regarded as generic vectors, then we say that
%${\bf L}$ is a \emph{generic subspace}. Again, we use ${\bf L}$ to
%indicate a column vector, a collection of equations, as well as the
%vector subspace. Then we define the following operations in the
%vector space which will be often used throughout this paper.
%Assuming ${\bf L}^{[1]}$, ${\bf L}^{[2]}$ are two subspaces of ${\bf
%X}$, we denote by ${\bf L}^{[1]}\cap{\bf L}^{[2]}$ their
%intersection, and ${\bf L}^{[1]}\!\setminus\!{\bf L}^{[2]}$ the
%subspace of ${\bf L}^{[1]}$ which has null intersection with ${\bf
%L}^{[2]}$. In addition, we use ${\bf L}^c$ to represent the
%$(M-|{\bf L}|)$ dimensional subspace of ${\bf X}$ which has null
%intersection with ${\bf L}$, i.e., ${\bf L}^c$ is the null space of
%${\bf L}^T$. Then the two operations can be computed as follows.

It is useful to further familiarize ourselves with the vector space representations, for instance, with unions and intersection operations. Once again, we illustrate these with a simple example. Consider the following subspaces:
\begin{eqnarray}
{\bf L}_1^{[1]} &\!\!\!\!=\!\!\!\!& \mbox{span}([1~~1~~0]^T), \\
{\bf L}_2^{[1]} &\!\!\!\!=\!\!\!\!& \mbox{span}([2~~0~~3]^T), \\
{\bf L}_1^{[2]} &\!\!\!\!=\!\!\!\!& \mbox{span}([2~~-1~~4]^T), \\
{\bf L}_2^{[2]} &\!\!\!\!=\!\!\!\!& \mbox{span}([-2~~-3~~1]^T),
%{\bf
%L}_1^{[1]}=\left[\begin{array}{c}1\\1\\0\end{array}\right],~~~{\bf
%L}_2^{[1]}=\left[\begin{array}{r}2\\0\\3\end{array}\right],~~~{\bf
%L}_1^{[2]}=\left[\begin{array}{r}2\\-1\\4\end{array}\right],~~~{\bf
%L}_2^{[2]}=\left[\begin{array}{r}-2\\-3\\1\end{array}\right],
\end{eqnarray}
and let ${\bf L}^{[1]}, {\bf L}^{[2]}$ be defined as the vector spaces spanned by the unions:
\begin{eqnarray}
{\bf L}^{[1]}&\!\!\!\!=\!\!\!\!& \{{\bf L}^{[1]}_1,{\bf L}^{[1]}_2\},\\
{\bf L}^{[2]}&\!\!\!\!=\!\!\!\!& \{{\bf L}^{[2]}_1,{\bf L}^{[2]}_2\}.
\end{eqnarray}
Note that since the union of vector spaces is not generally a vector space, what is  meant here is that ${\bf L}^{[i]}$ is the vector space spanned by the union of the vector subspaces ${\bf L}^{[i]}_1, {\bf L}^{[i]}_2$.

Next let us consider the intersection of ${\bf L}^{[1]}$ and ${\bf L}^{[2]}$.  Note that given ${\bf L}^{[i]}$, we can compute ${\bf L}^{[i]^c}$
which is the subspace orthogonal to  the span of $({\bf L}^{[i]}_1,{\bf L}^{[i]}_2)$.
That is,
\begin{eqnarray}
{\bf L}^{[1]^c} &\!\!\!\!=\!\!\!\!&\mbox{span}([3~~-3~~-2]^T), \\
{\bf L}^{[2]^c} &\!\!\!\!=\!\!\!\!& \mbox{span}([-5.5~~5~~4]^T).
%{\bf L}^{[1]^c}=\left[\begin{array}{r}3\\-3\\-2\end{array}\right],~~~{\bf
%L}^{[2]^c}=\left[\begin{array}{r}-5.5\\5\\4\end{array}\right].
\end{eqnarray}
Thus, the intersection ${\bf L}^{[1]}\cap{\bf L}^{[2]}$ can be
obtained by computing the subspace orthogonal to  both ${\bf L}^{[1]^c}$ and
${\bf L}^{[2]^c}$, and thus it can be written as:
\begin{eqnarray}
{\bf L}^{[1]}\cap{\bf L}^{[2]}=([{\bf L}^{[1]^c}~{\bf
L}^{[2]^c}])^c=\mbox{span}([4~~2~~3]^T).
\end{eqnarray}
Similarly, we define ${\bf L}^{[1]}\!\setminus\!{\bf L}^{[2]}$ to be the subspace of ${\bf L}^{[1]}$ which is orthogonal to
${\bf L}^{[1]}\cap{\bf L}^{[2]}$, i.e.,
\begin{eqnarray}
{\bf L}^{[1]}\setminus{\bf L}^{[2]}&\!\!\!\!=\!\!\!\!&{\bf L}^{[1]}\setminus({\bf L}^{[1]}\cap{\bf L}^{[2]})={\bf L}^{[1]}\cap({\bf L}^{[1]}\cap{\bf L}^{[2]})^c\\
&\!\!\!\!=\!\!\!\!&{\rm span}([5~~17~~-\!18]^T).
\end{eqnarray}
With this definition, we can also write ${\bf L}^{[1]}$ as
\begin{eqnarray}\label{eqn:union_two_parts}
{\bf L}^{[1]}=\{{\bf L}^{[1]}\cap{\bf L}^{[2]},~{\bf
L}^{[1]}\!\setminus\!{\bf L}^{[2]}\}.
\end{eqnarray}

A set of $M \times 1$ vectors is generic if and only if any $m$ of them are linearly independent whenever $m \leq M$. Generic subspaces are those spaces whose basis vectors are generic.

In this paper, because we are primarily interested in the notion of DoF,
%i.e., the pre-log coefficient when SNR goes into the infinity, the bounded variance noise terms are inconsequential. Thus, in the remaining of this paper,
we will use the notations $x(\rho,n)=:y(\rho,n)$, $x(\rho,n)\leq:y(\rho,n)$, $x(\rho,n)\geq:y(\rho,n)$ to represent $x(\rho,n)=y(\rho,n)+n~o(\log\rho)$, $x(\rho,n)\leq y(\rho,n)+n~o(\log\rho)$, $x(\rho,n)\geq y(\rho,n)+n~o(\log\rho)$,
%by meaning that
%\begin{eqnarray*}
%\lim_{\rho\rightarrow +\infty}\lim_{n\rightarrow +\infty}x(\rho,n)&\!\!=\!\!&\lim_{\rho\rightarrow +\infty}\lim_{n\rightarrow +\infty}y(\rho,n),\\
%\lim_{\rho\rightarrow +\infty}\lim_{n\rightarrow +\infty}x(\rho,n)&\!\!\leq \!\!&\lim_{\rho\rightarrow +\infty}\lim_{n\rightarrow +\infty}y(\rho,n),\\
%\lim_{\rho\rightarrow +\infty}\lim_{n\rightarrow +\infty}x(\rho,n)&\!\!\geq \!\!&\lim_{\rho\rightarrow +\infty}\lim_{n\rightarrow +\infty}y(\rho,n),
%\end{eqnarray*}
respectively. Next we summarize the basic properties associated with the vector subspace representations. The properties are stated in the multi-letter form, which is used in the information theoretic proofs. As such, we extend the vector space terminologies introduced above to their corresponding multi-letter forms.

${\bf L}^n \triangleq {\bf L}(1) \times {\bf L}(2) \times \cdots \times {\bf L}(n)$ is used to represent the collection of $n$ subspaces ${\bf L}(1), {\bf L}(2), \cdots, {\bf L}(n)$ of the $M$-dimensional vector space $\mathbb{C}^M$. If the dimension of the $n$ subspaces ${\bf L}(t), t \in \{1,2,\ldots,n\}$ is the same, we will denote it as $|{\bf L}|$. The basis representation $B({\bf L}^n)$ of ${\bf L}^n$ is the collection of the basis representations of each subspace, i.e., $B({\bf L}^n) \triangleq B({\bf L}(1)) \times B({\bf L}(2)) \times \cdots \times B({\bf L}(n))$. Also,
%$\mathbb{C}^{M^n}$ is the collection of $n$ $\mathbb{C}^M$ space,
$\mathbb{C}^{M^n} \triangleq \mathbb{C}^{M} \times  \mathbb{C}^{M} \times \cdots \times \mathbb{C}^{M}$. For two multi-letter subspaces ${\bf L}^{[1]^n}$ and ${\bf L}^{[2]^n}$, their intersection ${\bf L}^{[1]^n} \!\cap {\bf L}^{[2]^n}$ is defined as%the collection of component-wise intersection of ${\bf L}^{[1]^n}$ and ${\bf L}^{[2]^n}$, i.e.,
\begin{equation}
{\bf L}^{[1]^n} \!\cap {\bf L}^{[2]^n} \triangleq {\bf L}^{[1]}(1) \!\cap {\bf L}^{[2]}(1) \times {\bf L}^{[1]}(2) \!\cap {\bf L}^{[2]}(2) \times \cdots \times {\bf L}^{[1]}(n) \!\cap {\bf L}^{[2]}(n).
\end{equation}
Similar definitions are employed for ${\bf L}^{[1]^n} \!\setminus {\bf L}^{[2]^n}$, ${\bf L}^n\!\circ\!{\bf X}^n$ and $B({\bf L}^n)^T {\bf X}^n$,
\begin{eqnarray}
{\bf L}^{[1]^n} \!\setminus {\bf L}^{[2]^n} &\!\!\!\!\triangleq\!\!\!\!& {\bf L}^{[1]}(1) \! \setminus {\bf L}^{[2]}(1) \times {\bf L}^{[1]}(2) \!\setminus {\bf L}^{[2]}(2) \times \cdots \times {\bf L}^{[1]}(n) \!\setminus {\bf L}^{[2]}(n), \\
{\bf L}^{n}\!\circ\!{\bf X}^n &\!\!\!\!\triangleq\!\!\!\!& {\bf L}(1)\!\circ\!{\bf X}(1) \times {\bf L}(2)\!\circ\!{ \bf X}(2) \times \cdots \times {\bf L}(n)\!\circ\!{\bf X}(n), \\
B({\bf L}^n)^T {\bf X}^n &\!\!\!\!\triangleq\!\!\!\!& B({\bf L}(1))^T {\bf X}(1) \times B({\bf L}(2))^T {\bf X}(2) \times \cdots \times B({\bf L}(n))^T {\bf X}(n).
\end{eqnarray}
Equipped with these definitions, following (\ref{eqn:union_two_parts}), we can write
\begin{eqnarray}\label{eqn:union_two_parts_n}
{\bf L}^{[1]^n}=\{{\bf L}^{[1]^n}\cap{\bf L}^{[2]^n},~{\bf
L}^{[1]^n}\!\setminus\!{\bf L}^{[2]^n}\}.
\end{eqnarray}
Next, we proceed to the statement of the properties.
%circ )= \hbar({\bf L}^n)+n\log|\det(A)|
\begin{lemma}\label{lemma:property}
We have the following properties:
\begin{itemize}
\item[(P1)] $\hbar(B_i({\bf L}^n)^T {\bf X}^n) =: \hbar(B_j({\bf L}^n)^T {\bf X}^n)$ for any basis representations $B_i({\bf L}^n),B_j({\bf L}^n)$ of ${\bf L}^n$.
\end{itemize}
\emph{Justified by this property, we will write $\hbar(B({\bf L}^n)^T {\bf X}^n)$ simply as $\hbar({\bf L}^n \!\circ\!{\bf X}^n)$.}
\begin{itemize}
\item[(P2)] $\hbar({\bf L}^n \!\circ\! {\bf X}^n)\leq: n|{\bf L}|\log\rho$.
\item[(P3)] For generic subspaces
%\footnote{Vector subspace ${\bf L}$ is called generic when its basis representation $B({\bf L})$ consists of generic vectors.}
${\bf L}^{[1]}(t),{\bf L}^{[2]}(t),t \in \{1,2,\cdots,n\}$ of $\mathbb{C}^M$ with $|{\bf L}^{[1]}|+|{\bf L}^{[2]}|\geq M$, we have:
%the following statements:
\end{itemize}
\begin{itemize}
\item[P3a)] $\hbar({\bf L}^{[1]^n}\!\circ\! {\bf X}^n, {\bf L}^{[2]^n}\!\circ\! {\bf X}^n)=: \hbar({\bf X}^n)$.
\item[P3b)] $\hbar({\bf L}^{[1]^n}\!\circ\!{\bf X}^n |{\bf L}^{[2]^n}\!\circ\!{\bf X}^n)=: \hbar((\mathbb{C}^{M^n}\setminus{\bf L}^{[2]^n})\!\circ\!{\bf X}^n |{\bf L}^{[2]^n}\!\circ\!{\bf X}^n)$.
\item[P3c)] $\min(\hbar({\bf L}^{[1]^n}\!\circ\!{\bf X}^n |{\bf L}^{[2]^n}\!\circ\!{\bf X}^n), \hbar({\bf L}^{[2]^n}\!\circ\!{\bf X}^n |{\bf L}^{[1]^n}\!\circ\!{\bf X}^n) )\leq: \frac{1}{2} \hbar({\bf X}^n)$.
\end{itemize}
\end{lemma}
{\it Proof:} We will show the proofs for each property sequentially.

Property \emph{(P1)}:%\vspace{-0.1in}

This property is the multi-letter version of Lemma \ref{lemma:noise_distortion}, whose proof follows directly.

Property \emph{(P2)}:\vspace{-0.1in}
\begin{eqnarray}
\hbar({\bf L}^n\!\circ\!{\bf X}^n)&\!\!\!\!=\!\!\!\!&\sum_{t=1}^n \hbar({\bf L}(t)\!\circ\! {\bf X}(t)|{\bf L}(1)\!\circ\! {\bf X}(1),\cdots,{\bf L}(t-1)\!\circ\! {\bf X}(t-1))\\
&\!\!\!\!\leq\!\!\!\!& \sum_{t=1}^n \hbar({\bf L}(t)\!\circ\!{\bf X}(t))\label{eqn:p1_cond}\\
&\!\!\!\!=\!\!\!\!&\sum_{t=1}^n\sum_{i=1}^{|{\bf L}|} \hbar({\bf L}_i(t) \!\circ\! {\bf X}(t)|{\bf L}_{i-1}(t)\!\circ\!{\bf X}(t),\cdots,{\bf L}_1(t)\!\circ\!{\bf X}(t))\label{eqn:p1_col}\\
&\!\!\!\!\leq\!\!\!\!& \sum_{t=1}^n\sum_{i=1}^{|{\bf L}|} \hbar({\bf L}_i(t) \!\circ\! {\bf X}(t))\\
&\!\!\!\!\leq:\!\!\!\!& n|{\bf L}|\log\rho\label{eqn:p1_last}
\end{eqnarray}
where (\ref{eqn:p1_cond}) follows from the fact that removing conditional terms does not decrease the differential entropy; (\ref{eqn:p1_col}) is obtained due to the chain rule and ${\bf L}_{i}(t)$ denotes the space spanned by the $i$-th basis vector of ${\bf L}(t)$; (\ref{eqn:p1_last}) is obtained because one dimension can contribute upto one $\log\rho+o(\log\rho)$ term.

In addition, incorporating (\emph{P1}), we can also see that if $|{\bf L}| = M$, then $\hbar({\bf X}^n) = \hbar({\mathbb{C}}^{M^n} \!\circ\! {\bf X}^n)  =: \hbar({\bf L}^n \!\circ\! {\bf X}^n) \leq: nM\log\rho$.
%\begin{eqnarray}
%\hbar(A{\bf L}^n)&\!\!=\!\!& \hbar(A{\bf L}(1),A{\bf L}(2),\cdots,A{\bf L}(n))\\
%&\!\!=\!\!& \hbar(\textrm{diag}([A,\cdots,A])[{\bf L}(1)^T~\cdots~{\bf L}(n)^T]^T)\\
%&\!\!=\!\!& \hbar([{\bf L}(1)^T~\cdots~{\bf L}(n)^T]^T)+\log|\det(\textrm{diag}([A,\cdots,A]))|\\
%&\!\!=\!\!& \hbar({\bf L}^n)+n\log|\det(A)|\\
%&\!\!=:\!\!& \hbar({\bf L}^n).
%\end{eqnarray}

Property \emph{(P3a)}:\vspace{-0.1in}
\begin{eqnarray}
\hbar({\bf L}^{[1]^n}\!\circ\! {\bf X}^n,{\bf L}^{[2]^n}\!\circ\! {\bf X}^n)&\!\!\!\!=\!\!\!\!& \hbar( \{{\bf L}^{[1]^n}\!\cap{\bf L}^{[2]^n}, {\bf L}^{[1]^n}\!\setminus\!{\bf L}^{[2]^n}\} \!\circ\! {\bf X}^n, {\bf L}^{[2]^n}\!\circ\! {\bf X}^n)\label{eqn:p3a_union}\\
&\!\!\!\!=\!\!\!\!& \hbar( ({\bf L}^{[1]^n}\!\cap{\bf L}^{[2]^n})\!\circ\! {\bf X}^n, ({\bf L}^{[1]^n}\!\setminus\!{\bf L}^{[2]^n}) \!\circ\! {\bf X}^n, {\bf L}^{[2]^n}\!\circ\! {\bf X}^n)\label{eqn:p3a_proj1}\\
&\!\!\!\!=\!\!\!\!& \hbar( ({\bf L}^{[1]^n}\!\cap{\bf L}^{[2]^n})\!\circ\! {\bf X}^n, \{ {\bf L}^{[1]^n}\!\setminus\!{\bf L}^{[2]^n}, {\bf L}^{[2]^n}\}\!\circ\! {\bf X}^n)\label{eqn:p3a_proj2}\\
&\!\!\!\!=:\!\!\!\!& \hbar( ({\bf L}^{[1]^n}\!\cap{\bf L}^{[2]^n})\!\circ\! {\bf X}^n,  \mathbb{C}^{M^n} \!\circ\! {\bf X}^n)\label{eqn:p3a_con}\\
&\!\!\!\!=\!\!\!\!& \hbar(({\bf L}^{[1]^n}\!\cap{\bf L}^{[2]^n})\!\circ\! {\bf X}^n, {\bf X}^n)\\
&\!\!\!\!=\!\!\!\!& \hbar({\bf X}^n)+ \hbar( ({\bf L}^{[1]^n}\!\cap{\bf L}^{[2]^n})\!\circ\!{\bf X}^n|{\bf X}^n)\\
&\!\!\!\!=\!\!\!\!& \hbar({\bf X}^n)+n~o(\log\rho)\label{eqn:p3a_last}\\
&\!\!\!\!=:\!\!\!\!& \hbar({\bf X}^n)
\end{eqnarray}
where (\ref{eqn:p3a_union}) follows directly from (\ref{eqn:union_two_parts_n}); both (\ref{eqn:p3a_proj1}) and (\ref{eqn:p3a_proj2}) are obtained due to the fact that the two subspaces participating the splitting or the union operations are orthogonal to each each other; (\ref{eqn:p3a_con}) follows from the assumption $|{\bf L}^{[1]}|+|{\bf L}^{[2]}|\geq M$ and Property (\emph{P1}); and (\ref{eqn:p3a_last}) is obtained
because the subspace ${\bf L}^{[1]^n}\!\cap{\bf L}^{[2]^n}$ is contained in $\mathbb{C}^{M^n}$. Basically, it implies that
%over any channel use,
the $M$ variables comprising the vector ${\bf X}$ can be used to construct any linear combination of ${\bf X}$ subject to the bounded noise distortion.
%from generic sets of linear equations subject to the noise distortion as long as the total number of equations is at least $M$.

Property \emph{(P3b)}:
\begin{eqnarray}
\hbar({\bf L}^{[1]^n}\!\circ\! {\bf X}^n|{\bf L}^{[2]^n}\!\circ\! {\bf X}^n)&\!\!\!\!=\!\!\!\!& \hbar(({\bf L}^{[1]^n}\setminus{\bf L}^{[2]^n})\!\circ\! {\bf X}^n, ({\bf L}^{[1]^n}\!\cap{\bf L}^{[2]^n}) \!\circ\! {\bf X}^n|{\bf L}^{[2]^n}\!\circ\! {\bf X}^n)\\
&\!\!\!\!=\!\!\!\!& \hbar(\{{\bf L}^{[1]^n}\setminus{\bf L}^{[2]^n},{\bf L}^{[2]^n}\}\!\circ\! {\bf X}^n, ({\bf L}^{[1]^n}\!\cap{\bf L}^{[2]^n})\!\circ\! {\bf X}^n|{\bf L}^{[2]^n} \!\circ\! {\bf X}^n)\\
&\!\!\!\!=:\!\!\!\!& \hbar({\bf X}^n, ({\bf L}^{[1]^n}\!\cap{\bf L}^{[2]^n})\!\circ\! {\bf X}^n|{\bf L}^{[2]^n} \!\circ\! {\bf X}^n)\\
&\!\!\!\!=\!\!\!\!& \hbar({\bf X}^n|{\bf L}^{[2]^n}\!\circ\! {\bf X}^n)+h(({\bf L}^{[1]^n}\!\cap{\bf L}^{[2]^n})\!\circ\! {\bf X}^n|{\bf L}^{[2]^n}\!\circ\! {\bf X}^n,{\bf X}^n)\\
&\!\!\!\!=\!\!\!\!& \hbar((\mathbb{C}^{M^n}\setminus{\bf L}^{[2]^n})\!\circ\! {\bf X}^n |{\bf L}^{[2]^n}\!\circ\! {\bf X}^n)+n~o(\log\rho)\\
&\!\!\!\!=:\!\!\!\!& \hbar((\mathbb{C}^{M^n}\setminus{\bf L}^{[2]^n}) \!\circ\! {\bf X}^n |{\bf L}^{[2]^n}\!\circ\! {\bf X}^n).
\end{eqnarray}
The intuition of this property is that adding $\hbar({\bf L}^{[2]^n} \!\circ\! {\bf X}^n )$ to both sides of the equation produces the $\hbar({\bf X}^n)$ term on both sides.

Property \emph{(P3c)}:
\begin{eqnarray}
&& \min(\hbar({\bf L}^{[1]^n}\!\circ\! {\bf X}^n|{\bf L}^{[2]^n}\!\circ\! {\bf X}^n), \hbar({\bf L}^{[2]^n}\!\circ\! {\bf X}^n|{\bf L}^{[1]^n} \!\circ\! {\bf X}^n)) \\
&\!\!\!\!\leq\!\!\!\!& \frac{1}{2}\big[\hbar({\bf L}^{[1]^n}\!\circ\! {\bf X}^n|{\bf L}^{[2]^n}\!\circ\! {\bf X}^n) + \hbar({\bf L}^{[2]^n}\!\circ\! {\bf X}^n|{\bf L}^{[1]^n} \!\circ\! {\bf X}^n)\big]\\
&\!\!\!\!\leq\!\!\!\!& \frac{1}{2}\big[\hbar({\bf L}^{[1]^n}\!\circ\! {\bf X}^n|{\bf L}^{[2]^n}\!\circ\! {\bf X}^n) + \hbar({\bf L}^{[2]^n}\!\circ\! {\bf X}^n)\big]\\
&\!\!\!\!=\!\!\!\!& \frac{1}{2}\hbar({\bf L}^{[1]^n}\!\circ\! {\bf X}^n, {\bf L}^{[2]^n}\!\circ\! {\bf X}^n)\\
&\!\!\!\!=:\!\!\!\!& \frac{1}{2}\hbar({\bf X}^n)\label{eqn:p3c_p3a}
\end{eqnarray}
where (\ref{eqn:p3c_p3a}) is obtained owing to Property (\emph{P3a}).
\hfill\QED

\subsection{Multiple Subspaces of the Vector Space}

In this section, we introduce important properties associated with vector subspaces. %through a lemma and corollaries.
%{\color{red} TBD: Lemma 3 is difficult to understand. Please re-write the lemma clearly.}

Suppose we have $K$ subspaces ${\bf L}^{[k]}(t), ~t \in \{1,2,\ldots,n\}, ~k\in\{1,2,\cdots,K\}$ of the $M$-dimensional vector space $\mathbb{C}^M$. The dimension of ${\bf L}^{[k]}(t)$ is $l_k,\forall t$ and we define $l^*\triangleq \sum_{k=1}^K l_k$. Over the $t^{th}$ channel use, we enumerate all the $l^*$ basis vectors contained in the $K$ subspaces, ${\bf L}^{[k]}(t)$ and denote the span of these vectors as
${\bf L}_1(t),{\bf L}_2(t),\cdots,
%{\bf L}_{\lfloor \frac{l^*}{M} \rfloor M}^n,{\bf L}_{\lfloor \frac{l^*}{M} \rfloor M+1}(t),\cdots,
{\bf L}_{l^*}(t)$, so that the basis representation of
$\mathbf{L}^{[1]}(t)$ is comprised of the first $l_1$ basis vectors, the basis representation of $\mathbf{L}^{[2]}(t)$ the next $l_2$ basis vectors and so forth. Repeating such enumeration for all channel uses, we have %the following collections of basis vectors,
\begin{eqnarray}
\mathbf{L}^{[1]^n} &\!\!\!\!=\!\!\!\!& \{{\bf L}_1^n,{\bf L}_2^n,\cdots,{\bf L}_{l_1}^n\}, \\
\mathbf{L}^{[2]^n} &\!\!\!\!=\!\!\!\!& \{{\bf L}_{l_1+1}^n,\cdots,{\bf L}_{l_1+l_2}^n\}, \\
&\!\!\!\!\vdots\!\!\!\!&\notag\\
%\mathbf{L}^{[k]^n} &\!\!=\!\!& \{{\bf L}_{\sum_{i=1}^{k-1}l_i+1}^n,\cdots,{\bf L}_{\lfloor \frac{L}{M} \rfloor M}^n,{\bf L}_{\lfloor \frac{L}{M} \rfloor M + 1}^n,\cdots,{\bf L}_{\sum_{i=1}^kl_i}^n\}\\
%&\!\!\!\!\vdots\!\!\!\!&\notag\\
\mathbf{L}^{[K]^n} &\!\!\!\!=\!\!\!\!& \left\{{\bf L}_{\sum_{i=1}^{K-1}l_i+1}^n,\cdots,{\bf L}_{l^*}^n\right\}.
\end{eqnarray}
Now, let us start sequentially in the order of ${\mathbf{L}^{[1]}}^n,{\mathbf{L}^{[2]}}^n,\ldots$ to collect subspaces into a set and go as far as we can without the total number of \emph{linear independent} basis vectors exceeding $M$. There are two possibilities. If we happen to collect exactly $M$ independent vectors then we set these aside and start building the next set of vectors, proceeding sequentially again from where the first set terminated. On the other hand, if we fall short of $M$ vectors, i.e., we cannot include the next subspace in the set without exceeding a total of $M$ independent vectors in the set, then we need to split the next subspace into two parts. This is done by taking the intersection of the next subspace in the sequence with the space spanned by the basis vectors in the current set to form the intersecting space. The intersecting part is separated out as the remainder of the subspace, and the non-intersecting part is incorporated into the set to complete the desired $M$ independent vectors. The process then continues with the remaining subspaces, starting with the remainder of the most recently split subspace. The process is terminated when we run out of basis vectors. The number of complete sets (sets of $M$ linearly independent basis vectors) that are generated through this process is denoted as $L_\Sigma$. The remaining basis vectors are discarded  if they are insufficient to create another complete basis.

%The whole process is terminated after we collect  $L_\Sigma$ sets, each with $M$ basis vectors. Finally, we throw away the rest equations.

We now proceed to the statement of the properties on vector subspaces.

%Then we throw away the rest
%$l^*-\lfloor\frac{l^*}{M}\rfloor M$
%equations whose entropy is no less than the noise contained thus its normalization by $n\log \rho$ is
%non-negative, completing the proof. \hfill\QED
\begin{lemma}\label{lemma:multiple_looks}
The following bound on the entropy holds:
\begin{eqnarray}
\sum_{k=1}^{K}\hbar\left({\bf L}^{[k]^n} \!\circ\! {\bf X}^n\right)\geq:
L_\Sigma \hbar({\bf X}^n).\label{eqn:multiple_looks}
\end{eqnarray}
\end{lemma}

%{\color{blue} Explanation to Prof in case it is needed. Note that here I emphasize that the matrices are created sequentially from the vectors. We cannot pick the vectors in arbitrary order as this does not match the direction of chain rule.}

Intuitively, Lemma \ref{lemma:multiple_looks} implies that when a collection of $l^*$ linear combinations of the $M$ variables
comprising ${\bf X}$ can reconstruct ${\bf X}$ $L_\Sigma$ times, the equations must carry at least their proportional share of the total entropy of ${\bf X}$. Note that when the subspaces are generic, $L_\Sigma = \big\lfloor \frac{l^*}{M}\big\rfloor $.

We start with two simple cases, and then present the general proof.

{\bf Case 1:} $K=1,~l_1=M$.%and the dimension of ${\bf L}^{[1]}(t)$ is $M$.

In this case, $l^*=l_1=M$ and $L_{\Sigma}$ = 1. Property (\emph{P1}) of Lemma \ref{lemma:property} gives us
\begin{eqnarray}
\hbar\left({\bf L}^{[1]^n}\!\circ\! {\bf X}^n\right) =:  \hbar\left( \mathbb{C}^{M^n} \!\circ\!\mathbf{X}^n\right) = \hbar(\mathbf{X}^n)
\end{eqnarray}
which implies (\ref{eqn:multiple_looks}).
%in
%Lemma \ref{lemma:multiple_looks} becomes
%\begin{eqnarray}
%\hbar({\bf L}^{[1]^n}\!\circ\! {\bf X}^n) \geq:  \hbar(\mathbf{X}^n)
%\end{eqnarray}
%which follows directly from Property (\emph{P2}) of
%Lemma \ref{lemma:property} and essentially, $\geq:$ can be replaced by $=:$.

{\bf Case 2:} $K=M+1,,~M>1,~l_k=1,~\forall k \in \{1,2,\ldots,K\}$ and all the subspaces are generic.

In this case, we want to create $L_{\Sigma} = 1$  set with $M$ basis vectors with $M+1$ generic vectors. %Each ${\bf L}^{[k]}(t)$ is a one-dimensional subspace,
Then we have
\begin{eqnarray}
\sum_{k=1}^{K} \hbar\left({\bf L}^{[k]^n} \!\circ\! {\bf X}^n\right) &\!\!\!\!=\!\!\!\!& \hbar\left({\bf L}^{[1]^n}\!\circ\! {\bf X}^n\right)+\cdots+\hbar\left({\bf L}^{[M]^n}\!\circ\! {\bf X}^n\right)+\hbar\left({\bf L}^{[M+1]^n}\!\circ\! {\bf X}^n\right)\\
&\!\!\!\!\geq\!\!\!\!& \hbar\left({\bf L}^{[1]^n}\!\circ\! {\bf X}^n,\ldots,{\bf L}^{[M]^n}\!\circ\! {\bf X}^n\right)+\hbar\left({\bf L}^{[M+1]^n}\!\circ\! {\bf X}^n\right) \label{eqn:lemma_multilook_c1}\\
&\!\!\!\!=\!\!\!\!& \hbar\left(\{{\bf L}^{[1]^n},\ldots,{\bf L}^{[M]^n}\}\!\circ\! {\bf X}^n\right)+\underbrace{\hbar\left({\bf L}^{[M+1]^n}\!\circ\! {\bf X}^n\right)}_{\geq: 0} \label{xz}\\
&\!\!\!\!\geq:\!\!\!\!& \hbar({\bf X}^n)%=\big\lfloor \frac{M+1}{M}\big\rfloor \hbar({\bf X}^n)
\label{eqn:lemma_multilook_c2}
\end{eqnarray}
where %(\ref{eqn:lemma_multilook_c1}) follows from chain rule and
(\ref{eqn:lemma_multilook_c2}) follows from Case 1, i.e., the space spanned by the union of $M$ generic vectors,
$\{{\bf L}^{[1]},\ldots,{\bf L}^{[M]}\}$, is the
$M$-dimensional vector space $\mathbb{C}^{M}$. Note that the second term of (\ref{xz}) contains no less differential entropy than the noise therein and the differential entropy of noise normalized by $n\log(\rho)$ is non-negative.
%Case 3: $M=3,K=3,l_k=2, \forall k \in \{1,2,3\}$.

Now we present the proof for the general setting of
Lemma \ref{lemma:multiple_looks}.

{\it Proof:} The collection of vectors and the splitting of the subspaces are consistent with the
chain rule of entropy, so that the same direction of inequalities is
obtained. In the end, we have collected  $L_{\Sigma}$ sets, each with $M$ basis vectors and the projection of ${\bf X}^n$ to the space spanned by the vectors in each set would contribute
entropy $\hbar({\bf X}^n)$, which is guaranteed by Property ($P1$) of Lemma 2.
Finally, the entropy of the discarded equations is no less than the entropy of the noise contained thus its normalization by $n\log \rho$ is
non-negative. This completes the proof. \hfill\QED

We illustrate this lemma with the following example.

{\bf Example:} $M=3,~K=6,~(l_1,l_2,l_3,l_4,l_5,l_6)=(1,2,1,1,3,2)$.
We assume $n=1$ and the subspace are given by:
\begin{subequations}
\begin{eqnarray}
{\bf L}^{[1]}&\!\!\!\!=\!\!\!\!& \mbox{span}\{[1~~1~~ 1]^T\},\\%\{X_1+X_2+X_3\},\\
{\bf L}^{[2]}&\!\!\!\!=\!\!\!\!& \mbox{span}\{[0~~2~~3]^T , [0~~1~~-1]^T\},\\%\{X_2+2X_3,~X_2-X_3\},\\
{\bf L}^{[3]}&\!\!\!\!=\!\!\!\!& \mbox{span}\{[1~~-1~~0]^T\},\\%{X_1-X_2\},\\
{\bf L}^{[4]}&\!\!\!\!=\!\!\!\!& \mbox{span}\{[1~~0~~1]^T\},\\%\{X_1+X_3\},\\
{\bf L}^{[5]}&\!\!\!\!=\!\!\!\!& \mbox{span}\{[1~~-1~~3]^T,[1~~0~~0]^T,[0~~1~~0]^T\},\\%\{X_1-X_2+3X_3,~X_1,~X_2\},\\
{\bf L}^{[6]}&\!\!\!\!=\!\!\!\!& \mbox{span}\{[0~~0~~1]^T,[1~~2~~-4]^T\}.%\{X_3,~X_1+2X_2-4X_3\}.
\end{eqnarray}
\end{subequations}
%, and from these $M=3$ equations we can resolve the three symbols comprising of ${\bf X}$ subject to the noise distortion
%The spaces chosen is not fully generic, but they satisfy certain linear independency requirements needed for the proof.
Note that $l^*=\sum_{k=1}^6 l_k=10$. It turns out that we can build 3 sets whose vector-elements collectively build full rank matrices.
%, so we need to build $\lfloor \frac{10}{3} \rfloor=3$ sets.
We start the proof by building the first set up to $\{\mathbf{L}^{[1]},\mathbf{L}^{[2]}\}$. At this stage we have collected exactly $l_1+l_2=3=M$ independent vectors. So we terminate this set and start building the next set. Now we can go up to $\{\mathbf{L}^{[3]},\mathbf{L}^{[4]}\}$ which contains $l_3+l_4=2$ independent vectors, i.e., short of $M=3$, but we cannot include
$\mathbf{L}^{[5]}$ entirely because $l_3+l_4+l_5=5$ will exceed $M=3$. So we will split $\mathbf{L}^{[5]}$ into a part, $\mathbf{L}^{[5]}_a$ that overlaps with $\{\mathbf{L}^{[3]},\mathbf{L}^{[4]}\}$ and the remainder that does not overlap with $\{\mathbf{L}^{[3]},\mathbf{L}^{[4]}\}$. Specifically,
\begin{eqnarray}
|\mathbf{L}^{[5]}| &\!\!\!\!=\!\!\!\!& l_5 = 3, \\
|\{\mathbf{L}^{[3]},\mathbf{L}^{[4]}\}| &\!\!\!\!=\!\!\!\!& l_3+l_4=2,\\
M &\!\!\!\!=\!\!\!\!& 3, \\
|\mathbf{L}^{[5]}_a| = |\mathbf{L}^{[5]} \cap \{\mathbf{L}^{[3]},\mathbf{L}^{[4]}\}| &\!\!\!\!=\!\!\!\!& 3+2-M = 2, \\
|\mathbf{L}^{[5]}\setminus\mathbf{L}^{[5]}_a| &\!\!\!\!=\!\!\!\!& 3-2 = 1, \\
|\{\mathbf{L}^{[3]},\mathbf{L}^{[4]},\mathbf{L}^{[5]}\setminus \mathbf{L}^{[5]}_a\}| &\!\!\!\!=\!\!\!\!& M, \\
\mathbf{L}^{[5]} &\!\!\!\!=\!\!\!\!& \{\mathbf{L}^{[5]}_a,\mathbf{L}^{[5]}\!\setminus\!\mathbf{L}^{[5]}_a\}.
%\hbar(\mathbf{L}^{[5]}) &\!\!=\!\!&
%\hbar(\mathbf{L}^{[5]}_a,(\mathbf{L}^{[5]}\!\setminus\!\mathbf{L}^{[5]}_a)).
\end{eqnarray}
Thus, we obtain
\begin{eqnarray}
\mathbf{L}^{[5]}_a&\!\!\!\!=\!\!\!\!&{\rm
span}\left\{\left[\begin{array}{rr}1&1\\-1&0\\0&1\end{array}\right]\right\}\cap{\rm
span}\left\{\left[\begin{array}{rrr}1&1&0\\-1&0&1\\3&0&0\end{array}\right]\right\}={\rm
span}\left\{\left[\begin{array}{rr}1&1\\-1&0\\0&1\end{array}\right]\right\},\\
\mathbf{L}^{[5]}\!\setminus\!\mathbf{L}^{[5]}_a &\!\!\!\!=\!\!\!\!& {\rm
span}\{[1~~1~~-1]^T\}.
\end{eqnarray}
%can resolve the three symbols comprising ${\bf X}$ again
%We observe that
%$\mathbf{L}^{[5]}_a$ is still generic and can be included into the third set.
The union of these three vectors spans $\mathbb{C}^3$, i.e., $\{ {\bf L}^{[3]}, {\bf L}^{[4]}, \mathbf{L}^{[5]}\!\setminus\!\mathbf{L}^{[5]}_a \} = \mathbb{C}^3$. Thus, our second set becomes $\{\mathbf{L}^{[3]},\mathbf{L}^{[4]},\mathbf{L}^{[5]} \!\setminus\! \mathbf{L}^{[5]}_a\}$
which contains $M=3$ linearly independent vectors. Finally, we start to build the third set starting with $\mathbf{L}^{[5]}_a$ and
continuing on to $\mathbf{L}^{[6]}$. Again, since $|\mathbf{L}^{[5]}_a|<3$ and $|\mathbf{L}^{[5]}_a| +l_6>3$, we need to split $\mathbf{L}^{[6]}$ into two parts, one $\mathbf{L}^{[6]}_a$ that overlaps with $\mathbf{L}_a^{[5]}$ and the other that does not
overlap with $\mathbf{L}_a^{[5]}$. Therefore, our final set becomes $\{\mathbf{L}^{[5]}_a,\mathbf{L}^{[6]}\!\setminus\!\mathbf{L}^{[6]}_a\}$
with dimension $M=3$, where $\mathbf{L}^{[6]}_a = \mathbf{L}^{[6]}\cap \mathbf{L}^{[5]}_a$ has dimension $2+2-M=1$, and is given by
\begin{eqnarray}
\mathbf{L}^{[6]}_a&\!\!\!\!=\!\!\!\!&{\rm
span}\left\{\left[\begin{array}{rr}1&1\\-1&0\\0&1\end{array}\right]\right\}\cap{\rm
span}\left\{\left[\begin{array}{rr}0&1\\0&2\\1&-4\end{array}\right]\right\}={\rm
span}\left\{\left[\begin{array}{r}1\\2\\3\end{array}\right]\right\},\\
\mathbf{L}^{[6]}\!\setminus\!\mathbf{L}^{[6]}_a &\!\!\!\!=\!\!\!\!& {\rm
span}\{[3~~6~~-5]^T\}.
\end{eqnarray}
This construction can be translated to the following information
theoretical proof:
\begin{eqnarray}
&& \sum_{k=1}^6 \hbar(\mathbf{L}^{[k]}  \!\circ\! {\bf X})\\
&\!\!\!\!\geq\!\!\!\!& \hbar(\mathbf{L}^{[1]} \!\circ\! {\bf X},\mathbf{L}^{[2]} \!\circ\! {\bf X}) + \sum_{k=3}^6 \hbar(\mathbf{L}^{[k]} \!\circ\! {\bf X})\\
&\!\!\!\!=:\!\!\!\!& \hbar({\bf X}) + \sum_{k=3}^6 \hbar({\bf L}^{[k]}  \!\circ\! {\bf X}) \label{eqn:multilook_1}\\
&\!\!\!\!\geq\!\!\!\!& \hbar({\bf X}) + \hbar({\bf L}^{[3]}  \!\circ\! {\bf X},{\bf L}^{[4]} \!\circ\! {\bf X}) + \hbar({\bf L}^{[5]} \!\circ\! {\bf X}) + \hbar({\bf L}^{[6]} \!\circ\! {\bf X})\\
&\!\!\!\!=\!\!\!\!& \hbar({\bf X}) + \hbar({\bf L}^{[3]} \!\circ\! {\bf X},{\bf L}^{[4]} \!\circ\! {\bf X}) + \hbar({\bf L}_a^{[5]} \!\circ\! {\bf X}, ({\bf L}^{[5]}\!\setminus\!{\bf L}_a^{[5]}) \!\circ\! {\bf X}) + \hbar({\bf L}^{[6]} \!\circ\! {\bf X})\\
&\!\!\!\!=\!\!\!\!& \hbar({\bf X}) + \hbar({\bf L}^{[3]} \!\circ\! {\bf X},{\bf L}^{[4]} \!\circ\! {\bf X})+\hbar({\bf L}_a^{[5]} \!\circ\! {\bf X}) + \hbar( ({\bf L}^{[5]}\!\setminus{\bf L}_a^{[5]}) \!\circ\! {\bf X}|{\bf L}_a^{[5]} \!\circ\! {\bf X})+\hbar({\bf L}^{[6]} \!\circ\! {\bf X})\\
&\!\!\!\!\geq\!\!\!\!& \hbar({\bf X}) + \hbar({\bf L}^{[3]} \!\circ\! {\bf X},{\bf L}^{[4]} \!\circ\! {\bf X})+\hbar({\bf L}_a^{[5]} \!\circ\! {\bf X}) + \hbar(({\bf L}^{[5]}\!\setminus\!{\bf L}_a^{[5]}) \!\circ\! {\bf X}|{\bf L}_a^{[5]} \!\circ\! {\bf X},{\bf L}^{[3]} \!\circ\! {\bf X},{\bf L}^{[4]} \!\circ\! {\bf X})\notag\\
&&+~\hbar({\bf L}^{[6]} \!\circ\! {\bf X})\label{eqn:multilook_eg_cond1}\\
&\!\!\!\!=\!\!\!\!& \hbar({\bf X}) + \hbar({\bf L}^{[3]} \!\circ\! {\bf X},{\bf L}^{[4]}  \!\circ\! {\bf X})+\hbar({\bf L}_a^{[5]}  \!\circ\! {\bf X}) + \hbar(({\bf L}^{[5]}\!\setminus\!{\bf L}_a^{[5]}) \!\circ\! {\bf X}|{\bf L}^{[3]}  \!\circ\! {\bf X},{\bf L}^{[4]} \!\circ\! {\bf X})\notag\\
&&+~\hbar({\bf L}^{[6]} \!\circ\! {\bf X})\label{eqn:multilook_eg_cond2}\\
&\!\!\!\!=\!\!\!\!& \hbar({\bf X}) + \hbar(\{{\bf L}^{[3]},{\bf L}^{[4]},{\bf L}^{[5]}\!\setminus\!{\bf L}_a^{[5]}\}  \!\circ\! {\bf X}) + \hbar({\bf L}_a^{[5]} \!\circ\! {\bf X}) + \hbar({\bf L}^{[6]} \!\circ\! {\bf X})\\
&\!\!\!\!=:\!\!\!\!& 2\hbar({\bf X}) + \hbar({\bf L}_a^{[5]}  \!\circ\! {\bf X}) + \hbar({\bf L}^{[6]}  \!\circ\! {\bf X}) \label{eqn:multilook_2}\\
&\!\!\!\!=\!\!\!\!& 2\hbar({\bf X}) + \hbar({\bf L}_a^{[5]} \!\circ\! {\bf X}) + \hbar({\bf L}_a^{[6]} \!\circ\! {\bf X},({\bf L}^{[6]}\!\setminus\!{\bf L}_a^{[6]}) \!\circ\! {\bf X})\\
&\!\!\!\!\geq\!\!\!\!& 2\hbar({\bf X}) + \hbar({\bf L}_a^{[5]} \!\circ\! {\bf X}) + \hbar(({\bf L}^{[6]}\!\setminus\!{\bf L}_a^{[6]}) \!\circ\! {\bf X}|{\bf L}_a^{[6]} \!\circ\! {\bf X},{\bf L}_a^{[5]} \!\circ\! {\bf X}) + \hbar({\bf L}_a^{[6]} \!\circ\! {\bf X})\\
&\!\!\!\!=\!\!\!\!& 2\hbar({\bf X}) + \hbar({\bf L}_a^{[5]} \!\circ\! {\bf X}) + \hbar(({\bf L}^{[6]}\!\setminus\!{\bf L}_a^{[6]}) \!\circ\! {\bf X}|{\bf L}_a^{[5]} \!\circ\! {\bf X}) + \hbar({\bf L}_a^{[6]} \!\circ\! {\bf X})\\
&\!\!\!\!=\!\!\!\!& 2\hbar({\bf X}) + \hbar(\{ {\bf L}_a^{[5]},{\bf L}^{[6]}\!\setminus\!{\bf L}_a^{[6]} \} \!\circ\! {\bf X}) + \hbar({\bf L}_a^{[6]} \!\circ\! {\bf X})\\
&\!\!\!\!\geq:\!\!\!\!& 3\hbar({\bf X}) \label{eqn:multilook_3}
\end{eqnarray}
where (\ref{eqn:multilook_eg_cond1}) follows from the property that adding conditioning terms does not increase the entropy;
(\ref{eqn:multilook_eg_cond2}) is obtained because ${\bf L}_a^{[6]}$ is the intersection of ${\bf L}^{[5]}$ and $\{{\bf L}^{[3]},{\bf
L}^{[4]}\}$, thus it is also contained in $\{{\bf L}^{[3]},{\bf L}^{[4]}\}$. This means ${\bf L}_a^{[5]} \!\circ\! {\bf X}$ can be reconstructed from $\{{\bf L}^{[3]} \!\circ\! {\bf X},{\bf L}^{[4]} \!\circ\! {\bf X}\}$, within bounded noise distortion. In (\ref{eqn:multilook_1}), (\ref{eqn:multilook_2}) and (\ref{eqn:multilook_3}), we use the fact that the construction produces each set with $M$ independent vectors and the argument that their differential entropy is no less than $\hbar({\bf X})$ follows from Case 1. Thus we have the desired result. Note that the derivations from (\ref{eqn:multilook_2}) to (\ref{eqn:multilook_3}) also follow from Property (\emph{P3}) of Lemma 2.
\hfill\QED

When each subspace ${\bf L}^{[k]}(t)$ is generic, we have the following corollary.
%The following useful corollaries directly follow from Lemma \ref{lemma:multiple_looks}.
\begin{corollary} For generic subspaces ${\bf L}^{[k]}(t)$ of $\mathbb{C}^M$, we have
\begin{eqnarray}
\sum_{k=1}^{K}\hbar\left({\bf L}^{[k]^n}\!\circ\! {\bf X}^n\right) \geq: %\frac{N}{\gcd(N,M)}
\lfloor \frac{l^*}{M} \rfloor \hbar({\bf X}^n).
\end{eqnarray}
\end{corollary}
{\it Proof:} According to Lemma \ref{lemma:multiple_looks}, with $l^* = \sum_{k=1}^K l_k$ generic vectors, we can build $\lfloor \frac{l^*}{M} \rfloor$ sets, each with $M$ basis vectors.
\hfill\QED

In case that one may be also interested in Lemma \ref{lemma:multiple_looks} with conditional terms, we have the following corollary.
\begin{corollary}
For an arbitrary random variable $Q$, we have
\begin{eqnarray}
\sum_{k=1}^{K}\hbar\left({\bf L}^{[k]^n} \!\circ\! {\bf X}^n | Q\right)\geq: L_\Sigma \hbar({\bf X}^n|Q).
\end{eqnarray}
\end{corollary}
{\it Proof:} The proof follows along the same lines as Lemma \ref{lemma:multiple_looks} and thus we omit it here. \hfill\QED

\section{Four Ideas Comprising the Genie Chains Approach}\label{sec:4ideas}

To keep the presentation complete and as more intuitive as possible, we need additional terminologies, particularly the notion of the \emph{exposed subspace}, some notations for the \emph{generic subspace} and the \emph{interference subspace} available to a RX after decoding and removing the signal carrying its desired message. Here, we remind the reader of that although they are called subspaces for convenience, in fact they represent the linear combinations of signals projected to those subspaces.
%While the bounded variance AWGN terms are always present in any received signal, since they are inconsequential for DoF arguments, they are omitted for simplicity of exposition.
%And all the equations are noisy, although the bounded variance noise term is not shown for simplicity.}

\begin{itemize}
\item {\bf Exposed Subspace:} An exposed subspace, e.g., from TX 1 to RX 2, denoted as ${\bf \bar{X}}^{[1\sim 2]}$, refers to the
linear combinations involving only ${\bf X}^{[1]}$ variables that are obtained at RX 2 after subtracting the signal carrying its desired message (for RX 2 this would be ${\bf X}^{[2]}$) and zero forcing (i.e., projecting into the null space, or simply using
Gaussian elimination to remove) the other interference (in this case ${\bf X}^{[3]}, {\bf X}^{[4]}$). For example, consider the exposed subspace ${\bf \bar{X}}^{[1\sim 2]}$ in the $(M_T,M_R)=(2,5)$ setting. At RX 2, after removing desired signal ${\bf X}^{[2]}$, we have 5 equations involving 6 variables ${\bf X}^{[1]}, {\bf X}^{[3]}, {\bf X}^{[4]}$ (Since $M_T=2$, each ${\bf X}^{[k]}$ represents $2$ variables). Eliminating 4 variables, ${\bf X}^{[3]}, {\bf X}^{[4]}$, leaves only one equation involving the two variables ${\bf X}^{[1]}$. This remaining linear combination, involving ${\bf X}^{[1]}$ only is the exposed subspace at RX 2 from TX 1. The dimensionality of the exposed space is indicated with a subscript, e.g., ${\bf \bar{X}}^{[1\sim 2]}_1$ in this
example. As the AWGN terms are always presented in the received signal, the exposed subspace is always noisy. When we refer to the noise-free exposed subspace, we omit the bar notation on the top, e.g., ${\bf {X}}^{[1\sim 2]}$ is the noise-free version of ${\bf \bar{X}}^{[1\sim 2]}$.

\item {\bf Generic Subspace:}  We use ${\bf X}^{[k]}_{(m)}$ to denote $m$ generic linear combinations of
the $M$ variables in ${\bf X}^{[k]}$. When the linear combinations are added with bounded variance independent noise, we denote them as ${\bf \bar{X}}^{[k]}_{(m)}$.

\item {\bf Interference Subspace:} We use the notation ${\bf \bar{S}}^{[k]}$ to
refer to the received signal at RX $k$, after the
\emph{desired} variables ${\bf X}^{[k]}$ are set to zero. This is
meaningful because the RX is always guaranteed to be able to
reliably decode, and therefore subtract out, its desired signals,
leaving it with a view of only the interference subspace from which
it may attempt to resolve undesired signal dimensions. Similarly, the noise-free interference space is
denoted as ${\bf {S}}^{[k]}$.
\end{itemize}
Note that the extension of these subspaces from the single-letter to the multi-letter form is immediate.%, i.e., combine the corresponding subspaces got over each channel use.

\subsection{Ideas Illustrated through the 4-User MIMO Interference Channel}

The starting point of our outer bound is the common principle of providing a RX enough additional linear combinations of transmitted symbols to allow it to resolve all of the interferers, so that subject to the noise distortion (which is inconsequential for DoF), it can decode all the messages. In general, because we are proving a converse, which means that we start with a reliable coding scheme,
a RX is already guaranteed to reliably decode its desired message, which also allows the RX to subtract its desired symbols from its received signal. Now, the question remains whether the RX can decode \emph{all} messages. For the $K=4$ user MIMO interference channel, with $M_R$ receive antennas, if $3M_T>M_R$, then we have fewer equations and more unknowns, so that resolution of interfering symbols is not guaranteed. So, we provide $3M_T-M_R$ genie dimensions, i.e., $|{\bf \bar{G}}|=3M_T-M_R$ linearly independent combinations of interference symbols where ${\bf \bar{G}}$ represents the genie symbols set. ${\bf G}$ denotes the noise-free version of ${\bf \bar{G}}$. This provides the RX enough equations to resolve all transmitted symbols. Equivalently, the undesired signal vectors
${\bf X}^{[i]},~i\!\in\!\mathcal{K}\!\setminus\!\{k\}$ are now invertible (within noise distortion) from the RX's own observations combined with the genie dimensions. Since noise distortion is irrelevant for DoF arguments, the ability to resolve all symbols is equivalent to the ability to decode all symbols for DoF purposes. This forms the general basis for the outer bound, and is so far not a novel concept at all.

The challenging aspect, and where the novelty of our approach comes in, is to determine \emph{which genie dimensions} to provide so that
a useful DoF outer bound results. We propose a series of steps where we continue to cycle through various RXs in a chain of genie
aided outer bounds containing entropies of various subspace equations introduced above, following four basic principles, that lead us to a cancelation of successive entropy terms, producing the desired outer bound. The four basic principles of the ``genie chains" approach are highlighted next through simple examples.

\begin{idea}
{\bf Use the exposed space from one RX as a genie for the next.}
\end{idea}

{\bf Example 1:} $(M_T,M_R)=(2,5)\Rightarrow d\leq 10/7$

In this example, $|{\bf \bar{G}}|=3M_T\!-\!M_R=1$, so we need to provide a
one-dimensional genie. Suppose we start with the generic subspace
${\bf \bar{G}}_1={\bf \bar{X}}^{[1]}_{(1)}$ and give it as  genie to RX 2. Since
this genie allows RX 2 to decode all the messages subject to the
noise distortion, we have
\begin{eqnarray}
&& n(R_1\!+\!R_2\!+\!R_3\!+\!R_4)-n\epsilon_n\notag\\
&\!\!\!\!\leq:\!\!\!\!& I(W_1,W_2,W_3,W_4; {\bf \bar{Y}}^{[2]^n},{\bf \bar{G}}_1^n)\label{eqn:2by5_fano}\\
&\!\!\!\!=\!\!\!\!& \hbar({\bf Y}^{[2]^n},{\bf X}^{[1]^n}_{(1)})-\hbar({\bf Y}^{[2]^n},{\bf X}^{[1]^n}_{(1)}|W_1,W_2,W_3,W_4)\\
&\!\!\!\!=:\!\!\!\!& \hbar({\bf Y}^{[2]^n})+\hbar({\bf X}^{[1]^n}_{(1)}|{\bf Y}^{[2]^n},W_2)\label{eqn:2by5_dec_desire}\\
&\!\!\!\!=\!\!\!\!& \hbar({\bf Y}^{[2]^n})+\hbar({\bf X}^{[1]^n}_{(1)}|{\bf S}^{[2]^n})\label{eqn:2by5_inf_space}\\
&\!\!\!\!=\!\!\!\!& \hbar({\bf Y}^{[2]^n})+\hbar({\bf X}^{[1]^n}_{(1)}|[{\bf H}^{[23]^n}~{\bf H}^{[24]^n}]^T{\bf S}^{[2]^n},([{\bf H}^{[23]^n}~{\bf H}^{[24]^n}]^c)^T{\bf S}^{[2]^n})\label{eqn:2by5_zf}\\
&\!\!\!\!\leq \!\!\!\!& \hbar({\bf Y}^{[2]^n})+\hbar({\bf X}^{[1]^n}_{(1)}|([{\bf H}^{[23]^n}~{\bf H}^{[24]^n}]^c)^T{\bf S}^{[2]^n})\\
&\!\!\!\!\leq \!\!\!\!& \hbar({\bf Y}^{[2]^n})+\hbar({\bf X}^{[1]^n}_{(1)}|{\bf X}^{[1\sim 2]^n}_1)\label{eqn:2by5_remove_constraint}\\
&\!\!\!\!= \!\!\!\!& \hbar({\bf Y}^{[2]^n})+\hbar({\bf X}^{[1]^n}_{(1)},{\bf X}^{[1\sim 2]^n}_1)-\hbar({\bf X}^{[1\sim 2]^n}_1)\\
&\!\!\!\!\leq: \!\!\!\!& 5n\log(\rho) + nR - \hbar({\bf X}^{[1\sim 2]^n}_1)\label{eqn:2by5_recover_R}.
\end{eqnarray}
In the derivations above, (\ref{eqn:2by5_fano}) follows from
Fano's inequality. (\ref{eqn:2by5_dec_desire}) is obtained because
from all the four messages, one can reconstruct the received
signal vector and the genie signal subject to bounded noise distortion.
Since RX 2 can decode its own message $W_2$, it can subtract the signal contributed by ${\bf X}^{[2]}$ from ${\bf \bar{Y}}^{[2]}$,
and then produce the interference space ${\bf \bar{S}}^{[2]}$. Note that the subtracted part is a linear function of ${\bf X}^{[2]}$ and thus is independent with ${\bf X}^{[1]}_{(1)}$, as shown in (\ref{eqn:2by5_inf_space}). (\ref{eqn:2by5_zf}) follows from the fact that we can separate the space ${\bf \bar{S}}^{[2]}$ into two orthogonal subspaces, a one-dimensional projection that is orthogonal to the channels from TX 3 and TX 4 to RX 2, and  the remaining four-dimensional subspace. Thus, RX 2
%zero force the first subspace, and
obtains the one-dimensional exposed subspace ${\bf \bar{X}}_1^{[1\sim2]}$ in (\ref{eqn:2by5_remove_constraint}). Finally, the first term in (\ref{eqn:2by5_recover_R}) follows from Property (\emph{P2}) of Lemma 2 and the second term  in
(\ref{eqn:2by5_recover_R}) is obtained because of the fact that we can use the two dimensional
observations $\{{\bf \bar{X}}^{[1]^n}_{(1)},{\bf \bar{X}}^{[1\sim 2]^n}_1\}$ to
recover the transmitted signal vector ${\bf X}^{[1]}$ within noise distortion, thus
contributing the term $nR$ subject to the noise distortion, as proved in Property (\emph{P3a}) of Lemma 2.
Because of the symmetry of the problem, there is no loss of generality in focusing on symmetric rates $R_1=R_2=R_3=R_4=R$, which allows us to re-write (\ref{eqn:2by5_recover_R}) as follows,
%Therefore, we can rewrite the inequality above as:
\begin{eqnarray}
4nR-n \epsilon_n\leq: 5n\log(\rho) + nR - \hbar({{\bf X}^{[1\sim 2]}_1}^n).\label{eqn:eg_2by5_ob1_light}
\end{eqnarray}

Note that the exposed space has appeared as a negative entropy term.
As a rule, in our approach, the negative entropy terms will become
the genie signals for the subsequent bounds, leading to their
eventual cancellation. Also, we will attempt to obtain a total of $M_T$ useful bounds. In
this case, $M_T=2$, so we move to our final bound, and to the next
RX, RX 3. The genie, as just mentioned, will be in the
previous negative entropy term ${\bf \bar{G}}_2={\bf \bar{X}}^{[1\sim 2]}_1$.
As ${\bf \bar{G}}_2$ is the exposed subspace at RX 2, which is independent with the channels associated with RX 3, almost surely ${\bf \bar{G}}_2$ is independent with ${\bf \bar{Y}}_3$ such that RX 3 can now decode all the messages with this genie.
The resulting bound is the following
\begin{eqnarray}
n(R_1+R_2+R_3+R_4)-n\epsilon_n&\!\!\!\!\leq:\!\!\!\!& I(W_1,W_2,W_3,W_4; {{\bf \bar{Y}}^{[3]^n}}, {\bf \bar{G}}_2^n)\\
&\!\!\!\!\leq:\!\!\!\!& \hbar({{\bf Y}^{[3]}}^n, {\bf X}^{[1\sim 2]^n}_1)\\
&\!\!\!\!\leq \!\!\!\!& \hbar( {{\bf Y}^{[3]}}^n)+\hbar( {{\bf X}^{[1\sim 2]}_1}^n)\label{eqn:2by5_entropy_ineq}\\
\Longrightarrow 4nR-n\epsilon_n&\!\!\!\!\leq:\!\!\!\!& 5n\log(\rho) +
\hbar({{\bf X}^{[1\sim 2]}_1}^n)\label{eqn:eg_2by5_ob2_light}
\end{eqnarray}
where (\ref{eqn:2by5_entropy_ineq}) follows from chain rule and the fact that removing the conditional terms
does not decrease the differential entropy.

Adding up the two inequalities (\ref{eqn:eg_2by5_ob1_light}) and
(\ref{eqn:eg_2by5_ob2_light}), we obtain %the following inequality:
\begin{eqnarray}
8nR-n\epsilon_n\leq: 10n\log(\rho)+nR\Longrightarrow 7nR\leq
10n\log(\rho)+n~o(\log\rho)+n\epsilon_n.
\end{eqnarray}
By letting $n\rightarrow \infty$ first and then $\rho\rightarrow
\infty$, we obtain the desired outer bound on DoF per user:
\begin{eqnarray}
d\leq 10/7.
\end{eqnarray}
\hfill\QED

In order for the reader to have a more intuitive idea about the associated subspaces in each step, we provide an numerical example in the following.

\verb"Initialization:" $M_T=2$, $M_R=5$, randomly generate $5\times
2$ channel matrices from each TX to each RX. For
example, we randomly generate the following associated channel
realizations that are relevant in the proof:
\begin{small}
\begin{eqnarray*}
{\bf
H}^{[21]}=\left[\begin{array}{rr}0.5888&-0.3927\\1.0095&-1.5730\\-0.4297&-1.3400\\0.3536&0.4674\\-1.4046&0.6240\end{array}\right],~~~
{\bf H}^{[23]}=\left[\begin{array}{rr}
-2.4617&0.1171\\1.9378&1.5657\\0.8237&0.5253\\-0.8099&1.5186\\0.4344&-0.6581\end{array}\right],~~~
{\bf
H}^{[24]}=\left[\begin{array}{rr}-0.5819&-1.4890\\0.2349&0.1483\\-0.0988&0.9539\\-0.1352&2.2932\\-1.8865&-0.1452\end{array}\right],\\
{\bf
H}^{[31]}=\left[\begin{array}{rr}0.0720&-1.9399\\0.7140&2.4346\\1.2446&0.3470\\0.4961&-0.9756\\0.5580&0.4654\end{array}\right],~~~
{\bf H}^{[32]}=\left[\begin{array}{rr}
-0.0999&-0.9784\\-0.2805&-1.1571\\0.4136&-0.0548\\0.2967&1.1387\\1.1556&0.7722\end{array}\right],~~~
{\bf
H}^{[34]}=\left[\begin{array}{rr}0.6760&0.0171\\-0.8062&-0.3684\\0.0049&-0.3526\\0.8783&0.3086\\-0.9020&0.3290\end{array}\right].
\end{eqnarray*}
\end{small}

\verb"Step 1:" We randomly generate a vector ${\rm rand}(2,1)$ which captures the direction of the genie signal ${\bf \bar{G}}_1$. For
example, the vector is $[0.6109~~0.0712]^T$ and thus ${\bf
\bar{G}}_1=0.6109X_1^{[1]}+0.0712X_2^{[1]} + Z_1$, where $Z_1$ is an independent noise with bounded variance. Then by zero forcing the
interference from TX 3 and TX 4, RX 2
obtains a one-dimensional observation, say $\mathcal{O}_2$, of the transmitted signals from
TX 1. That is,
\begin{eqnarray*}
{\bf X}_{1}^{[1\sim2]} = \mathcal{O}_2 = {\bf X}^{[1]^T} \mathbf{H}^{[21]^T}([{\bf H}^{[23]}~~{\bf
H}^{[24]}])^c= 0.3227 X_1^{[1]} + 1.2639 X_2^{[1]},
\end{eqnarray*}
which is linearly independent with ${\bf G}_1$ because the two linear combinations are not collinear.%$\det([{\bf
%G}_1~\mathcal{O}_2])=0.7491\neq 0$.

%[0.3227~~1.2639]^T$ or \mathcal{O}_2=
\verb"Step 2:" We provide ${\bf
\bar{G}}_2= \mathcal{O}_2 + Z_2 =0.3227X_1^{[1]}+1.2639X_2^{[1]} + Z_2$ as genie to
RX 3, where $Z_2$ is another independent noise. Then by zero forcing the interference from TX 2
and TX 4, RX 3 also obtains a one-dimensional
observation, say $\mathcal{O}_3$, of the transmitted signals from TX 1. That is,
\begin{eqnarray*}
\mathcal{O}_3 = {\bf X}^{[1]^T} \mathbf{H}^{[31]^T}([{\bf H}^{[32]}~~{\bf
H}^{[34]}])^c= 0.7366 X_1^{[1]} + 1.0464 X_2^{[1]},
%[0.7366~~1.0464]^T,
\end{eqnarray*}
which is also linearly independent with ${\bf G}_2$ so that we can provide ${\bf \bar{G}}_2$ as genie to ensure RX 3 can decode all the messages. %because $\det([{\bf G}_2~\mathcal{O}_3])=-0.5933\neq 0$.
\hfill\QED

{\it Remark:} The $(M_T,M_R)=(2,5)$ example is perhaps a bit
serendipitous because the size of the exposed space exactly matches
the required size of the genie at the next RX. In general, the
two will not be the same, and we need to create either a bigger or a
smaller genie. How to achieve a larger or smaller genie is the
subject of the remaining three ideas.

{\it Remark:} Notice that for each sum rate inequality, we always
start from Fano's inequality by providing enough dimensional
genie signals ${\bf \bar{G}}$ to RX $k$ such that it can decode all the
messages subject to the noise distortion. RX $k$ can
obtain the interference space ${\bf \bar{S}}^{[k]}$ by decoding $W_k$
first and then subtracting out the signal carrying $W_k$ from the
observations ${\bf \bar{Y}}^{[k]}$. Thus, we will always omit the
derivations from (\ref{eqn:2by5_fano}) to (\ref{eqn:2by5_inf_space}), and start directly from
\begin{eqnarray*}
KnR-n\epsilon_n\leq: \hbar({\bf Y}^{[k]^n})+\hbar({\bf G}^n|{\bf S}^{[k]^n})
\end{eqnarray*}
in the remainder of this paper.

\begin{idea}
{\bf Obtain a larger genie by exposing more dimensions.}
\end{idea}

{\bf Example 2:} $(M_T,M_R)=(3,7)\Rightarrow d\leq 21/10$

In this example, $|{\bf \bar{G}}|=3M_T\!-\!M_R=2$ so we need a
two-dimensional genie. However, $M_R-2M_T=1$, so the exposed space,
e.g., ${\bf \bar{X}}^{[1\sim 2]}$ is only one-dimensional. Similar to
Example 1, we start with a generic genie ${\bf \bar{G}}_1={\bf \bar{X}}^{[1]}_{(2)}$ at RX 2, which is linearly independent
with  ${\bf \bar{S}}^{[2]}$. Thus, RX 2 can decode all the messages and we have
%\vspace{-0.05in}
\begin{eqnarray}
4nR-n\epsilon_n &\!\!\!\!\leq:\!\!\!\!& \hbar({\bf Y}^{[2]^n})+\hbar({\bf G}_1^n|{{\bf S}^{[2]}}^n)\\
&\!\!\!\!\leq:\!\!\!\!& 7n\log(\rho)+\hbar({{\bf X}^{[1]}_{(2)}}^n|{{\bf S}^{[2]}}^n)\\
&\!\!\!\!\leq:\!\!\!\!& 7n\log(\rho)+nR-\hbar({{\bf X}^{[1\sim
2]}_1}^n).\label{eqn:eg_3by7_ob1_light}
\end{eqnarray}

For the next bound, we move to RX 3. We will use the genie
corresponding to the previous negative entropy term, ${\bf
\bar{X}}^{[1\sim 2]}_1$, but since this is only one-dimensional and we
need 2 genie dimensions, we will complement it with a generic
dimension from the next TX, ${\bf \bar{X}}^{[2]}_{(1)}$. That is,
${\bf \bar{G}}_2=\{{\bf \bar{X}}^{[1\sim 2]}_1,{\bf \bar{X}}^{[2]}_{(1)}\}$. The most
important element here is how a new dimension gets exposed. RX 3
originally has one exposed dimension from TX 2. However, when the genie provides ${\bf \bar{X}}^{[1\sim 2]}_1$,
it exposes one additional dimension from TX 2, so that the new
exposed space from TX 2 is denoted as ${\bf \bar{X}}^{[2\sim 3]}_2$. The
resulting bound is given by:
\begin{eqnarray}
4nR-n\epsilon_n&\!\!\!\!\leq:\!\!\!\!& \hbar({\bf Y}^{[3]^n})+\hbar({\bf G}_2^n|{{\bf S}^{[3]}}^n)\\
&\!\!\!\!\leq:\!\!\!\!& 7n\log(\rho)+\hbar({{\bf X}^{[1\sim 2]}_1}^n,{{\bf X}^{[2]}_{(1)}}^n|{{\bf S}^{[3]}}^n)\\
&\!\!\!\!=\!\!\!\!& 7n\log(\rho)+\hbar({{\bf X}^{[1\sim 2]}_1}^n)+\hbar({{\bf X}^{[2]}_{(1)}}^n|{{\bf S}^{[3]}}^n,{{\bf X}^{[1\sim 2]}_1}^n)\\
&\!\!\!\!\leq:\!\!\!\!& 7n\log(\rho)+\hbar({{\bf X}^{[1\sim 2]}_1}^n)+\hbar({{\bf X}^{[2]}_{(1)}}^n|{{\bf X}^{[2\sim 3]}_2}^n)\\
&\!\!\!\!\leq:\!\!\!\!& 7n\log(\rho)+\hbar({{\bf X}^{[1\sim 2]}_1}^n)+nR-\hbar({{\bf
X}^{[2\sim 3]}_2}^n).\label{eqn:eg_3by7_ob2_light}
\end{eqnarray}

Now, with the additional exposed dimension, the exposed space ${\bf \bar{X}}^{[2\sim 3]}_2$ is two-dimensional and matches the desired size of the genie. This gives us our third, and final, bound as we cyclically move on to the next RX, RX 4, with the genie ${\bf \bar{G}}_3={\bf \bar{X}}^{[2\sim 3]}_2$. Since the channel coefficients associated with RX 4 are generic, the one-dimensional observation available at RX 4 from TX 2 is linearly independent with ${\bf \bar{G}}_3$. Thus, RX 4 can decode all the messages subject to the noise distortion, and we have
\begin{eqnarray}
4nR-n\epsilon_n\leq: \hbar({\bf Y}^{[4]^n})+\hbar({\bf G}_3^n|{{\bf
S}^{[4]}}^n)\leq: 7n\log(\rho)+\hbar({{\bf X}^{[2\sim
3]}_2}^n).\label{eqn:eg_3by7_ob3_light}
\end{eqnarray}

Adding up the inequalities (\ref{eqn:eg_3by7_ob1_light}),
(\ref{eqn:eg_3by7_ob2_light}), (\ref{eqn:eg_3by7_ob3_light}), we
have %the following sum rate inequality:
\begin{eqnarray}
12nR-n\epsilon_n \leq 21n\log(\rho)+2nR+n~o(\log\rho)
\end{eqnarray}
which produces the desired outer bound
\begin{eqnarray}
d\leq 21/10.
\end{eqnarray}\hfill\QED

%Because the only condition to proceed to
%(\ref{eqn:eg_3by7_ob3_light}) is that ${\bf G}_3$ is light, we have
%the desired bound
%\begin{eqnarray}
%4d\leq N+2d/3\Rightarrow d\leq 3N/10=21/10.
%\end{eqnarray}

%\begin{remark}
%$|{\bf L}|<|{\bf G}|$ implies that we need to build up a larger
%genie to cancel out the entropy term of the observation we obtain
%until last inequality, thereby exposing new dimensions from the next
%transmitter to create new genies.
%\end{remark}

In the following we provide an {\em alternative} proof for this
example to shed light on the following idea.

\begin{idea}
{\bf Combine exposed subspaces from multiple RXs to create a
larger genie.}
\end{idea}

After obtaining (\ref{eqn:eg_3by7_ob1_light}) at RX 2,
similarly if a genie provides to RX 3 two random linear
combination of ${\bf X}^{[1]}$, i.e, ${\bf \bar{G}}'_2={\bf
\bar{X}}^{[1]'}_{(2)}$, we have another inequality at RX 3
\begin{eqnarray}
4nR-n\epsilon_n \leq:  7n\log(\rho)+nR-\hbar({{\bf X}^{[1\sim
3]}_1}^n)\label{eqn:eg2_3by7_ob2_light}
\end{eqnarray}
where ${\bf \bar{X}}^{[1\sim 3]}_1$ is the exposed one dimensional
observation available at RX 3 projecting from TX 1.
%
%Note that if any of ${\bf X}^{[1\sim 2]}_1,{\bf X}^{[1\sim 3]}_1$ is
%heavy, then we already obtained the desired bound. Thus, we assume
%they are {\em both} light, thereby $h({\bf X}^{[1\sim 2]}_1,{\bf
%X}^{[1\sim 3]}_1)\leq h({\bf X}^{[1\sim 2]}_1)+h({\bf X}^{[1\sim
%3]}_1)<2d/3$, i.e., $\{{\bf X}^{[1\sim 2]}_1,{\bf X}^{[1\sim
%3]}_1\}$ is light.

Finally, a genie provides ${\bf \bar{G}}'_3=\{{\bf \bar{X}}^{[1\sim 2]}_1,{\bf
\bar{X}}^{[1\sim 3]}_1\}$ to RX 4 where ${\bf \bar{G}}'_3$ is linearly
independent with the 7-dimensional ${\bf S}^{[4]^n}$ space. Thus, RX 4 can decode all the messages as well. So we have
\begin{eqnarray}
4nR-n\epsilon_n &\!\!\!\!\leq:\!\!\!\!& \hbar({\bf Y}^{[4]^n})+\hbar({\bf G}_3^{n'}|{\bf S}^{[4]^n})\\
&\!\!\!\!\leq:\!\!\!\!& 7n\log(\rho)+\hbar({{\bf X}^{[1\sim 2]}_1}^n)+\hbar({{\bf
X}^{[1\sim 3]}_1}^n).\label{eqn:eg2_3by7_ob3_light}
\end{eqnarray}

Adding (\ref{eqn:eg_3by7_ob1_light}), (\ref{eqn:eg2_3by7_ob2_light}) and (\ref{eqn:eg2_3by7_ob3_light}), we again obtain the desired
outer bound
\begin{eqnarray}
12nR-n\epsilon_n \leq: 21n\log(\rho)+2nR\Longrightarrow d\leq 21/10.
\end{eqnarray}
\hfill\QED

{\it Remark:} Idea 3 is especially useful in the $M_T>M_R$ settings.

\begin{idea}
{\bf Obtain a smaller size genie by intersections.}
\end{idea}

{\bf Example 3: $(M_T,M_R)=(3,8)\Rightarrow d\leq 24/11$}

Starting with a generic genie ${\bf \bar{G}}_1={\bf \bar{X}}^{[1]}_{(1)}$ at RX 2, we have the first inequality
\begin{eqnarray}
4nR-n\epsilon_n &\!\!\!\!\leq:\!\!\!\!& \hbar({\bf Y}^{[2]^n})+\hbar({\bf G}_1^n|{{\bf S}^{[2]}}^n)\\
&\!\!\!\!\leq:\!\!\!\!& 8n\log(\rho)+\hbar({{\bf X}^{[1]}_{(1)}}^n|{{\bf S}^{[2]}}^n)\\
&\!\!\!\!\leq:\!\!\!\!& 8n\log(\rho)+nR-\hbar({{\bf X}^{[1\sim
2]}_{2}}^n).\label{eqn:eg_3by8_ob1}
\end{eqnarray}

Now, the required size of the genie is $|{\bf \bar{G}}|=3M\!-\!N=1$ while
exposed spaces have size 2. How to create a smaller genie? We will
do that by creating multiple exposed spaces, each of which may be
too big to be an acceptable genie, but their intersection will turn
out to be an acceptable genie. A genie provides to RX 3
another random linear combination of ${\bf X}^{[1]}$, i.e., ${\bf
\bar{G}}_2={\bf \bar{X}}^{[1]'}_{(1)}$, so that
\begin{eqnarray}
4nR-n\epsilon_n \leq:  8n\log(\rho)+nR-\hbar({{\bf X}^{[1\sim
3]}_{2}}^n)\label{eqn:eg_3by8_ob2}
\end{eqnarray}
where $\mathcal{O}_3={\bf X}^{[1\sim 3]}_{2}$ is the two-dimensional exposed space of TX 1 at RX 3. Since the
construction of $\mathcal{O}_2={\bf X}^{[1\sim 2]}_2$ and
$\mathcal{O}_3$ only involve the channel coefficients associated
with their own RXs, they are generic and have $2+2-3=1$
dimensional intersection, denoted as
$\mathcal{I}=\mathcal{O}_2\cap\mathcal{O}_3$. Thus, we can rewrite
(\ref{eqn:eg_3by8_ob2}) as
\begin{eqnarray}
4nR-n\epsilon_n &\!\!\!\!\leq: \!\!\!\!&8n\log(\rho)+nR-\hbar({{\bf X}^{[1\sim 3]}_{2}}^n)-\hbar({{\bf X}^{[1\sim 2]}_{2}}^n)+\hbar({{\bf X}^{[1\sim 2]}_{2}}^n)\\
&\!\!\!\!=\!\!\!\!& 8n\log(\rho)+\hbar({{\bf X}^{[1\sim 2]}_{2}}^n)+nR-\hbar(\mathcal{O}_3^n\setminus\mathcal{I}^n,\mathcal{I}^n)-\hbar(\mathcal{O}_2^n)\\
&\!\!\!\!=\!\!\!\!& 8n\log(\rho)+\hbar({{\bf X}^{[1\sim 2]}_{2}}^n)+nR-\hbar(\mathcal{I}^n)-\hbar(\mathcal{O}_3^n\setminus\mathcal{I}^n|\mathcal{I}^n)-\hbar(\mathcal{O}_2^n)\\
&\!\!\!\!\leq \!\!\!\!& 8n\log(\rho)+\hbar({{\bf X}^{[1\sim 2]}_{2}}^n)+nR-\hbar(\mathcal{I}^n)-\hbar(\mathcal{O}_3^n\setminus\mathcal{I}^n|\mathcal{I}^n,\mathcal{O}_2^n)-\hbar(\mathcal{O}_2^n)\\
&\!\!\!\!=: \!\!\!\!& 8n\log(\rho)+\hbar({{\bf X}^{[1\sim 2]}_{2}}^n)+nR-\hbar(\mathcal{I}^n)-\hbar(\mathcal{O}_2^n)-\hbar(\mathcal{O}_3^n\setminus\mathcal{I}^n|\mathcal{O}_2^n)\label{eqn:eg_3by8_ob2_intsect}\\
&\!\!\!\!= \!\!\!\!& 8n\log(\rho)+\hbar({{\bf X}^{[1\sim 2]}_{2}}^n)+nR-\hbar(\mathcal{I}^n)-\hbar(\mathcal{O}_2^n,\mathcal{O}_3^n\setminus\mathcal{I}^n)\\
&\!\!\!\!=: \!\!\!\!& 8n\log(\rho)+\hbar({{\bf X}^{[1\sim
2]}_{2}}^n)-\hbar(\mathcal{I}^n)\label{eqn:eg_3by8_interbound}
\end{eqnarray}
where (\ref{eqn:eg_3by8_ob2_intsect}) is obtained because
$\mathcal{I}$ is included in $\mathcal{O}_2$, and
(\ref{eqn:eg_3by8_interbound}) follows from that
$\{\mathcal{O}_2,\mathcal{O}_3\setminus \mathcal{I}\}$ are three linear independent equations in ${\bf X}^{[1]}$ so that we can use Property (\emph{P3}) of Lemma 2.
%contributes $nR$ differential entropy subject to the noise distortion.
We call the inequality (\ref{eqn:eg_3by8_interbound}) the ``intermediate
bound" which is constructed by intersecting two subspaces at
different RXs. The derivations above are the same as that in Lemma 3.

Finally, we should provide the observations we obtain in the last
step as the genie to RX 4, i.e., ${\bf \bar{G}}_3=\mathcal{I} + Z$, where $Z$
is an independent noise. Since ${\bf \bar{G}}_3$ only involves the channel coefficients associated
with RX 2 and 3, it is linearly independent with the original
two dimensional observations from TX 1 at RX 4. Thus,
RX 4 can decode all the messages, and we have the last inequality
\begin{eqnarray}
4nR-n\epsilon_n \leq:\hbar({\bf Y}^{[4]^n})+\hbar({\bf G}_3^n|{{\bf
S}^{[4]}}^n)\leq
8n\log(\rho)+\hbar(\mathcal{I}^n).\label{eqn:eg_3by8_ob3}
\end{eqnarray}

Adding up the inequalities (\ref{eqn:eg_3by8_ob1}),
(\ref{eqn:eg_3by8_interbound}) and (\ref{eqn:eg_3by8_ob3}), we have
\begin{eqnarray}
12nR-n\epsilon_n\leq: 3NnR+nR\Longrightarrow d\leq 3N/11=24/11.
\end{eqnarray}
\hfill\QED

The three examples above show that our goal is to use a chain of
arguments, where we start with the exposed spaces and continue to
build new genies with more dimensions by peeling off overlaps, or
less dimensions by taking intersections, until we have the genie of
the correct size, which requires exactly $M_T$ steps, and produces
the bound $d\leq \frac{M_T M_R}{M_T+M_R}$, if all genies in this
process are acceptable, i.e., linearly independent of the space
already available to the RXs.

This is summarized in the following theorem in the context of $K$-user interference channel, which is the main result of this paper.
\begin{theorem}
For the $K$ user $M_T\times M_R$ MIMO interference channel
where each TX has $M_T$ and each RX has $M_R$
antennas, if we can create a genie chain with $M_T$ genie signal sets and each genie signal (with appropriate size) is linearly independent of the exposed subspace at each corresponding receiver, then the DoF value per user is given by $d=\frac{M_T M_R}{M_T+M_R}$.
%The decomposition bound is tight whenever the chain of M genie signals is linearly independent of previously
%exposed subspaces at each receiver.
\end{theorem}

Note that the genie chain technique is applicable to arbitrary
channel realizations, to the extent that the
genie signals remain linearly independent of
previously exposed spaces. Thus this
technique can be used to test arbitrary settings, although in this paper,
we focus exclusively on deriving DoF results that hold almost surely
for generic channels.

Now we have a general result for the $K$-user
MIMO interference channel, then building genie signals with appropriate sizes and testing the linear independence condition
are all that remain. This is the
problem that we will address for various cases,
and leave open for others.

Also, it should be noted that when the genie
signals start becoming linearly dependent,
one can terminate the chain by simply
replacing the entropy term by its maximum
signal dimension. The bound may be loose
but it is still the best bound we can get through
the genie chain approach, and likely better
than any other existing approach.

\section{Application: $K=4$ User MIMO Interference Channel}\label{sec:4user}

In this section, we apply the genie chains approach to investigate the DoF characterization for the $K=4$ user $M_T\times M_R$
MIMO interference channel. For brevity, let $M = \min(M_T,M_R)$ and $N = \max(M_T,M_R)$. Since the DoF results and corresponding proofs when $M/N\leq 3/8$  follow from the $K=3$ user case \cite{Wang_Gou_Jafar_3userMxN} but requires much more complicated analysis, we will consider this regime later in Section \ref{sec:big_picture}.1.
In this section, we only consider the setting $M_T/M_R>3/8$. The main result for this regime is presented in the theorem.

\begin{theorem}\label{theorem:4user}
{\it For the $K= 4$ user $M_T\times M_R$ MIMO interference channel
where each TX has $M_T$ and each RX has $M_R$
antennas, if $M_T/M_R>3/8$, then the DoF value per user is given by $d=\frac{M_T M_R}{M_T+M_R}$ for every $(M_T,M_R)$ where $\frac{M_T}{M_R}\in \mathcal{P}_1\cup \mathcal{P}_2\cup \mathcal{P}_3\triangleq \mathcal{P}$ where $\mathcal{P}_1=\{\frac{M_T}{M_R}|\frac{1}{2}\leq \frac{M_T}{M_R}<1,~M_T,M_R\in\mathbb{Z}^+,~M_R\leq 20\}$, $\mathcal{P}_2=[2/5,1/2)$, and $\mathcal{P}_3=\{\frac{8}{21}\}\cup\{\frac{2c-1}{5c-2}|c\in
\mathbb{Z}^+,c\geq 2\}$.}
\end{theorem}

As reported in \cite{Ghasemi_Motahari_Khandani_MIMO}, $\frac{MN}{M+N}$ DoF per user are achievable almost surely by using the rational alignment framework. To establish the DoF result implied by Theorem \ref{theorem:4user}, it suffices to show that $d=\frac{MN}{M+N}$ is also the information theoretic DoF outer bound per user.
In the remainder of this section, we will propose a systematic approach,
labeled as the ``genie chains" approach, based on four central ideas that we show in Section \ref{sec:4ideas}. We will provide generally two-layer type proofs through specific algorithms for
$M_T/M_R$ belonging to $\mathcal{P}_1$, $\mathcal{P}_2$, $\mathcal{P}_3$ sequentially. In addition, note
that we only consider the $M_T<M_R$ setting in this section, which means that $(M_T,M_R)=(M,N)$. Further
discussion on the results will be presented in Section
\ref{sec:big_picture}.
%, based on the four ideas highlighted in Section \ref{sec:4ideas}.

{\it Remark:} Theorem \ref{theorem:4user} replaces a stronger version previously reported in  the conference version of this paper (Theorem 1 of \cite{Wang_Sun_Jafar_ISIT}). The stronger claim, that $d=\frac{M_T M_R}{M_T+M_R}$ whenever $M_T/M_R>3/8$, was based on a  linear independence argument, which we have since discovered to be incomplete. While we still conjecture that the  claim is correct, and have proved it in various regimes, e.g., $\mathcal{P}_1, \mathcal{P}_2,\mathcal{P}_3$, unfortunately we have not  been able to resolve this conjecture for arbitrary $M,N$.

\subsection{$M/N\in [1/2,1)$ Case}

Since $N-2M\leq 0$, each RX cannot directly obtain exposed
subspaces from any interferer by zero forcing the signals from the
other two interferers. For brevity we let $M_0\!=\!N\!-\!M$ where
$M_0$ is a positive integer. Also, note that the random linear
combinations provided by a genie in each step are generic, although
we may (have to) use the same notations.

{\it Proof:} The general proof for this setting is given by the
following algorithm.

{\bf \verb"Algorithm 1"} ($M/N\in [1/2,1)$)

\begin{itemize}
\item{\verb"Step 1":}

Start from RX $k\!=\!2$. A genie provides RX $k$ the signal set
${\bf \bar{G}}\!=\!\{{\bf \bar{X}}^{[k+1]},{\bf \bar{X}}^{[k-1]}_{(M-M_0)}\}$, where
${\bf X}^{[k-1]}_{(M\!-\!M_0)}$ are $M-M_0$ random linear
combinations of the transmit signals from TX $k-1$. In the absence
of the interference
from TX $k+1$, %after zero forcing the interference from TX
%$k\!+\!2$,
RX $k$ has $M_0$ dimensional observations of the transmit signals
from TX $k-1$ by zero-forcing the signals from TX $k+2$. We denote
by $\mathcal{O}$ the $M_0$ dimensional observations. This process
produces the first sum rate inequality:
\begin{eqnarray*}
4nR-n\epsilon_n&\!\!\!\!\leq:\!\!\!\!& \hbar({\bf Y}^{[k]^n})+\hbar({\bf G}^n|{\bf S}^{[k]^n})\\
&\!\!\!\!\leq:\!\!\!\!& Nn\log\rho+\hbar({\bf X}^{[k+1]^n})+\hbar({\bf X}^{[k-1]^n}_{(M-M_0)}|{\bf S}^{[k]},{\bf X}^{[k+1]^n})\\
&\!\!\!\!\leq:\!\!\!\!& Nn\log\rho+nR+\hbar({\bf X}^{[k-1]^n}_{(M-M_0)}|\mathcal{O}^n)\\
&\!\!\!\!\leq:\!\!\!\!& Nn\log\rho+2nR-\hbar(\mathcal{O}^n).
\end{eqnarray*}

\item{\verb"Step 2:"}

If $|\mathcal{O}|=|{\bf G}|-M=M-M_0$, go to Step 3.

If $|\mathcal{O}|<|{\bf G}|-M=M-M_0$, go to Step 4.

If $|\mathcal{O}|>|{\bf G}|-M=M-M_0$, go to Step 5.

\item{\verb"Step 3:"}

A genie provides RX $k+1$ the set ${\bf \bar{G}}=\{{\bf
\bar{X}}^{[k+2]},\mathcal{O} + {\bf Z}\}$. Note that in the absence of interference
from TX $k+2$, RX $k+1$ originally has $M_0$ dimensional
observations of TX $k-1$ after zero-forcing the interference from TX
$k$, which is denoted as $\mathcal{O'}$.
%Since $\mathcal{O}$ is only
%associated with TX $k-1$ and also $\mathcal{O'}$ is linearly
%independent with $\mathcal{O}$,
As $\mathcal{O}$ and $\mathcal{O'}$ are observations at different RXs, they are independent almost surely. So
from $\{\mathcal{O},\mathcal{O'}\}$,
RX $k+1$ is able to recover the transmit signal from TX $k-1$
subject to the noise distortion. This process produces one sum DoF
inequality
\begin{eqnarray*}
4nR-n\epsilon_n &\!\!\!\!\leq:\!\!\!\!& \hbar({\bf Y}^{[k+1]^n})+\hbar({\bf G}^n|{\bf S}^{[k+1]^n})\\
&\!\!\!\!\leq:\!\!\!\!& Nn\log\rho+\hbar({\bf X}^{[k+2]^n},\mathcal{O}^n|{\bf S}^{[k+1]^n})\\
&\!\!\!\!\leq:\!\!\!\!& Nn\log\rho+nR+\hbar(\mathcal{O}^n).
\end{eqnarray*}
Adding up all inequalities that we have so far produces the
inequality
\begin{eqnarray*}
4MnR-n\epsilon_n\leq: MNnR+(3M-N)nR
\end{eqnarray*}
which leads to our desired DoF outer bound $d\leq \frac{MN}{M+N}$.
Then we stop.

\item{\verb"Step 4:"}

A genie provides ${\bf \bar{G}}=\{{\bf \bar{X}}^{[k+2]},\mathcal{O} + {\bf Z}, {\bf
\bar{X}}^{[k]}_{(M-M_0-|\mathcal{O}|)}\}$ to RX $k+1$. In the absence of
interference from TX $k+2$, RX $k+1$ originally has $M_0$
dimensional observations of the transmit signals from TX $k$ after
zero-forcing the interference from TX $k-1$, which is denoted as
$\mathcal{O'}$. Since providing $\mathcal{O}$ to RX $k+1$ releases
other $M_0$ dimensional observations of the transmit signals from TX
$k$, which is denoted as $\mathcal{\tilde{O}}$, RX $k+1$ has a total
of $|\mathcal{\tilde{O}}|+M_0$ dimensional observations of ${\bf
X}^{[k]^n}$. This process produces the sum rate inequality
\begin{eqnarray*}
4nR-n\epsilon_n &\!\!\!\!\leq:\!\!\!\!& \hbar({\bf Y}^{[k+1]^n})+\hbar({\bf G}^n|{\bf S}^{[k+1]^n})\\
&\!\!\!\!\leq:\!\!\!\!& Nn\log\rho+\hbar({\bf X}^{[k+2]^n},\mathcal{O}^n, {\bf X}^{[k]^n}_{(M-M_0-|\mathcal{O}|)}|{\bf S}^{[k+1]^n})\\
&\!\!\!\!\leq:\!\!\!\!& Nn\log\rho+\hbar({\bf X}^{[k+2]^n})+\hbar(\mathcal{O}^n)+\hbar({\bf X}^{[k]^n}_{(M-M_0-|\mathcal{O}|)}|\mathcal{\tilde{O}}^n,\mathcal{O'}^n)\\
&\!\!\!\!\leq:\!\!\!\!&
Nn\log\rho+nR+\hbar(\mathcal{O}^n)+nR-\hbar(\mathcal{\tilde{O}}^n,\mathcal{O'}^n).
\end{eqnarray*}
Now we update $\mathcal{O}=\{\mathcal{\tilde{O}},\mathcal{O'}\}$ and
$k=k+1$. Go back to Step 2.

\item{\verb"Step 5:"}

A genie provides ${\bf \bar{G}}\!=\!\{\mathcal{O} + {\bf Z},{\bf
\bar{X}}^{[k]}_{(2M-M_0-|\mathcal{O}|)}\}$ to RX $k+1$. In the $N$
dimensional observation ${\bf S}^{[k+1]^n}$, after zero-forcing the
interference from TX $k-2$, we still have $N\!-\!M=M_0$ observations
of the interference from TX $k-1$ and TX $k$. After providing
$\mathcal{O}$ to RX $k+1$, now we have a total of
$M_0+|\mathcal{O}|>M$ dimensions of the interference from TX $k-1$
and TX $k$. Therefore, we continue to zero force the interference
from TX $k-1$, thus only leaving $M_0\!+|\mathcal{O}|\!-M$
dimensional observations of ${\bf X}^{[k]}$, denoted as
$\mathcal{O'}$. Note that $\mathcal{O'}$ is linearly independent
with ${\bf \bar{X}}^{[k]^n}_{(2M-M_0-|\mathcal{O}|)}$ provided by the
genie, and from them together RX $k+1$ is able to recover the
transmit signal from TX $k$ subject to the noise distortion. This
process produces the inequality
\begin{eqnarray*}
4nR-n\epsilon_n&\!\!\!\!\leq:\!\!\!\!& \hbar({\bf Y}^{[k+1]^n})+\hbar({\bf G}^n|{\bf S}^{[k+1]^n})\\
&\!\!\!\!\leq:\!\!\!\!& Nn\log\rho+\hbar(\mathcal{O}^n,{\bf X}^{[k]^n}_{(2M-M_0-|\mathcal{O}|)}|{\bf S}^{[k+1]^n})\\
&\!\!\!\!\leq:\!\!\!\!& Nn\log\rho+\hbar(\mathcal{O}^n)+nR-\hbar(\mathcal{O'}^n).
\end{eqnarray*}
Now we update $\mathcal{O}=\mathcal{O'}$ and $k=k+1$. Go back to Step 2.
\end{itemize}
\hfill\QED

{\it Remark:} Note that we do not need the intermediate DoF outer
bound in this case. In contrast, for $M/N\in[2/5,1/2)$ or
$[3/8,2/5)$ cases the intermediate bound is necessary, as we have
shown in Example 3 in Section \ref{sec:4ideas}.

In this algorithm, the genie signal ${\bf \bar{G}}$ always contains $3M-N$
dimensions in each step. If we want to recover all the interference
symbols from $\{{\bf \bar{G}},{\bf \bar{S}}^{[k]}\}$ subject to the noise
distortion, it remains to be shown that the ${\bf \bar{G}}$
is linearly independent with ${\bf \bar{S}}^{[k]}$, which is a linear algebra problem now.
We are able to verify the linear independence through numerical tests when $M_R \leq 20$. This completes the proof for this regime.

%\textcolor{red} {As a matter of fact, it is not difficult to see that it is true. We provide a proof of the linearly independence for each step of Algorithm 1 in Appendix \ref{app:subsec_alg1}.

%It seems that we can not prove this rigorously as in the theorem we have the condition $M_R \leq 20$. But this is not mentioned in Appendix C.1 or here. Moreover, the red sentences say \emph{"it is not difficult to it is true"}. Chenwei, maybe you can fix this part. We should mention that here the problem becomes a linear algebra problem, as said in the abstract.

%Also maybe we need to mention somewhere our conference version is \emph{wrong and different with the current theorem.}
%}

\subsection{$M/N\in [2/5,1/2)$ Case}

Since $N-2M> 0$, each RX obtains a fixed $N-2M$ dimensional
clean observations from each interferer, by simply zero forcing the
signals from the other two interferers. For brevity we let
$M_0\!=\!N\!-\!2M$ where $M_0$ is a positive integer.

{\it Proof:} The general proof for this setting is given by the
following algorithm.

{\bf \verb"Algorithm 2"} ($M/N\in [2/5,1/2)$)
\begin{itemize}
\item{\verb"Step 1:"}

Start from RX $k=2$. A genie provides signals ${\bf \bar{G}}={\bf
\bar{X}}^{[k-1]}_{(M-M_0)}$ to RX $k$ which originally has $M_0$
dimensional observations of transmit signals from TX $k-1$,
after it zero-forces the interference of the other two users. We
denote by $\mathcal{O}$ these $M_0$ dimensional observations. This
process produces the first sum rate inequality
\begin{eqnarray*}
4nR-n\epsilon_n &\!\!\!\!\leq:\!\!\!\!& \hbar({\bf Y}^{[k]^n})+\hbar({\bf G}^n|{\bf S}^{[k]^n})\\
&\!\!\!\!\leq:\!\!\!\!& Nn\log\rho+\hbar({\bf X}^{[k-1]^n}_{(M\!-\!M_0)}|\mathcal{O}^n)\\
&\!\!\!\!\leq:\!\!\!\!& Nn\log\rho+nR-\hbar(\mathcal{O}^n).
\end{eqnarray*}
% Then go to Step 2.

\item{\verb"Step 2:"}

If $|\mathcal{O}|=|{\bf G}|=M-M_0$, go to Step 3.

If $|\mathcal{O}|<|{\bf G}|=M-M_0$, go to Step 4.

If $|\mathcal{O}|>|{\bf G}|=M-M_0$, go to Step 5.

\item{\verb"Step 3:"}

A genie provides ${\bf \bar{G}}=\mathcal{O} + {\bf Z}$ to RX $k+1$. RX
$k+1$ originally has $M_0$ dimensional observations of the transmit
signals from TX $k$ after it zero-forces the interference
from the other two users. We denote by $\mathcal{O'}$ these $M_0$
dimensional observations, which combined with the $|\mathcal{O}|$
dimensional observations of transmit signals from TX $k$
opened up by $\mathcal{O}$ allows RX $k+1$ to recover signals
from TX $k$ subject to the noise distortion. This process
produces the inequality
\begin{eqnarray*}
4nR-n\epsilon_n &\!\!\!\!\leq:\!\!\!\!& \hbar({\bf Y}^{[k+1]^n})+\hbar({\bf G}^n|{\bf S}^{[k+1]^n})\\
&\!\!\!\!\leq:\!\!\!\!& Nn\log\rho+\hbar(\mathcal{O}^n).
\end{eqnarray*}
Adding up the inequalities we have so far produces the inequality
\begin{eqnarray*}
4MnR-n\epsilon_n\leq: MNnR+(3M-N)nR
\end{eqnarray*}
which leads to our desired DoF outer bound $d\leq \frac{MN}{M+N}$.
Then we stop.

\item{\verb"Step 4:"}

A genie provides the set ${\bf \bar{G}}=\{\mathcal{O} + {\bf Z}, {\bf
\bar{X}}^{[k]}_{(M-M_0-|\mathcal{O}|)}\}$ to RX $k+1$. RX
$k+1$ originally has $M_0$ dimensional observations of transmit
signals from TX $k$ after zero forcing the interference
from the other two suers. Denote these $M_0$ dimensional
observations as $\mathcal{O'}$. Since providing $\mathcal{O}$ to
RX $k+1$ releases other $M_0$ observations of transmit signals
from TX $k$, denoted as $\mathcal{\tilde{O}}$, RX $k+1$
has a total of $|\mathcal{\tilde{O}}|+M_0$ dimensional observations of
${\bf X}^{[k]}$. This process produces the inequality
\begin{eqnarray*}
4nR-n\epsilon_n &\!\!\!\!\leq:\!\!\!\!& \hbar({\bf Y}^{[k+1]^n})+\hbar({\bf G}^n|{\bf S}^{[k+1]^n})\\
&\!\!\!\!\leq:\!\!\!\!& Nn\log\rho+\hbar(\mathcal{O}^n, {\bf X}^{[k]^n}_{(M\!-\!M_0\!-\!|\mathcal{O}|)}|{\bf S}^{[k+1]^n})\\
&\!\!\!\!\leq:\!\!\!\!& Nn\log\rho+\hbar(\mathcal{O}^n)+\hbar({\bf X}^{[k]^n}_{(M\!-\!M_0\!-\!|\mathcal{O}|)}|\mathcal{\tilde{O}}^n,\mathcal{O'}^n)\\
&\!\!\!\!\leq:\!\!\!\!&
Nn\log\rho+\hbar(\mathcal{O}^n)+nR-\hbar(\mathcal{\tilde{O}}^n,\mathcal{O'}^n).
\end{eqnarray*}
Now we update $\mathcal{O}=\{\mathcal{\tilde{O}},\mathcal{O'}\}$ and
$k=k+1$. Go back to Step 2.

\item{\verb"Step 5:"}

A genie provides ${\bf \bar{G}}\!=\!{\bf \bar{X}}^{[k-1]}_{(M\!-\!M_0)}$ to
RX $k+1$, which originally has $M_0$ dimensional observations
of transmit signals from TX $k-1$ after zero forcing the
interference from the other two users. Denote these $M_0$
dimensional observations as $\mathcal{O'}$. Because
$|\mathcal{O}|+|\mathcal{O'}|>M$, the subspaces $\mathcal{O}$ and
$\mathcal{O'}$ have an $|\mathcal{O}|+M_0-M$ dimensional
intersection. We denote this intersection by $\mathcal{I}$. Note
that $\{\mathcal{O},\mathcal{O'}\setminus\mathcal{I}\}$ is already
the whole $M$ dimensional space, thus contributing $R+o(\log\rho)$
differential entropy. Note that we still have the remaining
$|\mathcal{O}|\!+\!M_0\!-\!M$ dimensional observations of ${\bf
X}^{[k-1]}$, i.e., the intersection $\mathcal{I}$. This process
produces the intermediate bound
\begin{eqnarray*}
4nR-n\epsilon_n &\!\!\!\!\leq:\!\!\!\!& \hbar({\bf Y}^{[k+1]^n})+\hbar({\bf G}^n|{\bf S}^{[k+1]^n})\\
&\!\!\!\!\leq:\!\!\!\!& Nn\log\rho+\hbar({\bf X}^{[k-1]^n}_{(M\!-\!M_0)}|\mathcal{O'}^n)\\
&\!\!\!\!\leq:\!\!\!\!& Nn\log\rho+nR-\hbar(\mathcal{O'}^n)-\hbar(\mathcal{O}^n)+\hbar(\mathcal{O}^n)\\
&\!\!\!\!\leq:\!\!\!\!& Nn\log\rho+\hbar(\mathcal{O}^n)+nR-\hbar(\mathcal{O'}^n\setminus\mathcal{I}^n,\mathcal{I}^n)-\hbar(\mathcal{O}^n)\\
&\!\!\!\!\leq:\!\!\!\!& Nn\log\rho+\hbar(\mathcal{O}^n)-\hbar(\mathcal{I}^n)+nR-\hbar(\mathcal{O'}^n\setminus\mathcal{I}^n|\mathcal{I}^n)-\hbar(\mathcal{O}^n)\\
&\!\!\!\!\leq:\!\!\!\!& Nn\log\rho+\hbar(\mathcal{O}^n)-\hbar(\mathcal{I}^n)+nR-\hbar(\mathcal{O'}^n\setminus\mathcal{I}^n|\mathcal{I}^n,\mathcal{O}^n)-\hbar(\mathcal{O}^n)\\
&\!\!\!\!\leq:\!\!\!\!& Nn\log\rho+\hbar(\mathcal{O}^n)-\hbar(\mathcal{I}^n)+nR-\hbar(\mathcal{O'}^n\setminus\mathcal{I}^n|\mathcal{O}^n)-\hbar(\mathcal{O}^n)\\
&\!\!\!\!\leq:\!\!\!\!& Nn\log\rho+\hbar(\mathcal{O}^n)-\hbar(\mathcal{I}^n)+nR-\hbar(\mathcal{O'}^n\setminus\mathcal{I}^n,\mathcal{O}^n)\\
&\!\!\!\!\leq:\!\!\!\!& Nn\log\rho+\hbar(\mathcal{O}^n)-\hbar(\mathcal{I}^n).
\end{eqnarray*}
Now we update $\mathcal{O}=\mathcal{I}$, $k=k+1$. Go back to Step 2.
\end{itemize}
\hfill\QED

{\it Remark:} The phrase ``intermediate" implies that the
observations associated with the negative entropy term are obtained
by intersecting the two subspaces available at successive two
RXs looking at the same interferer. Moreover, in this case we
only need intermediate bounds (with respect to two RXs) by
intersecting for at most once.

Similarly to Algorithm 1, In this algorithm, we also need to show
that the $3M-N$ dimensional genie signal ${\bf \bar{G}}$ is linearly
independent with ${\bf \bar{S}}^{[k]}$. Equivalently, since each RX
has a clean $N-2M$ dimensional subspace of each interferer, we need
to guarantee that ${\bf \bar{G}}$ is linearly independent with these
subspaces. In order to avoid the complicated mathematics equations,
we defer the rigorous proof to Appendix \ref{app:subsec_alg2}.

\subsection{$M/N\in [3/8,2/5)$ Case}

When $M/N$ falls into $[3/8,2/5)$ regime, we show the proofs
when $M/N=(2c-1)/(5c-2),~c\in \mathbb{Z}^+\!\setminus\!\{1\}$ and
$M/N=8/21$.

{\it Proof:} Let us first consider $M/N=(2c-1)/(5c-2),~c\in
\mathbb{Z}^+\!\setminus\!\{1\}$ cases. It can be checked that for
these cases we only need \emph{one successive intermediate bound}.
Thus, we can still use Algorithm 2 to produce the information
theoretic outer bound proofs for these cases. Similarly, what
remains to be shown is that at each step the clean $N-2M$
dimensional observations of the associated TX are linearly
independent with the $|{\bf \bar{G}}|=3M-N$ dimensional observations of
that TX opened up by the provided genie ${\bf \bar{G}}$, i.e., we
will show that the resulting $M\times M$ square matrix has full
rank. The proof in detail is deferred to Appendix
\ref{app:subsecalg2set}.

Next, we prove when $M/N=8/21$. Note that this does not fall
into the category that $M/N=(2c-1)/(5c-2)$. What is special
for this case is that we need \emph{two successive
intermediate bounds}.
%if we use the tool of genie chains and follow from the four ideas that we introduced in Section \ref{sec:4ideas}.
Suppose $(M,N)=(8a,21a),~a\in\mathbb{Z}^+$, then the proof is
shown through the following eight steps.
\begin{itemize}
\item{Step 1:} Start from RX 2 and TX 1. A genie provides ${\bf
\bar{G}}_1\!=\!{\bf \bar{X}}^{[1]}_{(3a)}$ to RX $2$ such that it can
decode all the messages subject to the noise distortion. After zero
forcing the interference from TX 3 and TX 4,
RX 2 originally has $5a$ dimensional observations of the
signals sent from TX 1, which is denoted as $\mathcal{O}$
where $|\mathcal{O}|=5a$. This process produces the first sum rate
inequality
\begin{eqnarray}
4nR-n\epsilon_n&\!\!\!\!\leq:\!\!\!\!& \hbar({\bf Y}^{[2]^n})+\hbar({\bf G}_1^n|{\bf S}^{[2]^n})\notag\\
&\!\!\!\!\leq:\!\!\!\!& Nn\log\rho+\hbar({\bf G}_1^n|\mathcal{O}^n)\notag\\
&\!\!\!\!\leq:\!\!\!\!& Nn\log\rho+nR-\hbar(\mathcal{O}^n).\label{eqn:8by21_ob1}
\end{eqnarray}

\item{Step 2:} Since $|\mathcal{O}|+5a>M$, we go to RX 3 looking at TX 1 for an
intermediate bound. A genie provides ${\bf \bar{G}}_2={\bf
\bar{X'}}^{[1]}_{(3a)}$ to RX 3, which originally has $5a$
dimensional observations of signals sent from TX 1, after
zero forcing these $5a$ dimensional observations as $\mathcal{O'}$.
Because $|\mathcal{O}|+|\mathcal{O'}|=10a>8a=M$, they have a
$2a$-dimensional intersection. We denote this intersection by
$\mathcal{I}$. This process produces the intermediate bound
\begin{eqnarray}
4nR-n\epsilon_n &\!\!\!\!\leq:\!\!\!\!& \hbar({\bf Y}^{[3]^n})+\hbar({\bf G}_2^n|{\bf S}^{[3]^n})\notag\\
&\!\!\!\!\leq:\!\!\!\!& Nn\log\rho+\hbar(\mathcal{O}^n|\mathcal{O'}^n)\notag\\
&\!\!\!\!\leq:\!\!\!\!& Nn\log\rho+nR-\hbar(\mathcal{O'}^n)-\hbar(\mathcal{O}^n)+\hbar(\mathcal{O}^n)\notag\\
&\!\!\!\!\leq:\!\!\!\!& Nn\log\rho+\hbar(\mathcal{O}^n)-\hbar(\mathcal{I}^n).
\end{eqnarray}
Then we update $\mathcal{O}=\mathcal{I}$ and $|\mathcal{O}|=2a$.

\item{Step 3:} Since $|\mathcal{I}|+5a=7a<M$, we do not need an intermediate bound
here. We go to RX 4 looking at TX 3. A genie provides
${\bf \bar{G}}_3=\{\mathcal{O} + {\bf Z}, {\bf \bar{X}}^{[3]}_{(a)}\}$ to RX 4,
which originally has $5a$ dimensional observations of signals sent
from TX 3, denoted as $\mathcal{O'}$. Since providing
$\mathcal{O}$, which is associated with User 1, to RX $4$
releases other $|\mathcal{O}|=2a$ observations of signals sent from
TX 3, denoted as $\mathcal{\tilde{O}}$, RX 4 has a
total of $|\mathcal{\tilde{O}}|+M_0=7a$ dimensional observations of
${\bf X}^{[3]}$. This process produces
\begin{eqnarray}
4nR-n\epsilon_n&\!\!\!\!\leq:\!\!\!\!& \hbar({\bf Y}^{[4]^n})+\hbar({\bf G}_3^n|{\bf S}^{[4]^n})\notag\\
&\!\!\!\!\leq:\!\!\!\!& Nn\log\rho+\hbar(\mathcal{O}^n,{\bf X}^{[3]^n}_{(a)}|{\bf S}^{[4]^n})\notag\\
&\!\!\!\!\leq\!\!\!\!& Nn\log\rho+\hbar(\mathcal{O}^n)+\hbar({\bf X}^{[3]^n}_{(a)}|{\bf S}^{[4]^n},\mathcal{O}^n,\mathcal{O'}^n)\notag\\
&\!\!\!\!\leq\!\!\!\!& Nn\log\rho+\hbar(\mathcal{O}^n)+\hbar({\bf X}^{[3]^n}_{(a)}|\mathcal{\tilde{O}}^n,\mathcal{O'}^n)\notag\\
&\!\!\!\!\leq:\!\!\!\!&
Nn\log\rho+\hbar(\mathcal{O}^n)+nR-\hbar(\mathcal{\tilde{O}}^n,\mathcal{O'}^n).
\end{eqnarray}
Now we update $\mathcal{O}=\{\mathcal{\tilde{O}},\mathcal{O'}\}$ and
$|\mathcal{O}|=7a$.

\item{Step 4:} Now since $|\mathcal{O}|+5a>M$, we need an
intermediate bound. So we go to RX 1 still looking at
TX 3. A genie provides ${\bf \bar{G}}_4={\bf \bar{X}}^{[3]}_{(3a)}$ to
RX 1, which originally has $5a$ dimensional observations of
signals sent from TX 1, denoted as $\mathcal{O'}$ and
$|\mathcal{O'}|=5a$. Because $|\mathcal{O}|+|\mathcal{O'}|=12a>M$,
they have a $4a$-dimensional intersection. We denote this
intersection by $\mathcal{I}$. This process produces the
intermediate bound as follows:
\begin{eqnarray}
4nR-n\epsilon_n &\!\!\!\!\leq:\!\!\!\!& \hbar({\bf Y}^{[1]^n})+\hbar({\bf G}_4^n|{\bf S}^{[1]^n})\notag\\
&\!\!\!\!\leq:\!\!\!\!& Nn\log\rho+\hbar({\bf X}^{[3]^n}_{(3a)}|\mathcal{O'}^n)\notag\\
&\!\!\!\!\leq:\!\!\!\!& Nn\log\rho+nR-\hbar(\mathcal{O'}^n)-\hbar(\mathcal{O}^n)+\hbar(\mathcal{O}^n)\notag\\
&\!\!\!\!\leq:\!\!\!\!& Nn\log\rho+\hbar(\mathcal{O}^n)-\hbar(\mathcal{I}^n).
\end{eqnarray}
Then we update $\mathcal{O}=\mathcal{I}$ and $|\mathcal{O}|=4a$.

\item{Step 5:} Because we have again $|\mathcal{O}|+5a>M$, we need to resort to the next RX, RX 2, looking at TX 3 for
another intermediate bound. Now a genie provides the set ${\bf
\bar{G}}_3={\bf \bar{X'}}^{[3]^n}_{(3a)}$ to RX 2, which also has $5a$
dimensional observations of ${\bf X}^{[3]}$, denoted as
$\mathcal{O'}$. We denote by $\mathcal{I}$ the intersection of
$\mathcal{O}$ and $\mathcal{O'}$, and
$|\mathcal{I}|\!=\!|\mathcal{O}|\!+5a\!-8a=a$. This process produces
the intermediate bound
\begin{eqnarray}
4nR-n &\!\!\!\!\leq:\!\!\!\!& \hbar({\bf Y}^{[2]^n})+\hbar({\bf G}_5^n|{\bf S}^{[2]^n})\notag\\
&\!\!\!\!\leq:\!\!\!\!& Nn\log\rho+\hbar({\bf X'}^{[3]^n}_{(3a)}|\mathcal{O'}^n)\notag\\
&\!\!\!\!\leq:\!\!\!\!& Nn\log\rho+nR-\hbar(\mathcal{O'}^n)-\hbar(\mathcal{O}^n)+\hbar(\mathcal{O}^n)\notag\\
&\!\!\!\!\leq:\!\!\!\!& Nn\log\rho+\hbar(\mathcal{O}^n)-\hbar(\mathcal{I}^n).
\end{eqnarray}
Then we update $\mathcal{O}=\mathcal{I}$ and $|\mathcal{O}|=a$.

\item{Step 6:} Now since $|\mathcal{O}|+5a=6a<M$, we do not need an intermediate bound
here. Now let us recall how we obtain the observations (subspaces)
$\mathcal{O}$ here. We start from RX 4 in Step 3 where we have
$7a$ dimensional observations of ${\bf X}^{[3]}$, $5a$ dimensions
that RX 4 originally has and the other $2a$ dimensions opened
up by the genie. Then we intersect these $7a$ dimensional
observations with the $5a$ dimensional observations at RX 1
and RX 2 in Step 4 and Step 5, respectively, to produce
$\mathcal{O}$. That is to say, the $a$ dimensional observations
$\mathcal{O}$ are already contained in the clean observations at
RX 1 and RX 2. Therefore, we cannot provide
$\mathcal{O}$ as a genie to those two RXs. Also, $\mathcal{O}$
is the observations of ${\bf X}^{[3]}$ and thus cannot be a genie
provided to RX 3. Thus, we can only provide it as a genie to
RX 4. Moreover, we want to use $\mathcal{O}$ as a genie to
open up the dimensions of signals sent from other TX, i.e.,
not TX 3 or TX 4. Suppose a genie provides ${\bf
\bar{G}}_4=\{\mathcal{O} + {\bf Z}, {\bf \bar{X}}^{[2]}_{(2a)}\}$ to RX 2, which
originally has $5a$ dimensional observations of signals sent from
TX 2, denoted as $\mathcal{O'}$. As what we have described,
providing $\mathcal{O}$ to RX $4$ releases other
$|\mathcal{O}|=a$ observations of signals sent from TX 2,
denoted as $\mathcal{\tilde{O}}$, RX 4 has a total of
$|\mathcal{\tilde{O}}|+M_0=6a$ dimensional observations of ${\bf
X}^{[2]}$. This process produces the following inequality:
\begin{eqnarray}
4nR-n\epsilon_n &\!\!\!\!\leq:\!\!\!\!& \hbar({\bf Y}^{[4]^n})+\hbar({\bf G}_6^n|{\bf S}^{[4]^n})\notag\\
&\!\!\!\!\leq:\!\!\!\!& Nn\log\rho+\hbar(\mathcal{O}^n,{\bf X}^{[2]^n}_{(2a)}|{\bf S}^{[4]^n})\notag\\
&\!\!\!\!\leq\!\!\!\!& Nn\log\rho+\hbar(\mathcal{O}^n)+\hbar({\bf X}^{[2]^n}_{(2a)}|{\bf S}^{[4]^n},\mathcal{O}^n,\mathcal{O'}^n)\notag\\
&\!\!\!\!\leq\!\!\!\!& Nn\log\rho+\hbar(\mathcal{O}^n)+\hbar({\bf X}^{[2]^n}_{(2a)}|\mathcal{\tilde{O}}^n,\mathcal{O'}^n)\notag\\
&\!\!\!\!\leq:\!\!\!\!&
Nn\log\rho+\hbar(\mathcal{O}^n)+nR-\hbar(\mathcal{\tilde{O}}^n,\mathcal{O'}^n).
\end{eqnarray}
Now we update $\mathcal{O}=\{\mathcal{\tilde{O}},\mathcal{O'}\}$ and
$|\mathcal{O}|=6a$.

\item{Step 7:} Since $|\mathcal{O}|+5a>M$, we again need an
intermediate bound. Consider RX 1 looking at TX 2. A
genie provides ${\bf \bar{G}}={\bf \bar{X}}^{[2]}_{(3a)}$ to RX 1, which
originally has $5a$ dimensional observations of signals sent from
TX 2, denoted as $\mathcal{O'}$ and $|\mathcal{O'}|=5a$.
Because $|\mathcal{O}|+|\mathcal{O'}|=11a>M$, they have a
$3a$-dimensional intersection. We denote this intersection by
$\mathcal{I}$. Then this process produces the intermediate bound
\begin{eqnarray}
4nR-n\epsilon_n &\!\!\!\!\leq:\!\!\!\!& \hbar({\bf Y}^{[1]^n})+\hbar({\bf G}_7^n|{\bf S}^{[1]^n})\notag\\
&\!\!\!\!\leq:\!\!\!\!& Nn\log\rho+\hbar({\bf X}^{[2]^n}_{(3a)}|\mathcal{O'}^n)\notag\\
&\!\!\!\!\leq:\!\!\!\!& Nn\log\rho+nR-\hbar(\mathcal{O'}^n)-\hbar(\mathcal{O}^n)+\hbar(\mathcal{O}^n)\notag\\
&\!\!\!\!\leq:\!\!\!\!& Nn\log\rho+\hbar(\mathcal{O}^n)-\hbar(\mathcal{I}^n).
\end{eqnarray}
Then we update $\mathcal{O}=\mathcal{I}$ and $|\mathcal{O}|=3a$.

\item{Step 8:} Finally, consider RX 3 looking at TX 2. A genie provides ${\bf
\bar{G}}_8=\mathcal{O} + {\bf Z}$ to RX 3, which originally has $5a$
dimensional observations of TX 2, denoted as
$\mathcal{O'}$, which combined with the $|\mathcal{O}|=3a$
dimensional genie signals from TX 2 allows RX 3 to
recover ${\bf X}^{[2]}$ subject to the noise distortion. This
process produces the inequality
\begin{eqnarray}
4nR\leq: \hbar({\bf Y}^{[3]^n})+\hbar({\bf G}_8^n|{\bf S}^{[3]^n})\leq
Nn\log\rho+\hbar(\mathcal{O}^n).\label{eqn:8by21_ob8}
\end{eqnarray}

Notice that in each of the eight inequalities from
(\ref{eqn:8by21_ob1}) to (\ref{eqn:8by21_ob8}), the differential
entropy term with the negative sign always appears with the positive
sign in the next inequality. Thus, adding up all the eight sum rate
inequalities from (\ref{eqn:8by21_ob1}) to (\ref{eqn:8by21_ob8}),
all the negative terms on the right-hand side are fully canceled
out, thus producing the following inequality:
\begin{eqnarray}
32nR-n\epsilon_n\leq: 8Nn\log\rho+3nR+n~o(\log\rho)+o(n)\Rightarrow
d\leq \frac{8N}{29}=\frac{8\times 21}{29}.
\end{eqnarray}
\end{itemize}

Note that in each step we still need to ensure at each step the
clean $N-2M$ dimensional observations of the associated TX
are linearly independent with the $|{\bf \bar{G}}|=3M-N$ dimensional
observations of that TX opened up by the provided genie
${\bf \bar{G}}$. Since $M/N=8/21$ is only a special example, we rely on
the numerical test by randomly generating the channel matrices, and
finally the test result shows that it is true. \hfill\QED

%We still let $M_0\!=\!N\!-\!2M$ for brevity where $M_0$ is a
%positive integer. Here, we define the index vector
%$U^{(j,i)}=[j\!+\!1,\!\cdots\!,i\!-\!1,i\!+\!1,\!\cdots\!,j\!+\!4]^T$,
%i.e., a vector with three entries from $j\!+\!1$ to $j\!+\!4$ by
%deleting $i$. Also, we define
%$V^{(j,i)}=[j\!-\!1,\!\cdots\!,i\!+\!1,i\!-\!1,\!\cdots\!,j\!-\!4]^T$,
%i.e., a vector with three entries from $j\!-\!1$ to $j\!-\!4$ by
%deleting $i$. In addition, we denote by $U^{(j,i)}_k$ and
%$V^{(j,i)}_k$ to represent the $k^{th}$ entry of $U^{(j,i)}$ and
%$V^{(j,i)}$, respectively. Notice that as what have mentioned
%before, all entries are interpreted as the \emph{modulo 4} sense.

{\it Remark:} For any $(M,N)$ pair where $M/N\in [3/8,2/5)$, after
running the algorithm we obtain a series of inequalities, in which
the intermediate bounds can appear successively for at most twice.
In addition, we cannot derive more than two {\em successively}
intermediate bounds. The reason is the following. At any step, if we
provide $\mathcal{O}$ as a genie where $|\mathcal{O}|+(N-2M)>M$,
then we need an intermediate bound. After we deriving that
inequality, we provide a $|\mathcal{O}|+(N-2M)-M$ dimensional genie
to the RX we concern in the next. Again, if
$|\mathcal{O}|+(N-2M)-M+(N-2M)>M$, we need immediately another
intermediate bound. With the same analysis, suppose we need the
third successive intermediate bound, we have to have:
\begin{eqnarray}
|\mathcal{O}|+(N-2M)-M+(N-2M)-M+(N-2M)>M
\end{eqnarray}
which, due to $|\mathcal{O}|<M$, implies that
\begin{eqnarray}
M<M+(N-2M)-M+(N-2M)-M+(N-2M)\Longrightarrow \frac{M}{N}<\frac{3}{8}
\end{eqnarray}
which is contradictive. Intuitively, this conclusion implies that
the three $N-2M$ dimensional clean observations of one interferer at
all undesired RXs have only null intersection in common after
projecting the clean subspaces back to that TX.
Furthermore, for $K$ user $M\times N$ MIMO interference channel, the
clean observations of one interferer at all undesired RXs,
after we project them back to that transmit space, will have a
common intersection with $[(K-1)(N-(K-2)M)-(K-2)M]^+$ dimension.
This intersection would be the null space as long as
\begin{eqnarray}
(K-1)(N-(K-2)M)\leq (K-2)M\Longrightarrow \frac{M}{N}\geq \frac{K-1}{K(K-2)}.
\end{eqnarray}
%which is consistent with the cross point of linear counting DoF bound and the decomposition DoF bound, as what we will show later.
In general, in the $K$-user case, we may have up to $K-2$ successive intermediate bound.

\section{Examples of Applications of Genie Chains}\label{section:app}

Aside from the application of the genie chains in $K=4$ user
semi-symmetric $M_T\times M_R$ MIMO interference channel where $M_T<
M_R$, the tool of genie chains can also be applied to the reciprocal
$M_T>M_R$ setting and many other wireless networks to produce the
desired information theoretic DoF outer bound. In this section, we
will provide four specific examples to show the application of genie
chains.

\subsection{DoF of the $K=4$ User Reciprocal Setting}

We consider $(M_T,M_R)=(8,3)=(N,M)$ MIMO interference channel, as an
example of the reciprocal $M_T>M_R$ setting in this section. The
channel model and associated definitions and notations are identical
to that in Section \ref{sec:model}. We are going to show that each
user in this channel has $24/11$ DoF. Since the achievability has
already been shown in \cite{Ghasemi_Motahari_Khandani_MIMO}, we
focus on the information theoretic DoF outer bound.

{\it Proof:} As what we have shown in previous examples,
intuitively, we need a total of $M_T=8$ sum rate bounds, which can
be produced through the following eight steps. Note that in each
step, the genie should have at least $|{\bf \bar{G}}|=3N-M=21$ dimensions.
\begin{itemize}
\item{Step 1:} Consider RX 2 and TX 1. After decoding its desired message
$W_2$, RX 2 can reconstruct ${\bf {X}}^{[2]}$ and subtract it from its
received signal vector. Thus, RX 2 has 3 linear combinations of
$8\times 3=24$ interference symbols from TX1, TX 3 and TX 4. A genie
provides ${\bf \bar{G}}_1\!=\!\{{\bf \bar{X}}^{[1]}_{(5)},{\bf \bar{X}}^{[3]},{\bf
\bar{X}}^{[4]}\}$ to RX 2. From provided ${\bf \bar{X}}^{[3]},{\bf \bar{X}}^{[4]}$, RX
2 can decode $W_3,W_4$. After RX 2 subtract ${\bf {X}}^{[2]},{\bf
{X}}^{[3]},{\bf {X}}^{[4]}$, it has three dimensional observations of
interference from TX 1, which are linearly independent with the
provided genie signals ${\bf \bar{X}}^{[1]}_{(5)}$. By inverting the
channel matrix associated with TX 1, RX 2 can decode $W_1$ as well
subject to the noise distortion. Hence, we obtain the first sum rate
inequality:
\begin{eqnarray}
4nR-n\epsilon_n&\!\!\!\!\leq:\!\!\!\!& \hbar({\bf Y}^{[2]^n})+\hbar({\bf G}_1^n|{\bf S}^{[2]^n})\\
&\!\!\!\!\leq:\!\!\!\!& nM_R\log\rho+\hbar({\bf X}^{[1]^n}_{(5)},{\bf X}^{[3]^n},{\bf X}^{[4]^n}|{\bf Y}^{[2]^n})\\
&\!\!\!\!\leq:\!\!\!\!& nM_R\log\rho+\hbar({\bf X}^{[3]^n},{\bf X}^{[4]^n})+\hbar({\bf X}^{[1]^n}_{(5)}|{\bf Y}^{[2]^n},{\bf X}^{[3]^n},{\bf X}^{[4]^n})\\
&\!\!\!\!\leq:\!\!\!\!& nM_R\log\rho+2nR+\hbar({\bf X}^{[1]^n}_{(5)}|{\bf X}_3^{[1\sim 2]^n})\\
&\!\!\!\!\leq:\!\!\!\!& nM_R\log\rho+2nR+\hbar({\bf X}^{[1]^n}_{(5)},{\bf X}_3^{[1\sim 2]^n})-\hbar({\bf X}_3^{[1\sim 2]^n})\\
&\!\!\!\!\leq:\!\!\!\!& nM_R\log\rho+3nR-\hbar({\bf X}_3^{[1\sim 2]^n}).
\end{eqnarray}

\item{Step 2:} Consider RX 3 and TX 1. Similar to Step 1, a genie
provides ${\bf \bar{G}}_2\!=\!\{{\bf \bar{X}}^{[1]}_{(5)},{\bf \bar{X}}^{[2]},{\bf
\bar{X}}^{[4]}\}$ to RX 2, such that it can decode all the messages as well
subject to the noise distortion. Thus step produces the second
inequality as follows:
\begin{eqnarray}
4nR-n\epsilon_n&\!\!\!\!\leq:\!\!\!\!& \hbar({\bf Y}^{[3]^n})+\hbar({\bf G}_2^n|{\bf S}^{[3]^n})\\
&\!\!\!\!\leq:\!\!\!\!& nM_R\log\rho+3nR-\hbar({\bf X}_3^{[1\sim 3]^n}).
\end{eqnarray}

\item{Step 3:} Consider RX 4 and TX 2. A genie
provides ${\bf \bar{G}}_3\!=\!\{{\bf \bar{X}}_3^{[1\sim 2]},{\bf \bar{X}}_3^{[1\sim
3]},{\bf \bar{X}}^{[2]}_{(7)},{\bf \bar{X}}^{[3]}\}$ to RX 4. Note that RX 4
originally is able to decode $W_4$ and reconstruct ${\bf X}^{[4]}$,
and then subtract it from its received signal vector. Thus, RX 4 has
three dimensional observations of the 16 interference symbols from
TX 1 and TX 2. With provided genie ${\bf \bar{X}}_3^{[1\sim 2]},{\bf
\bar{X}}_3^{[1\sim 3]},{\bf \bar{X}}^{[2]}_{(7)}$, RX 4 can invert the square
channel matrix associated with TX 1 and TX 2, and thus can decode
the other two messages as well subject to the noise distortion.
Therefore, we have the second inequality as follows:
%\begin{small}
\begin{eqnarray}
4nR-n\epsilon_n&\!\!\!\!\leq:\!\!\!\!& \hbar({\bf Y}^{[4]^n})+\hbar({\bf G}_3^n|{\bf S}^{[4]^n})\\
&\!\!\!\!\leq:\!\!\!\!& nM_R\log\rho+\hbar({\bf X}_3^{[1\sim 2]^n},{\bf X}_3^{[1\sim 3]^n},{\bf X}^{[2]^n}_{(7)},{\bf X}^{[3]^n}|{\bf Y}^{[4]^n})\\
&\!\!\!\!\leq:\!\!\!\!& nM_R\log\rho+\hbar({\bf X}_3^{[1\sim 2]^n}\!\!,{\bf X}_3^{[1\sim 3]^n})+\hbar({\bf X}^{[3]^n})+\hbar({\bf X}^{[2]^n}_{(7)}\!|{\bf S}^{[4]^n}\!\!,{\bf X}_3^{[1\sim 2]^n}\!\!,{\bf X}_3^{[1\sim 3]^n}\!\!,{\bf X}^{[3]^n})\ \ \ \ \\
&\!\!\!\!\leq:\!\!\!\!& nM_R\log\rho+\hbar({\bf X}_3^{[1\sim 2]^n})+\hbar({\bf X}_3^{[1\sim 3]^n})+nR+\hbar({\bf X}^{[2]^n}_{(7)}|{\bf X}_1^{[2\sim 4]^n})\\
&\!\!\!\!\leq:\!\!\!\!& nM_R\log\rho+\hbar({\bf X}_3^{[1\sim 2]^n})+\hbar({\bf X}_3^{[1\sim 3]^n})+nR+\hbar({\bf X}^{[2]^n}_{(7)},{\bf X}_1^{[2\sim 4]^n})-\hbar({\bf X}_1^{[2\sim 4]^n})\\
&\!\!\!\!\leq:\!\!\!\!& nM_R\log\rho+2nR+\hbar({\bf X}_3^{[1\sim 2]^n})+\hbar({\bf
X}_3^{[1\sim 3]^n})-\hbar({\bf X}_1^{[2\sim 4]^n}).
\end{eqnarray}
%\end{small}
\item{Step 4:} Consider RX 1 and TX 2. Similar to Step 1, a genie
provides ${\bf \bar{G}}_4\!=\!\{{\bf \bar{X}}^{[2]}_{(5)},{\bf \bar{X}}^{[3]},{\bf
\bar{X}}^{[4]}\}$ to RX 1, such that RX 1 can decode all the messages subject
to the noise distortion. Following the derivations in Step 1, we
obtain the fourth sum rate inequality:
\begin{eqnarray}
4nR-n\epsilon_n&\!\!\!\!\leq:\!\!\!\!& \hbar({\bf Y}^{[1]^n})+\hbar({\bf G}_4^n|{\bf S}^{[1]^n})\\
&\!\!\!\!\leq:\!\!\!\!& nM_R\log\rho+3nR-\hbar({\bf X}_3^{[2\sim 1]^n}).
\end{eqnarray}

\item{Step 5:} Consider RX 3 and TX 2. Similar to Step 2, a genie
provides ${\bf \bar{G}}_5\!=\!\{{\bf \bar{X}}^{[2]}_{(5)},{\bf \bar{X}}^{[1]},{\bf
\bar{X}}^{[4]}\}$ to RX 2, such that it can decode all the messages subject to
the noise distortion. Thus step produces the fifth inequality as
follows:
\begin{eqnarray}
4nR-n\epsilon_n&\!\!\!\!\leq:\!\!\!\!& \hbar({\bf Y}^{[3]^n})+\hbar({\bf G}_2^n|{\bf S}^{[3]^n})\\
&\!\!\!\!\leq:\!\!\!\!& nM_R\log\rho+3nR-\hbar({\bf X}_3^{[2\sim 3]^n}).
\end{eqnarray}

\item{Step 6:} Consider RX 2 and TX 4. A genie
provides ${\bf \bar{G}}_6\!=\!\{{\bf \bar{X}}^{[4]}_{(5)},{\bf \bar{X}}^{[1]},{\bf
\bar{X}}^{[3]}\}$ to RX 2, such that RX 2 is able to decode all the messages
subject to the noise distortion. The reason and derivations of this
step is similar to Step 1 and Step 4, and thus we have the following
inequality:
\begin{eqnarray}
4nR-n\epsilon_n&\!\!\!\!\leq:\!\!\!\!& \hbar({\bf Y}^{[2]^n})+\hbar({\bf G}_6^n|{\bf S}^{[2]^n})\\
&\!\!\!\!\leq:\!\!\!\!& nM_R\log\rho+3nR-\hbar({\bf X}_3^{[4\sim 2]^n}).
\end{eqnarray}

\item{Step 7:} Consider RX 3 and TX 4. A genie provides ${\bf \bar{G}}_7\!=\!\{{\bf
\bar{X}}^{[4]}_{(5)},{\bf \bar{X}}^{[1]},{\bf \bar{X}}^{[2]}\}$ to RX 3, such that RX
3 can decode all the messages subject to the noise distortion. This step
is similar to Step 2 and Step 5 that we have shown. This step
produces the seventh inequality as follows:
\begin{eqnarray}
4nR-n\epsilon_n&\!\!\!\!\leq:\!\!\!\!& \hbar({\bf Y}^{[3]^n})+\hbar({\bf G}_7^n|{\bf S}^{[3]^n})\\
&\!\!\!\!\leq:\!\!\!\!& nM_R\log\rho+3nR-\hbar({\bf X}_3^{[4\sim 3]^n}).
\end{eqnarray}

\item{Step 8:} Consider RX 1 and TX 4. A genie
provides ${\bf \bar{G}}_8\!=\!\{{\bf \bar{X}}_3^{[4\sim 2]},{\bf \bar{X}}_3^{[4\sim
3]},{\bf \bar{X}}^{[2]}_{(7)},{\bf \bar{X}}^{[3]}\}$ to RX 1, such that RX 1 is
able to decode all the messages subject to the noise distortion. This
step is similar to Step 3. Therefore, we have the eighth inequality
as follows:
\begin{eqnarray}
4nR-n\epsilon_n&\!\!\!\!\leq:\!\!\!\!& \hbar({\bf Y}^{[1]^n})+\hbar({\bf G}_8^n|{\bf S}^{[1]^n})\\
&\!\!\!\!\leq:\!\!\!\!& nM_R\log\rho+2nR+\hbar({\bf X}_3^{[4\sim 2]^n})+\hbar({\bf
X}_3^{[4\sim 3]^n})-\hbar({\bf X}_1^{[2\sim 1]^n}).
\end{eqnarray}
\end{itemize}

Finally, adding up all the eight sum rate inequalities we have so
far, we obtain the following inequality:
\begin{eqnarray}
32nR-n\epsilon_n&\!\!\!\!\leq\!\!\!\!& 8M_Rn\log\rho+22nR-\hbar({\bf X}_1^{[2\sim 4]^n})-\hbar({\bf X}_3^{[2\sim 1]^n})-\hbar({\bf X}_3^{[2\sim 3]^n})-\hbar({\bf X}_1^{[2\sim 1]^n})\\
&\!\!\!\!\leq\!\!\!\!& 8M_Rn\log\rho+22nR-nR.
\end{eqnarray}
where $R-\epsilon_n=\hbar({\bf X}_1^{[2\sim 4]},{\bf X}_3^{[2\sim
1]},{\bf X}_3^{[2\sim 3]},{\bf X}_1^{[2\sim 1]})\leq \hbar({\bf
X}_1^{[2\sim 4]})+\hbar({\bf X}_3^{[2\sim 1]})+\hbar({\bf X}_3^{[2\sim
3]})+\hbar({\bf X}_1^{[2\sim 1]})$. By letting $\rho\rightarrow \infty$
and $n\rightarrow \infty$ we obtain the desired outer bound:
\begin{eqnarray}
d\leq \frac{8M_R}{11}=\frac{24}{11}.
\end{eqnarray}
\hfill\QED

\subsection{DoF of the $K$-User MIMO Interference Channel}

In this section, we take one simple example of the MIMO interference
channel beyond the $K=4$ user setting to convey the idea of genie
chains. Consider the $(K,M,N)=(5,4,15)$ setting, we will show that $d=\frac{MN}{M+N}=\frac{60}{19}$.
%and the assumptions associated with the channel knowledge is the same as that in Section \ref{sec:model}. As what we will discuss later in Section \ref{sec:big_picture}, our goal is to show each user has $d=\frac{MN}{M+N}=\frac{60}{19}$ DoF, i.e., the decomposition DoF value is optimal.

{\it Proof:} %The value of DoF $d=\frac{MN}{M+N}$ implies that the DoF outer bound should be consistent with the decomposition (inner) bound. Intuitively, as what we have shown in previous examples,
We need $M=4$ sum rate bounds, which can be produced through
the following four steps.
\begin{itemize}
\item{Step 1:} Start from RX 2 and TX 1. After decoding its desired message
$W_2$, RX 2 can reconstruct ${\bf X}^{[2]}$ and subtract it from its
received signal vector. Thus, RX 2 has 15 linear combinations of
$4\times 4=16$ interference symbols from TX 2 to TX 5. A genie
provides ${\bf \bar{G}}_1\!=\!{\bf \bar{X}}^{[1]}_{(1)}$ to RX 2. Since ${\bf
X}^{[1]}_{(1)}$ is linearly independent with all the other 15
dimensions, RX 2 can invert the $16\times 16$ square matrix to
reconstruct the interference vectors sent from all interferers, and
thus RX can decode all the messages. Thus, we obtain the first sum rate
inequality:
\begin{eqnarray}
5nR-n\epsilon_n&\!\!\!\!\leq:\!\!\!\!& \hbar({\bf Y}^{[2]^n})+\hbar({\bf G}_1^n|{\bf S}^{[2]^n})\\
&\!\!\!\!\leq:\!\!\!\!& Nn\log\rho+\hbar({\bf X}^{[1]^n}_{(1)}|\mathcal{O}_2^n)\\
&\!\!\!\!=:\!\!\!\!& Nn\log\rho+nR-\hbar(\mathcal{O}_2^n)
\end{eqnarray}
where $\mathcal{O}_2$ denotes the $3$ dimensional observations
(space) of TX 1 after RX 2, after zero-forcing the interference from
TX 3, TX 4 and TX 5.

\item{Step 2:} Since $|\mathcal{O}|+3>M$, we go to RX 3 looking at TX 1 for an
intermediate bound. Similar to Step 1, a genie provides ${\bf
\bar{G}}_2={\bf \bar{X}}^{[1]}_{(1)}$ to RX 3 such that it can decode all
messages. Again, RX 3 originally has $3$ dimensional observations of
TX 1 after it zero-forces the interference from TX 2, TX 4 and TX 5.
We denote by $\mathcal{O}_3$ the 3 dimensional observations at RX 3.
Because $|\mathcal{O}_2|\!+\!|\mathcal{O}_3|\!=\!6\!>\!M$, they have
an a $2$ dimensional intersection. We denote this intersection by
$\mathcal{I}_3=\mathcal{O}_2\cap\mathcal{O}_3$ at RX 3. Thus, we
have the intermediate bound as follows:
\begin{eqnarray}
5nR-n\epsilon_n&\!\!\!\!\leq:\!\!\!\!& \hbar({\bf Y}^{[3]^n})+\hbar({\bf G}_2^n|{\bf S}^{[3]^n})\\
&\!\!\!\!\leq:\!\!\!\!& Nn\log\rho+\hbar({\bf X}^{[1]^n}_{(1)}|\mathcal{O}_3^n)\\
&\!\!\!\!\leq:\!\!\!\!& Nn\log\rho+\hbar({\bf X}^{[1]^n}_{(1)},\mathcal{O}_3^n)-\hbar(\mathcal{O}_3^n)\\
&\!\!\!\!\leq:\!\!\!\!& Nn\log\rho+nR+\hbar(\mathcal{O}_2^n)-\hbar(\mathcal{I}_3^n,\mathcal{O}_3^n\!\setminus\!\mathcal{I}_3^n)-\hbar(\mathcal{O}_2^n)\\
&\!\!\!\!\leq:\!\!\!\!& Nn\log\rho+nR+\hbar(\mathcal{O}_2^n)-\hbar(\mathcal{I}_3^n)-\hbar(\mathcal{O}_3^n\!\setminus\!\mathcal{I}_3^n|\mathcal{I}_3^n)-\hbar(\mathcal{O}_2^n)\\
&\!\!\!\!\leq:\!\!\!\!& Nn\log\rho+nR+\hbar(\mathcal{O}_2^n)-\hbar(\mathcal{I}_3^n)-\hbar(\mathcal{O}_3^n\!\setminus\!\mathcal{I}_3^n|\mathcal{I}_3^n,\mathcal{O}_2^n)-\hbar(\mathcal{O}_2^n)\\
&\!\!\!\!\leq:\!\!\!\!& Nn\log\rho+nR+\hbar(\mathcal{O}_2^n)-\hbar(\mathcal{I}_3^n)-\hbar(\mathcal{O}_3^n\!\setminus\!\mathcal{I}_3^n|\mathcal{O}_2^n)-\hbar(\mathcal{O}_2^n)\\
&\!\!\!\!\leq:\!\!\!\!& Nn\log\rho+nR+\hbar(\mathcal{O}_2^n)-\hbar(\mathcal{I}_3^n)-nR\\
&\!\!\!\!\leq:\!\!\!\!& Nn\log\rho+\hbar(\mathcal{O}_2^n)-\hbar(\mathcal{I}_3^n).
\end{eqnarray}

\item{Step 3:} Next, because we still have $|\mathcal{I}_3|+3>M$, we go to RX 4 looking at
TX 1 for another intermediate bound. A genie provides ${\bf
\bar{G}}_3={\bf \bar{X}}^{[1]}_{(1)}$ to RX 4, which originally also has $3$
dimensional observations of TX 1, denoted as $\mathcal{O}_4$.
Because $|\mathcal{I}_3|\!+\!|\mathcal{O}_4|\!=\!5\!>\!M$ again,
they have an one-dimensional intersection. We denote this
intersection by $\mathcal{I}_4=\mathcal{I}_3\cap\mathcal{O}_4$ at RX
4. Similar to Step 2, this process produces the intermediate bound
\begin{eqnarray}
5nR-n\epsilon_n&\!\!\!\!\leq:\!\!\!\!& \hbar({\bf Y}^{[4]^n})+\hbar({\bf G}_3^n|{\bf S}^{[4]^n})\\
&\!\!\!\!\leq:\!\!\!\!&Nn\log\rho+\hbar(\mathcal{I}_3^n)-\hbar(\mathcal{I}_4^n).
\end{eqnarray}

\item{Step 4:} Finally, consider RX 5 looking at RX 1. A genie provides the one-dimensional symbol ${\bf
\bar{G}}_4=\mathcal{I}_4 + {\bf Z}$ to RX 5, which again originally has $3$
dimensional observations of TX 1, denoted as $\mathcal{O}_5$, which
combined with $\mathcal{I}_4$ allows RX 5 to recover ${\bf X}^{[1]}$
subject to the noise distortion. Once again, this process produces
the inequality
\begin{eqnarray}
5nR-n\epsilon_n&\!\!\!\!\leq:\!\!\!\!& \hbar({\bf Y}^{[5]^n})+\hbar({\bf G}_4^n|{\bf S}^{[5]^n})\\
&\!\!\!\!\leq:\!\!\!\!& Nn\log\rho+\hbar(\mathcal{I}_4^n).
\end{eqnarray}

Adding up all the four sum rate inequalities we have so far, we
obtain the following inequality:
\begin{eqnarray}
20nR-n\epsilon_n\leq: 4Nn\log\rho+nR.
\end{eqnarray}
By letting $\rho\rightarrow \infty$ and $n\rightarrow \infty$ we
obtain the desired bound:
\begin{eqnarray}
d\leq \frac{4N}{19}=\frac{60}{19}.
\end{eqnarray}
\end{itemize}
\hfill\QED

\subsection{DoF of the Many-to-One MIMO Interference Channel}

\begin{figure}[!ht] \vspace{-0.1in}
\centering
\includegraphics[width=3.0in]{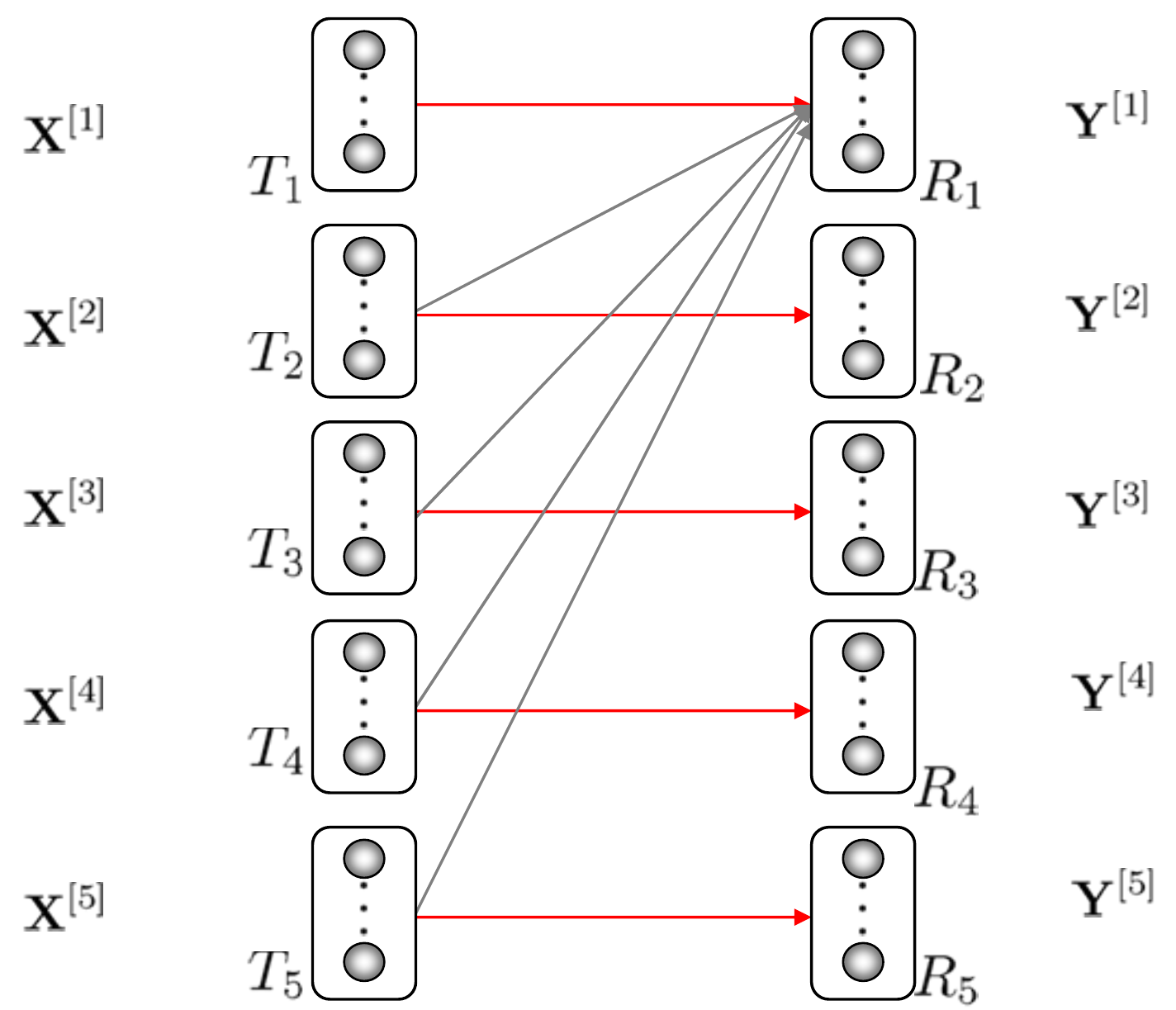}\vspace{-0.1in}
\caption{$(M,N)$ MIMO Five-to-One Interference Channel}
\vspace{-0.1in} \label{fig:many_to_one}
\end{figure}

In this section, we take a look at an example of Many-to-One
MIMO interference channel, where in a $K$
user interference channel, only one RX hears from all
TXs while the other RXs can only hear their own
desired signals. The Five-to-One MIMO interference channel
 is shown in Figure \ref{fig:many_to_one}, where each
TX has $M$ antennas and each RX has $N$ antennas.

\subsubsection{Five-to-One MIMO Interference Channel}

Consider the $(M,N)=(2,5)$ setting in Figure
\ref{fig:many_to_one}, we are interested in the DoF per user of this network.
Since RX $k,~k=2,3,4,5$ only hear their own desired signals
and $M<N$, they can decode their own messages respectively by the
reliable communications requirement. Thus, we need to ensure
that at RX 1 all interference is aligned among themselves as
much as possible. Note that since the DoF value implied by the linear
counting bound is less than that achieved by the decomposition
bound. Thus, we expect the decomposition DoF bound, i.e., $10/7$ DoF
per user, is also the information theoretic DoF outer bound. Next,
we will show it is true.

{\it Proof:} We need $M=2$ sum rate bounds,
which can be produced as follows.
\begin{itemize}
\item{Step 1:} A genie provides ${\bf
\bar{G}}_1\!=\!\{{\bf \bar{X}}^{[2]}_{(1)},{\bf \bar{X}}^{[3]}\}$ to RX 1. By the
assumption of reliable communications, RX 1 is able to decode its
desired message $W_1$. After decoding $W_1$, RX 1 can reconstruct
the signal vector ${\bf X}^{[1]}$ and then subtract it from ${\bf
\bar{Y}}^{[1]}$. Therefore, RX 1 has 5-dimensional observations of the
eight interference symbols from TX 2 to TX 5. Providing ${\bf \bar{G}}_1$
to RX 1 allows it to invert the square channel associated with the
interferers. Therefore, RX 1 can reconstruct all signal vectors sent
from all interferers, and thus RX 1 can decode all the messages. Thus
argument produces the
%zero-forcing the interference from TX 4 and TX 5, RX 1 still have
%one dimensional observation of the remaining interference, i.e., a
%linear combination of ${\bf X}^{[3]}, {\bf X}^{[2]}$. Providing
%${\bf X}^{[3]}$ as a genie to RX 1 can also let it remove ${\bf
%X}^{[3]}$, thus the one dimensional observation becomes a linear
%combination of ${\bf X}^{[2]}$. Since ${\bf X}^{[2]}_{(1)}$ is
%generic, which is also linearly independent with the one dimensional
%observation at RX 1, RX 1 is able to decode message $W_2$ subject to
%the noise distortion. Finally, RX 1 can decode all four interference
%messages, and thus we have the
following inequality:
\begin{eqnarray}
5nR-n\epsilon_n&\!\!\!\!\leq:\!\!\!\!& \hbar({\bf Y}^{[1]^n})+\hbar({\bf G}_1^n|{\bf S}^{[1]^n})\\
&\!\!\!\!\leq:\!\!\!\!& Nn\log\rho+\hbar({\bf X}^{[2]^n}_{(1)},{\bf X}^{[3]^n}|{\bf S}^{[1]^n})\\
&\!\!\!\!\leq:\!\!\!\!& Nn\log\rho+\hbar({\bf X}^{[3]^n})+\hbar({\bf X}^{[2]^n}_{(1)}|{\bf S}^{[1]^n},{\bf X}^{[3]^n})\\
&\!\!\!\!\leq:\!\!\!\!& Nn\log\rho+nR+\hbar({\bf X}^{[2]^n}_{(1)}|\mathcal{O}^n)\\
&\!\!\!\!\leq:\!\!\!\!& Nn\log\rho+nR+\hbar({\bf X}^{[2]^n}_{(1)},\mathcal{O}^n)-\hbar(\mathcal{O}^n)\\
&\!\!\!\!\leq:\!\!\!\!& Nn\log\rho+2nR-\hbar(\mathcal{O}^n).
\end{eqnarray}
where $\mathcal{O}$ is the one dimensional observation of TX 2 after
RX 1 removes its own signal ${\bf X}^{[1]}$, the provided genie
signal ${\bf \bar{X}}^{[3]}$ and zero-forces interference from TX 4 and TX
5.

\item{Step 2:} A genie provides ${\bf \bar{G}}_2\!=\!\{\mathcal{O} + {\bf Z},{\bf \bar{X}}^{[4]}\}$
to RX 1. Similar to Step 1, it can be easily seen that RX 1 can
decode all the messages as well. Thus, we obtain the second
inequality as follows:
\begin{eqnarray}
5nR-n\epsilon_n&\!\!\!\!\leq:\!\!\!\!& \hbar({\bf Y}^{[1]^n})+\hbar({\bf G}_2^n|{\bf S}^{[1]^n})\\
&\!\!\!\!\leq:\!\!\!\!& Nn\log\rho+\hbar(\mathcal{O}^n,{\bf X}^{[4]^n})\\
&\!\!\!\!\leq:\!\!\!\!& Nn\log\rho+\hbar(\mathcal{O}^n)+\hbar({\bf X}^{[4]^n})\\
&\!\!\!\!\leq:\!\!\!\!&  Nn\log\rho+nR+\hbar(\mathcal{O}^n).
\end{eqnarray}

Adding up all the four sum rate inequalities we have so far, we
obtain the following inequality:
\begin{eqnarray}
10nR-n\epsilon_n\leq: 2Nn\log\rho+3nR.
\end{eqnarray}
By letting first $n\rightarrow \infty$ and then $\rho\rightarrow
\infty$ we obtain the desired the DoF outer bound:
\begin{eqnarray}
d\leq \frac{2N}{7}=\frac{10}{7}.
\end{eqnarray}
\end{itemize}
\hfill\QED

\subsubsection{Four-to-One MIMO Interference Channel}

Besides the example of $K=5$ setting shown above, we also
present the DoF results of $K=4$ setting in the following, by
eliminating the fifth user from the network shown in Figure
\ref{fig:many_to_one}. In order to understand the interplay among
spatial signal dimensions projected from interferers without
zero-forcing at the TX side, we only consider the $M\leq N$
setting. The DoF results are included in the following theorem.
\begin{theorem}\label{theorem:4to1}
For a Four-to-One $M\times N$ MIMO Gaussian interference channel
where each TX has $M$ antennas, each RX has $N$
antennas and $M\leq N$, the DoF value per user is given by:
\begin{eqnarray}
d=\left\{\begin{array}{ccl}M,&&M/N\leq 1/4,\\
N/4,&&1/4\leq M/N\leq 1/3,\\
3M/4,&&1/3\leq M/N\leq 4/9,\\
N/3,&&4/9\leq M/N\leq 1/2,\\
2M/3,&&1/2\leq M/N\leq 3/5,\\
2N/5,&&3/5\leq M/N\leq 2/3,\\
3M/5,&&2/3\leq M/N\leq 5/6,\\
N/2,&&5/6\leq M/N\leq 1.
\end{array}\right.
\end{eqnarray}
\end{theorem}

{\it Proof:} The DoF achievability relies on linear interference
alignment schemes. In addition, since the rigorous proof still
follows the subspace alignment chains that we introduced in
\cite{Wang_Gou_Jafar_3userMxN} and genie chains that we primarily
investigate in this paper, we defer the proof to Appendix
\ref{app:4to1}, and only show the intuitions in this section.
\hfill\QED

The DoF results are shown in Figure \ref{fig:dof_4to1} where the red
line represents the DoF counting bound which is derived in Appendix
\ref{app:counting}. Notice that Theorem \ref{theorem:4to1} implies
that the DoF value is a piecewise linear function depending on $M$
and $N$ alternatively, which means that there are antenna
redundancies at either the TX side or the RX side.
While it is again similar to the DoF value of the three user MIMO
interference channel recently shown by Wang et. al. in
\cite{Wang_Gou_Jafar_3userMxN}, the DoF cruve only contains eight
pieces, in contrast to infinitely many of pieces in the three user
MIMO interference channel. In addition, because there is only one
RX, we only need to deal with the signal dimensions projected
from three interferers at that RX. Thus, understanding the
spatial signal dimensions of this network is helpful for us to learn
the spatial signal dimensions of more general networks.

\begin{figure}[!ht] \vspace{-0.1in}
\centering
\includegraphics[width=5.0in]{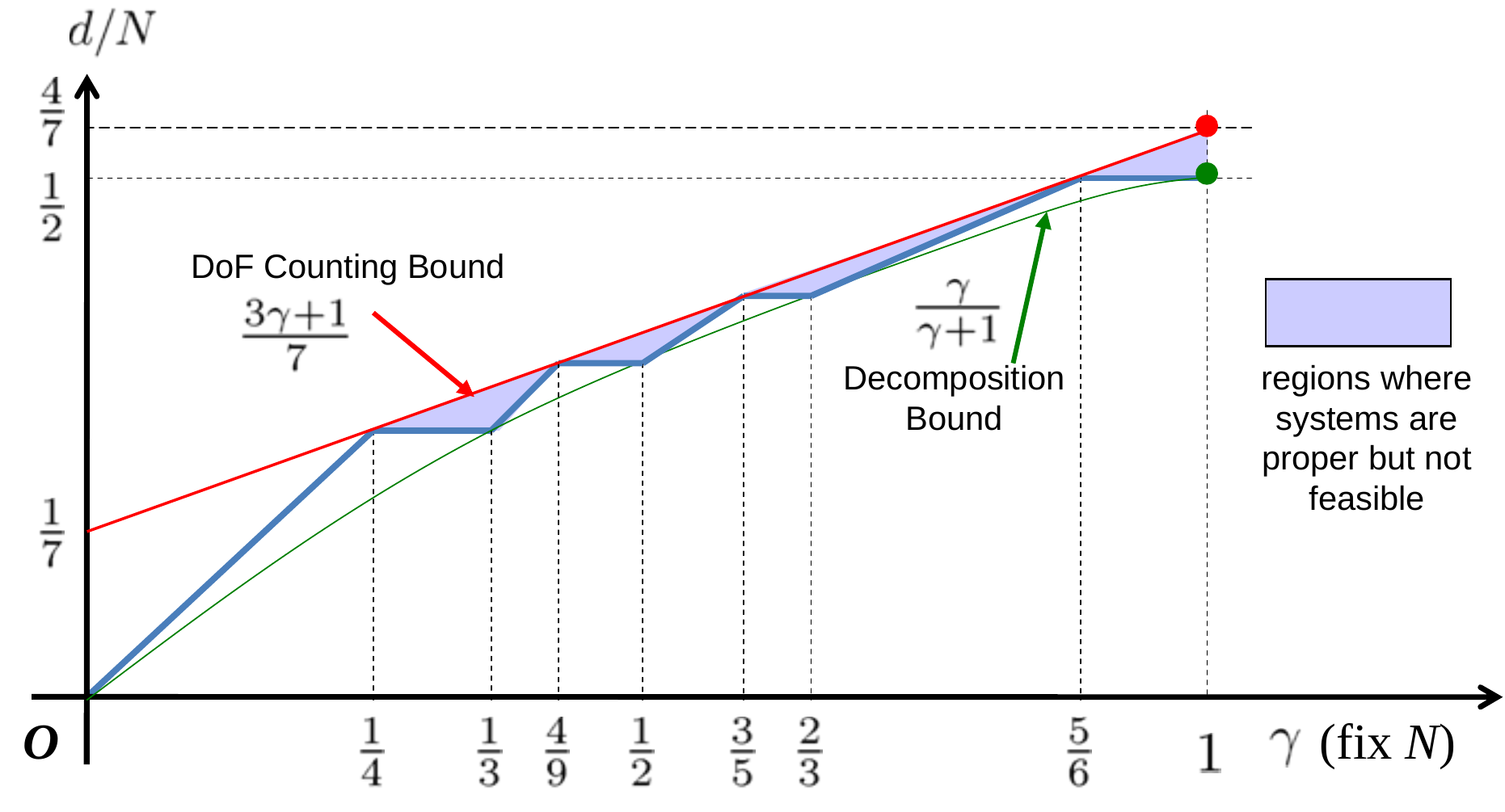}\vspace{-0.1in}
\caption{$d/N$ as a function of $\gamma=M/N$ for the Four-to-One
MIMO interference channel} \vspace{-0.1in} \label{fig:dof_4to1}
\end{figure}

In essence, Many-to-One MIMO interference channels are similar to cellular networks with two cells, as alignment is demanded to take place at only one interferer. Based on similar insights, the DoF value of MIMO two-cell cellular networks with 2 and 3 users per cell is found in \cite{Sridharan_Yu_2, Sridharan_Yu_3}.

\subsection{DoF of the MIMO $X$ Channel}

\begin{figure}[!ht]
\centering
\includegraphics[width=4.0in]{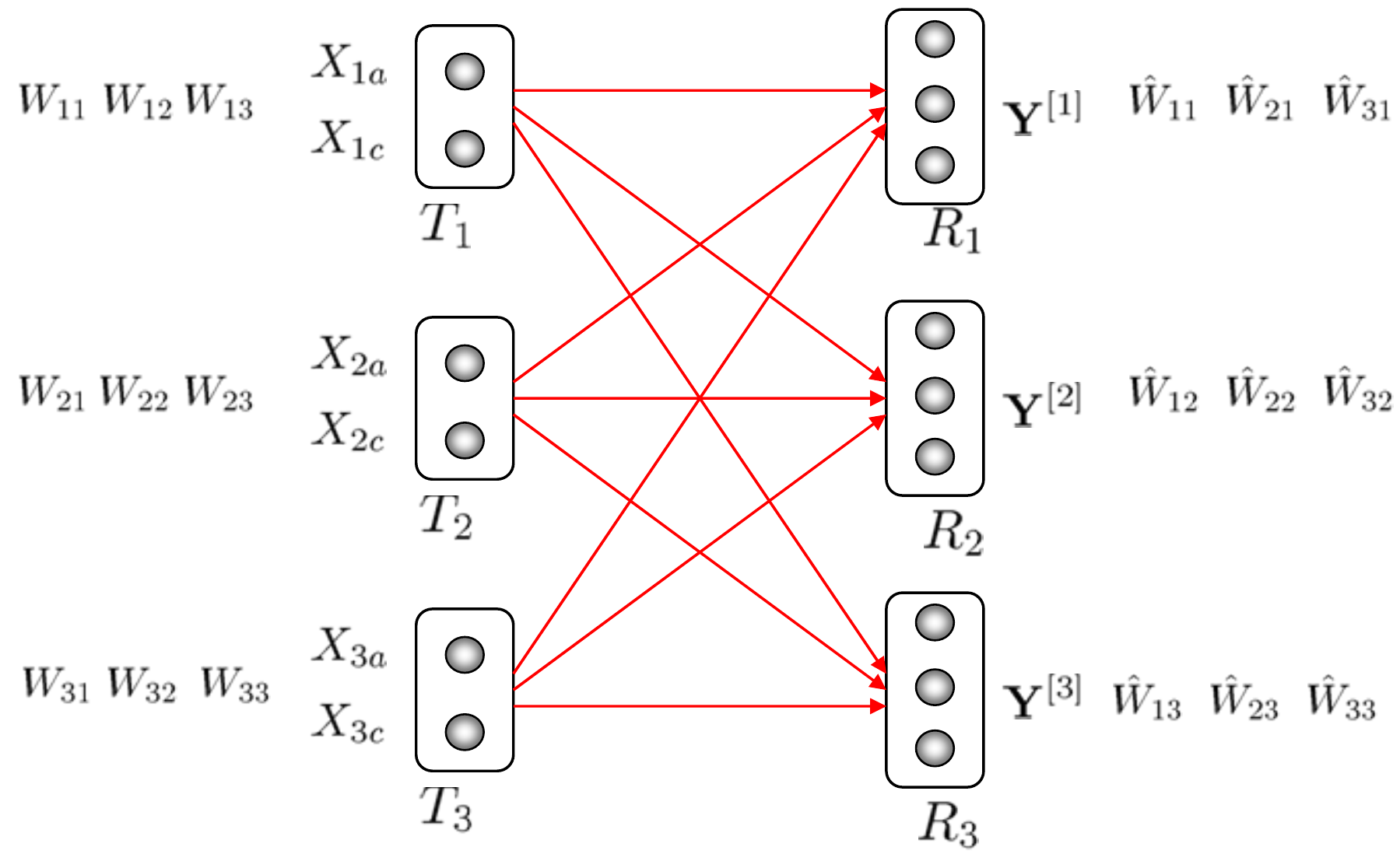}
\caption{$(M,N)=(2,3)$ MIMO $X$ Channel} \vspace{-0.1in}
\label{fig:2by3_mimoX}
\end{figure}

Besides the multiuser interference channel, the tool of genie chains
can also be applied in the MIMO $X$ channel as well. We show one
simple example in this section. Consider a $K=3$ user MIMO $X$
channel where each TX has $M=2$ antennas and each RX
has $N=3$ antennas, as shown in Figure \ref{fig:2by3_mimoX}. Each
TX $T_i$ sends one independent message $W_{ij}$ to RX
$R_j$, $i,j\in\{1,2,3\}$. Again, the constant complex channel coefficients are
assumed to be independently drawn from continuous distributions. Also, global channel knowledge is
assumed to be available at all nodes. %Other notations and definitions directly follow from that in Section \ref{sec:model}, by noting that
We refer to $R_{ij}$ and
$d_{ij}$ as the rate and DoF, respectively, of the message $W_{ij}$.
Again we are interested in the DoF of this network. Note that the
value of DoF implied by the linear counting bound is less than that
achieved by the decomposition bound. Thus, we expect the
decomposition DoF bound, i.e., $10/7$ DoF per user, is also the
information theoretic DoF outer bound. Next, we show it is
true.

{\it Proof:} We need $M_T=2$ sum rate
bounds, produced as follows.
\begin{itemize}
\item{Step 1:} A genie provides ${\bf
\bar{G}}_1\!=\!\{\bar{X}_{2a},W_{32},W_{33}\}$ to RX 1. By the assumption of
reliable communications, RX 1 is able to decode
$W_{11},W_{21},W_{31}$ from the observations ${\bf \bar{Y}}^{[1]}$. Also,
providing $W_{32},W_{33}$ to RX 1 allows it to reconstruct the
signal ${\bf X}^{[3]}=[X_{3a}~X_{3c}]^T$ and subtract it from ${\bf
\bar{Y}}^{[1]}$. Then the remaining interference comes from TX 1 and TX 2.
Since RX 1 has already three antennas, providing $X_{2a}$ to RX 1
allows it invert the square matrix and reconstruct the transmit
signal vectors from TX 1 and TX 2 subject to the noise distortion,
and thus RX 1 is able to decode all the messages subject to the
distortion. Therefore, we have the following inequality:
\begin{eqnarray}
9nR-n\epsilon_n&\!\!\!\!\leq:\!\!\!\!& \hbar({\bf Y}^{[1]^n})+\hbar({\bf G}_1^n|{\bf S}^{[1]^n})\\
&\!\!\!\!=:\!\!\!\!& Nn\log\rho+\hbar(X_{2a}^n,W_{32},W_{33}|{\bf S}^{[1]^n},W_{11},W_{21},W_{31})\\
&\!\!\!\!\leq:\!\!\!\!& Nn\log\rho+\hbar(W_{32},W_{33})+\hbar(X_{2a}|{\bf Y}^{[1]^n},W_{21},W_{31},W_{32},W_{33})\\
&\!\!\!\!\leq:\!\!\!\!& Nn\log\rho+2nR+\hbar(X_{2a}^n|\mathcal{O}^n,W_{21})\\
&\!\!\!\!=:\!\!\!\!& Nn\log\rho+2nR+\hbar(X_{2a}^n,\mathcal{O}^n,W_{21})-\hbar(W_{21})-\hbar(\mathcal{O}^n|W_{21})\\
&\!\!\!\!=:\!\!\!\!& Nn\log\rho+2nR+2nR-\hbar(\mathcal{O}^n|W_{21}).
\end{eqnarray}
where $\mathcal{O}$ is the one dimensional linear combination of
$X_{2a}$ and $X_{2c}$.

\item{Step 2:} A genie provides ${\bf \bar{G}}_2\!=\!\{\mathcal{O} + {\bf Z},W_{12},W_{13}\}$
to RX 1. Similar to Step 1, it can be easily seen that RX 1 can
decode all the messages as well. Thus, we obtain the second
inequality as follows:
\begin{eqnarray}
9nR-n\epsilon_n&\!\!\!\!\leq:\!\!\!\!& \hbar({\bf Y}^{[1]^n})+\hbar({\bf G}_2|{\bf S}^{[1]^n})\\
&\!\!\!\!\leq:\!\!\!\!& Nn\log\rho+\hbar(\mathcal{O}^n,W_{12},W_{13}|{\bf S}^{[1]^n},W_{21})\\
&\!\!\!\!\leq:\!\!\!\!& Nn\log\rho+\hbar(W_{12},W_{13})+\hbar(\mathcal{O}^n|W_{21})\\
&\!\!\!\!\leq:\!\!\!\!& Nn\log\rho+2nR+\hbar(\mathcal{O}^n|W_{21}).
\end{eqnarray}

Adding up all the four sum rate inequalities we have so far, we
obtain the following inequality:
\begin{eqnarray}
18nR-n\epsilon_n\leq: 2Nn\log\rho+6nR.
\end{eqnarray}
By letting $\rho\rightarrow \infty$ and $n\rightarrow \infty$ we
obtain the desired the DoF outer bound:
\begin{eqnarray}
d\leq \frac{2N}{12}=\frac{1}{2}=\frac{MN}{3M+2N}.
\end{eqnarray}
\end{itemize}
\hfill\QED

\section{Discussions on the DoF Characterization of the $K$-User MIMO Gaussian Interference
Channel}\label{sec:big_picture}

In this paper, since our primary goal is to introduce the genie
chains approach, and highlight the principles that could be applied
to not only MIMO interference channels, but also many other wireless
networks such as $X$ channel, etc., we only embark most effort to
the $K=4$ user case with $M_T\leq M_R$ to present the main ideas in
Section \ref{sec:4user}. In this section, we continue to discuss the
DoF results of the $K$-user $M_T\times M_R$ MIMO interference channel, not through rigorous proof
for each case but based on the available observations that we obtained so
far, to show a broad and fundamental DoF picture of the MIMO interference channel. Since the DoF
results of the $K=2$ and $K=3$ User $M_T\times M_R$ MIMO
Interference Channel have been reported in \cite{Jafar_Fakhereddin,
Wang_Gou_Jafar_3userMxN}, respectively, we begin with $K\geq 4$
cases.

\subsection{Unstructured Linear Schemes Achieving the Information Theoretic DoF Outer Bound}

In Section \ref{sec:4user}, we mention that the DoF result and corresponding proofs for $M/N\leq
3/8$ where $M=\min(M_T,M_R)$ and $N=\max(M_T,M_R)$ directly follows from the $K=3$ user case \cite{Wang_Gou_Jafar_3userMxN}. In fact, we can also extend the results for the general $K$ user case for $M/N\leq \frac{K-1}{K(K-2)}$. Similar to the $K=3$ user case, we show that linear interference alignment schemes are sufficient to
achieve the information theoretic DoF outer bound under the sense of spatial normalization.
Our results are presented in the following lemmas and theorem.

\begin{definition} We define the following quantity:
\begin{eqnarray}\label{eqn:lineardof}
d^*=\left\{\begin{array}{ccl}M,&& 0<\frac{M}{N}\leq \frac{1}{K},\\
\frac{N}{K},&& \frac{1}{K}\leq\frac{M}{N}\leq \frac{1}{K-1},\\
\frac{(K-1)M}{K},&& \frac{1}{K-1}\leq \frac{M}{N}\leq \frac{K}{K^2-K-1},\\
\frac{(K-1)N}{K^2-K-1},&& \frac{K}{K^2-K-1}\leq \frac{M}{N}\leq \frac{K-1}{K(K-2)}.\end{array}\right.
\end{eqnarray}
\end{definition}

\begin{lemma}\label{lemma:4user_linear_ob}
For the $K\geq 4$ user $M_T\times M_R$ MIMO interference
channel where each TX has $M_T$ and each RX has $M_R$
antennas, if $M/N\leq \frac{K-1}{K(K-2)}$, then the DoF per user are outer bounded by $d\leq d^*$.
\end{lemma}

{\it Proof:} Since the idea behind the proof for this lemma directly follows from the $K=3$ user case \cite{Wang_Gou_Jafar_3userMxN} yet requires much more complicated analysis,
we defer the proof of Lemma \ref{lemma:4user_linear_ob} to Appendix \ref{app:sec_linear}.1 and \ref{app:sec_linear}.2.

\begin{lemma}\label{lemma:4user_linear_ib}
For the $K\geq 4$ user $M_T\times M_R$ MIMO interference
channel where each TX has $M_T$ and each RX has $M_R$ antennas, if $M/N\leq
\frac{K-1}{K(K-2)}$, then $d^*$ DoF per user are achievable in the sense of spatial normalization.
\end{lemma}

{\it Proof:} Again, the idea behind the proof for this lemma directly follows from the $K=3$ user case \cite{Wang_Gou_Jafar_3userMxN}.
Thus, we defer the proof of Lemma \ref{lemma:4user_linear_ib} to Appendix \ref{app:sec_linear}.3.

\begin{theorem}\label{theorem_linear}
For the $K\geq 4$ user $M_T\times M_R$ MIMO interference
channel where each TX has $M_T$ and each RX has $M_R$ antennas, if $M/N\leq
\frac{K-1}{K(K-2)}$, then the spatial normalized DoF value per user is given by $d=d^*$.
\end{theorem}

{\it Proof:} The proof of this theorem directly follows Lemma \ref{lemma:4user_linear_ob} and Lemma \ref{lemma:4user_linear_ib}.

{\it Remark:} Basically, the DoF analysis above follows from the same intuition that we highlight in \cite{Wang_Gou_Jafar_3userMxN} that
reducing/increasing antenna dimension redundancies at any users does not increase/decrease channel capacity.

\subsection{The Decomposition DoF Bound Achieving the Information Theoretic DoF Outer Bound}

Next, let us consider if the value of $M/N$ falls into the interval
$\left(\frac{K-1}{K(K-2)},1\right)$. In Section \ref{sec:4user},
Theorem \ref{theorem:4user} already includes cases of $M_T/M_R\in
\mathcal{P}_1\cup \mathcal{P}_2\cup \mathcal{P}_3=\mathcal{P}$ for the $K=4$ setting,
where $\mathcal{P}_1=\{\frac{M_T}{M_R}|\frac{1}{2}\leq \frac{M_T}{M_R}<1,~M_T,M_R\in\mathbb{Z}^+,~M_R\leq 20\}$, $\mathcal{P}_2=[2/5,1/2)$, and $\mathcal{P}_3=\{\frac{8}{21}\}\cup\{\frac{2c-1}{5c-2}|c\in \mathbb{Z}^+,c\geq 2\}$. Next, we
will discuss the renaming cases of the $K=4$ setting not covered by the set $\mathcal{P}$ and the $K>4$
settings, to shed light on the insights behind DoF results of the
general $K$-user $M_T\times M_R$ MIMO Gaussian interference channel.
We begin with the $M_T<M_R$ setting.

First, it can be easily shown that any information theoretical DoF
outer bounds for the $K=K_0$ setting are valid information
theoretical DoF outer bounds for the $K>K_0$ settings as well.
Intuitively, this is because increasing the number of users in a
network cannot increase the symmetric capacity per user. Thus, for
$M_T/M_R\in \mathcal{P}$ cases, the DoF value per user
$d=\frac{MN}{M+N}$ is also optimal for $K>4$ settings.

Second, notice that for $K=4$ setting, the left boundary value of $\mathcal{P}_2$, i.e., $M_T/M_R=2/5$ is obtained in Section \ref{sec:4user} by showing that we never need \emph{successive two intermediate bounds}
by applying the genie chains approach for the $K=4,~M_T/M_R\geq 2/5$ setting. This
argument implies that at a given RX looking at one interferer,
we do not need to exhaust the other \emph{two unintended RXs}
to produce successive two intermediate bounds. Therefore, if we
apply the genie chains approach, then for the general $K\geq 4$ user
setting, the decomposition DoF bound $d=\frac{MN}{M+N}$ per user is
expected to be the information theoretic DoF outer bound as well, as
long as we never need \emph{successive $K-2$ intermediate bounds},
i.e., we only need up to successive $K-3$ intermediate bounds.
Equivalently, this implies that through the union of $N-(K-2)M$
dimensional observations of a given interferer at each unintended
RX, we have a total of $(K-2)(N-(K-2)M)$ dimensional
observations of that interferer, which contribute to recovery of the
transmit signal vector of that TX up to $K-3$ times. Thus,
we have the following inequality:
\begin{eqnarray}
(K-2)(N-(K-2)M)\leq (K-3)M\Rightarrow \frac{M}{N}\geq
\frac{K-2}{K^2-3K+1}.
\end{eqnarray}
Therefore, for any cases of $\frac{M_T}{M_R}\in
\left[\frac{K-2}{K^2-3K+1},1\right)$, we expect that the DoF value
per user $d=\frac{MN}{M+N}$ is also the information theoretic DoF
outer bound. Notice that when the value of $K$ grows,
$\frac{K-2}{K^2-3K+1}$ approaches zero, such that
$\left[\frac{K-2}{K^2-3K+1},1\right)$ becomes the dominant interval.

What remains to be shown is the regime $\frac{M_T}{M_R}\in\left(\frac{K-1}{K(K-2)},\frac{K-2}{K^2-3K+1}\right)$, which comprises of both regime 1 and regime 2 partially.
In Theorem \ref{theorem:4user}, we only show the DoF results of cases
$\frac{M_T}{M_R}\in \mathcal{P}_3$ for the $K=4$ setting, and the point sequence in
$\mathcal{P}_3$ converges to the boundary $2/5$. Although there
are infinitely many number of points in the set $\mathcal{P}_3$, they
have only zero measure, i.e., $\mathcal{P}_3$ is not a dense set.
Thus, the DoF characterization in this sub-regime is still open in general.
Although we conjectured $\frac{MN}{M+N}$ is still the DoF value per user in \cite{Wang_Sun_Jafar_ISIT}, feasibility analysis and numerical evidence in \cite{Feasibility} indicate this is not the case. Moreover, \cite{Feasibility} provides evidence that part of this regime behaves like the piece-wise linear regime discussed in the previous section.

%On the
%other hand, as what we have shown, for all cases of
%$\frac{M_T}{M_R}\in\left(\frac{K-1}{K(K-2)},\frac{K-2}{K^2-3K+1}\right)$,
%we may need to exhaust all the unintended receivers for up to $K-2$
%intermediate bounds by the genie chain approach. Since we do have
%$K-2$ unintended receivers, intuitively we could produce these
%intermediate bounds in some way, even combining with multiple genie
%chains. The $M_T/M_R=8/21$ for the $K=4$ setting is such an example
%which needs successive $K-2$ intermediate bounds.

Finally, for all known DoF results, the information theoretic DoF
satisfy  the principle of duality. That is,
the original channel and its reciprocal channel both have the same
number of DoF.

%Therefore, combining the results we establish and the intuitions we have so far, we have the following the following conjecture.

%\begin{conjecture}\label{conject}
%Consider the $K$ user $M_T\times M_R$ MIMO Gaussian interference
%network defined in Section \ref{sec:model} where $M=\min(M_T,M_R)$
%and $N=\max(M_T,M_R)$. If
%$\frac{M}{N}\geq \frac{K-1}{K(K-2)}$, then the DoF
%per user are given by $d=\frac{MN}{M+N}$, almost surely.
%\end{conjecture}

\subsection{Observations}\label{sec:observation}

\begin{figure}[!ht]
\centering
\includegraphics[width=6.0in]{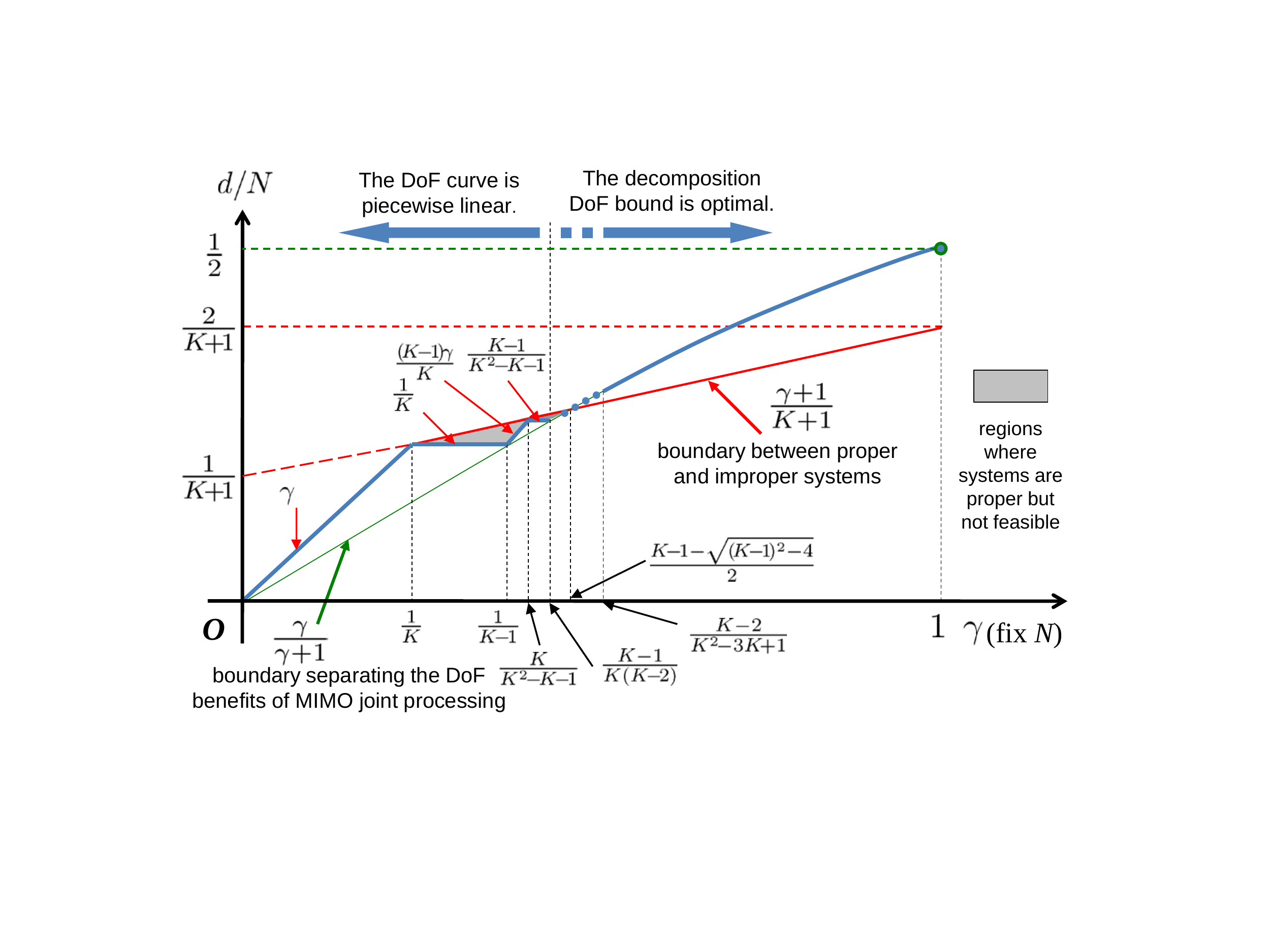} \vspace{-0.1in}
\caption{$d/N$ as a function of $\gamma=M/N$}
\label{fig:result}
\end{figure}

Now let us collect all the DoF results
%implied by Theorem \ref{theorem},
%Theorem \ref{theorem_linear} and Conjecture \ref{conject}
in Figure \ref{fig:result}.
There are three curves in Figure \ref{fig:result}. As we introduce in Section \ref{sec:introduction}, the red line and the green curve are the DoF counting bound and the DoF decomposition, respectively.
%The DoF results implied by Theorem \ref{theorem}, Theorem \ref{theorem_linear} and Conjecture \ref{conject} are shown in the blue curve.
It can be seen that if $M/N\leq \frac{K-1}{K(K-2)}=\gamma_0$, then the DoF curve is a
zigzag piecewise linear function depending on $M$ and $N$
alternately, which is similar to that of the $K=3$ setting
\cite{Wang_Gou_Jafar_3userMxN}. Intuitively, it means that there are
antenna redundancies at either TX or RX sides except
for $\frac{M}{N}=\frac{1}{K}$ and $\frac{K}{K^2-K-1}$. The
achievability relies on linear interference alignment schemes
without symbol extensions or with finite number of symbol
extensions (through numerical tests), i.e., asymptotical alignment is unnecessary. On the
other hand, if $\frac{M}{N}\geq \gamma_0$, then the
decomposition is expected to be optimal in many cases, and the
achievability relies on the asymptotic alignment. Note that when the
value of $K$ grows, this cross point moves towards to the left, so
that the interval $[\gamma_0,1)$ becomes the dominant, and the DoF
decomposition bound is optimal. Figure \ref{fig:result} also implies that
whenever the decomposition bound is larger than the counting bound,
the DoF decomposition bound is the information theoretic optimal.

%So far, from all DoF results we have obtained and all observations we introduced in this paper, we have the following conjecture.
%\begin{conjecture}
%For a general wireless interference channels, if the DoF decomposition bound is larger than
%the DoF counting bound, then the DoF decomposition bound is also the
%information theoretic DoF outer bound.
%\end{conjecture}

\section{Conclusion}

In this paper, we propose a novel tool, called genie chains, to study the information theoretic DoF outer bound of wireless interference networks, which essentially translates an information theoretic DoF outer bound problem into a much simpler linear algebraic problem. While this new tool has wide applications in various wireless interference networks, in this paper, we mainly study the MIMO interference channel as a typical example, followed by several other special examples including the many-to-one MIMO interference channel and the MIMO $X$ channel.

\appendix{}
\section*{Appendix}

\section{Proof of Lemma \ref{lemma:noise_distortion}}\label{app:lemma_noise_distortion}
Since $B_i({\bf L}), B_j({\bf L})$ both represent the basis of the same subspace ${\bf L}$, there exists an invertible $|{\bf L}|\times|{\bf L}|$ square matrix $A$ where $|{\bf L}|$ is the number of dimensions of the subspace ${\bf L}$, so that $B_i({\bf L})=A\cdot B_j({\bf L})$. Then we have
\begin{eqnarray}
\hbar(B_i({\bf L}){\bf X})&\!\!\!\!=\!\!\!\!&\hbar(A\cdot B_j({\bf L}){\bf X})\\
&\!\!\!\!=\!\!\!\!& h(A\cdot B_j({\bf L}){\bf X}+{\bf Z})\\
&\!\!\!\!=\!\!\!\!& h(A(B_j({\bf L}){\bf X}+A^{-1}{\bf Z}))\\
&\!\!\!\!=\!\!\!\!& h(B_j({\bf L}){\bf X}+A^{-1}{\bf Z})+\log|\det(A)|\\
&\!\!\!\!=\!\!\!\!& h(B_j({\bf L}){\bf X}+A^{-1}{\bf Z})+~o(\log\rho)\label{lemma:noise_distortion_eq1}
\end{eqnarray}
where $\log|\det(A)|$ is a constant which does not depend on the SNR, $\rho$. Notice that ${\bf Z}\sim\mathcal{CN}({\bf 0},{\bf I})$, thus the noise term $A^{-1}{\bf Z}\sim\mathcal{CN}({\bf 0}, {\bf K})$ where ${\bf K}=A^{-1}(A^{-1})^H$. Since ${\tilde{\bf Z}}$ is an independent noise vector, we have
\begin{eqnarray}
h(B_j({\bf L}){\bf X}+A^{-1}{\bf Z})=h(B_j({\bf L}){\bf X}+A^{-1}{\bf Z}+{\tilde{\bf Z}}|{\tilde{\bf Z}})\leq h(B_j({\bf L}){\bf X}+A^{-1}{\bf Z}+{\tilde{\bf Z}}).\label{lemma:noise_distortion_neq1}
\end{eqnarray}
On the other hand, since $B_j({\bf L}){\bf X}+A^{-1}{\bf Z}+{\tilde{\bf Z}}$ is a degraded version of $B_j({\bf L}){\bf X}+A^{-1}{\bf Z}$, we have
\begin{eqnarray}
0&\!\!\!\!\leq\!\!\!\!& I(B_j({\bf L}){\bf X};B_j({\bf L}){\bf X}+A^{-1}{\bf Z})-I(B_j({\bf L}){\bf X};B_j({\bf L}){\bf X}+A_{|{\bf L}|\times|{\bf L}|}^{-1}{\bf Z}+{\tilde{\bf Z}})\\
&\!\!\!\!=\!\!\!\!& h(B_j({\bf L}){\bf X}+A^{-1}{\bf Z})-h(B_j({\bf L}){\bf X}+A^{-1}{\bf Z}|B_j({\bf L}){\bf X})\notag\\
&\!\!\!\!&- h(B_j({\bf L}){\bf X}+A^{-1}{\bf Z}+{\tilde{\bf Z}})+h(B_j({\bf L}){\bf X}+A^{-1}{\bf Z}+{\tilde{\bf Z}}|B_j({\bf L}){\bf X})\\
&\!\!\!\!=\!\!\!\!& h(B_j({\bf L}){\bf X}+A^{-1}{\bf Z})-h(B_j({\bf L}){\bf X}+A^{-1}{\bf Z}+{\tilde{\bf Z}})-h(A^{-1}{\bf Z})+h(A^{-1}{\bf Z}+{\tilde{\bf Z}})\\
&\!\!\!\!=\!\!\!\!& h(B_j({\bf L}){\bf X}+A^{-1}{\bf Z})-h(B_j({\bf L}){\bf X}+A^{-1}{\bf Z}+{\tilde{\bf Z}})+\log(\det(\tilde{\bf K}{\bf K}^{-1}+{\bf I})).\label{lemma:noise_distortion_neq2}
\end{eqnarray}
Combining (\ref{lemma:noise_distortion_neq1}) and (\ref{lemma:noise_distortion_neq2}) produces
\begin{eqnarray}
h(B_j({\bf L}){\bf X}\!+A^{-1}{\bf Z})\leq h(B_j({\bf L}){\bf X}\!+A^{-1}{\bf Z}\!+{\tilde{\bf Z}})\leq h(B_j({\bf L}){\bf X}+A^{-1}{\bf Z}))\!+\log(\det(\tilde{\bf K}{\bf K}^{-1}\!+{\bf I}))
\end{eqnarray}
where $\log(\det(\tilde{\bf K}{\bf K}^{-1}+{\bf I}))$ is a constant which does not depend on $\rho$.
Thus we have
\begin{eqnarray}
h(B_j({\bf L}){\bf X}+A^{-1}{\bf Z})= h(B_j({\bf L}){\bf X}+A^{-1}{\bf Z}+{\tilde{\bf Z}})+~o(\log\rho).
\end{eqnarray}
Following the similar procedure, we also obtain
\begin{eqnarray}
h(B_j({\bf L}){\bf X}+{\tilde{\bf Z}})= h(B_j({\bf L}){\bf X}+A^{-1}{\bf Z}+{\tilde{\bf Z}})+~o(\log\rho).
\end{eqnarray}
Thus, we have
\begin{eqnarray}
h(B_j({\bf L}){\bf X}+A^{-1}{\bf Z})=h(B_j({\bf L}){\bf X}+{\tilde{\bf Z}})+~o(\log\rho).\label{lemma:noise_distortion_eq2}
\end{eqnarray}
Finally, substituting (\ref{lemma:noise_distortion_eq2}) into (\ref{lemma:noise_distortion_eq1}), we obtain
\begin{eqnarray}
\hbar(B_i({\bf L}){\bf X})= h(B_j({\bf L}){\bf X}+A^{-1}{\bf Z})+~o(\log\rho)= h(B_j({\bf L}){\bf X}+{\tilde{\bf Z}})+~o(\log\rho).
\end{eqnarray}
So far, we complete the proof of Lemma \ref{lemma:noise_distortion}.

\section{DoF of the $\frac{M}{N}\leq\frac{K-1}{K(K-2)}$ Setting for the $K$ User $M_T\times M_R$ MIMO Interference
Channel}\label{app:sec_linear}

For the $K$ user $M_T\times M_R$ MIMO interference channel, the DoF
value is a piecewise linear function of $M$ and $N$ alternately if
$\frac{M}{N}\leq\frac{K-1}{K(K-2)}$ where $M=\min(M_T,M_R)$ and
$N=\max(M_T,M_R)$. As shown in Theorem \ref{theorem_linear}, the DoF value
per user is given by:
\begin{eqnarray}\label{eqn:thm_linear}
d =\left\{\begin{array}{ccl}M,&& 0<\frac{M}{N}\leq \frac{1}{K},\\
\frac{N}{K},&& \frac{1}{K}\leq\frac{M}{N}\leq \frac{1}{K-1},\\
\frac{(K-1)M}{K},&& \frac{1}{K-1}\leq \frac{M}{N}\leq \frac{K}{K^2-K-1},\\
\frac{(K-1)N}{K^2-K-1},&& \frac{K}{K^2-K-1}\leq \frac{M}{N}\leq \frac{K-1}{K(K-2)}.\end{array}\right.
\end{eqnarray}
In this section, we investigate the DoF converse and the
achievability sequentially. Note that all the techniques applied in
proofs presented in this section follow similarly from the $K=3$
user interference channel setting that we have shown in
\cite{Wang_Gou_Jafar_3userMxN}, using the intuition of antenna
redundancies at either the TX side or the RX side. In
the following, we first show the information theoretical DoF outer
bound for the $M_T<M_R$ and $M_T>M_R$ settings, respectively, and
then provide the achievability proof.

\subsection{The Information Theoretical DoF Outer Bound for
$M_T<M_R$}\label{sec:appsub_linear}

We consider the $M_T<M_R$ setting in this section. $M_T<M_R$ implies
that $M=M_T,~N=M_R$.

Among the four regions shown in (\ref{eqn:thm_linear}), the DoF
outer bound of the first three regions can be established by the
single user DoF bound and the cooperation DoF outer bound.
Specifically, let us consider $ \frac{M}{N} \in (0,\frac{1}{K}]$
first. The DoF outer bound $d \leq M$ follows trivially from the
single user bound. Next, consider $\frac{M}{N} \in
[\frac{1}{K},\frac{1}{K-1}]$ and
$\frac{M}{N}\in[\frac{1}{K-1},\frac{K}{K^2-K-1}]$. Since
collaboration among the users does not decrease the capacity region,
we allow the $K-1$ users from User 2 to User $K$ to cooperate as one
user, such that the network becomes a two user MIMO interference
channel where the two TXs have $M$ and $(K-1)M$ antennas
respectively, and corresponding RXs have $N$ and $(K-1)N$
antennas, respectively. The sum DoF of this network, as reported in
\cite{Jafar_Fakhereddin}, is outer bounded by
$\min(\max((K-1)M,N),\max(M,(K-1)N)$, which produces the desired DoF
bound per user $d \leq \max((K-1)M,N)/K = \frac{N}{K}$ if $
\frac{M}{N} \in [\frac{1}{K},\frac{1}{K-1}]$, and $d \leq
\max((K-1)M,N)/K = \frac{(K-1)M}{K}$ if $ \frac{M}{N} \in
[\frac{1}{K-1},\frac{K}{K^2-K-1}]$.

\begin{figure}[h]
\centering
\includegraphics[width=5.1in]{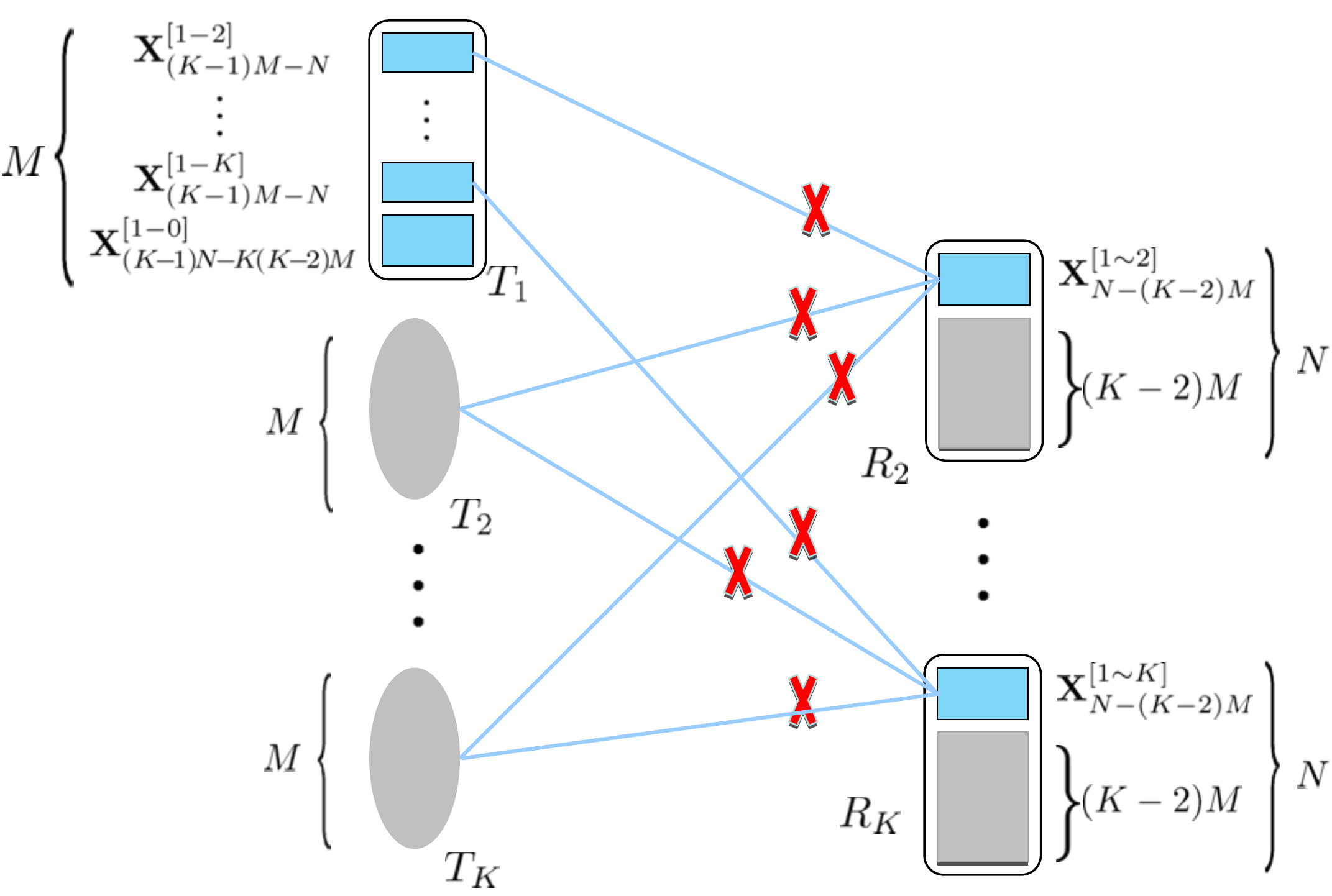}
\caption{Linear Transforms for the $M \times N$ Case, $\frac{M}{N}
\in [\frac{K}{K^2-K-1},\frac{K-1}{K(K-2)}]$ where the red cross
symbols stand for nulling. Note that we only show the transformed
channels for a clear presentation.} \label{fig:mnouter}
\end{figure}

Now let us focus on the remaining case $\frac{M}{N} \in
[\frac{K}{K^2-K-1},\frac{K-1}{K(K-2)}]$. We apply the similar linear
transformation approach as that we introduce in
\cite{Wang_Gou_Jafar_3userMxN}. Consider RX 2, which is able
to decode its own message $W_2$ due to the reliable communications
assumption. Thus, after removing the desired signal carrying message
$W_2$, RX 2 obtains an $N$-dimensional interference vector
space $\mathbf{S}^{[2]}$. By zero-forcing the interference from
TX 3 to TX $K$, RX 2 extracts the exposed
subspace $\mathbf{X}^{[1 \sim 2]}_{N-(K-2)M}$ from
$\mathbf{S}^{[2]}$. This can be done in the manner that left
multiply the received signal with an invertible $N \times N$ matrix
whose first $N-(K-2)M$ rows are orthogonal to the channel vectors
from each antenna of TX $3,\ldots,K$ to RX $2$, and
last $(K-2)M$ rows are the last $(K-2)M$ rows of the $N \times N$
identity matrix, as shown in Figure \ref{fig:mnouter}. After this
operation, the first $N-(K-2)M$ antennas at RX 2 only hear
TX $1$. Similarly, we proceed to RX $3,\ldots,K$
where we take linear transformations such that the first $N-(K-2)M$
antennas of each RX only hear TX 1. Therefore, we
obtain the exposed subspaces $\mathbf{X}^{[1 \sim
3]}_{N-(K-2)M},\ldots,\mathbf{X}^{[1 \sim K]}_{N-(K-2)M}$ at
RX $3,\ldots,K$, respectively. So far we complete the linear
transforms of the RX basis and then we switch to TX
1. We multiply an $M \times M$ matrix to the right-hand side of its
channel matrix such that the first $(K-1)M-N$ antennas of
TX 1 are \emph{not} heard by the first $N-(K-2)M$ antennas
at RX 2, the next $(K-1)M-N$ antennas are \emph{not} heard by
the first $N-(K-2)M$ antennas of RX 3 and so forth. This can
be done by choosing the first $(K-1)M-N$ columns of the
transformation matrix as the basis of the null space of the channel
matrix from the $M$ antennas of TX 1 to the first
$N-(K-2)M$ antennas at RX 2, the next $(K-1)M-N$ columns
associated with RX 3 and so forth. Note that the dimension
matches exactly as $(K-1)M-N = M - [N-(K-2)M]$, and thus
corresponding transmit signals of TX 1, not heard by
RX $k$, will be heard by all the other RX $k'$ where
$k'\in\mathcal{K}\setminus\{1,k\}$. Continuing to RX $K$, we
fix the directions of the first $(K-1)\times [(K-1)M-N]$ antennas of
TX $1$, and leave the last $M -[(K-1)^2M-(K-1)N] = (K-1)N -
K(K-2)M$ antennas which do not need change of basis and these
columns can be chosen as corresponding according columns of the
$M\times M$ identity matrix. Notice that $(K-1)N - K(K-2)M \geq 0$
because we are considering $\frac{M}{N} \leq \frac{K-1}{K(K-2)}$.
For brevity, we label the corresponding transmit signals from
TX 1 to RX $k$ after the invertible linear
transformations as $\mathbf{X}^{[1-k]}_{(K-1)M-N}$ where
$k\in\mathcal{K}\setminus\{1\}$ and the signals from the last
$(K-1)N - K(K-2)M$ antennas of TX 1 are represented as
$\mathbf{X}^{[1-0]}_{(K-1)N - K(K-2)M}$ (See Figure
\ref{fig:mnouter}). So far we also complete the linear
transformations at the TX side.

It can be seen that $\mathbf{X}^{[1 - 2]}_{(K-1)M-N}$ has $(K-1)M-N$
dimensions whose size is suitable for a genie to be provided to
Receive 2 to recover all $N + (K-1)M-N = (K-1)M$ dimensional
interference subject to the noise distortion. Also, since
$\mathbf{X}^{[1 - 2]}_{(K-1)M-N}$ is not heard at RX 2, it is
linearly independent with the exposed subspace $\mathbf{X}^{[1 \sim
2]}_{N-(K-2)M}$, thus making $\mathbf{X}^{[1 - 2]}_{(K-1)M-N}$ a
qualified genie. Note that $\mathbf{X}^{[1 - 2]}_{(K-1)M-N}$ is
totally determined by the channel matrices associated with RX
2, thus it is independent with any other RXs' channels.
Similarly, the genie $\mathbf{X}^{[1 - k]}_{(K-1)M-N}$ allows
RX $k$ where $k \in \{3,\ldots,K\}$, to decode all the messages
subject to the noise distortion. Starting from RX 2, by the
Fano's inequality, we have
\begin{eqnarray}
KnR-n\epsilon_n &\!\!\!\!\leq:\!\!\!\!& nN\log(\rho) + \hbar(\mathbf{X}^{[1 - 2]^n}_{(K-1)M-N} | \mathbf{S}^{[2]^n})\\
&\!\!\!\!\leq:\!\!\!\!& nN\log(\rho) + \hbar(\mathbf{X}^{[1 - 2]^n}_{(K-1)M-N} | \mathbf{X}^{[1 \sim 2]^n}_{N-(K-2)M})\\
&\!\!\!\!\leq:\!\!\!\!& nN\log(\rho) + \hbar(\mathbf{X}^{[1 -
2]^n}_{(K-1)M-N} |
\mathbf{X}^{[1-3]^n}_{(K\!-\!1)M\!-\!N},\ldots,\mathbf{X}^{[1-K]^n}_{(K\!-\!1)M\!-\!N},\mathbf{X}^{[1-0]^n}_{{(K\!-\!1)N\!-\!K(K\!-\!2)\!M}})\
\ \ \label{eqn:c1}
\end{eqnarray}
where (\ref{eqn:c1}) follows from Property (P2) in Lemma 2. For
compactness, we omit the dimension-denoting subscript which is clear
from the context. Thus, we can rewrite the equation above as
\begin{eqnarray}
KnR-n\epsilon_n\leq:
nN\log(\rho)+\hbar(\mathbf{X}^{[1-2]^n}|\mathbf{X}^{[1-3]^n},\ldots,\mathbf{X}^{[1-K]^n},
\mathbf{X}^{[1-0]^n}).
\end{eqnarray}
Similarly at RX $k$ where $k\in\{3,\cdots,K\}$, a genie
provides $\mathbf{X}^{[1 - i]^n}$ to RX $k$ such that it can
decode all the messages subject to the noise distortion. Thus we have
\begin{eqnarray}
KnR\!-\!n\epsilon_n\!\leq:\! nN\log(\rho) + \hbar(\mathbf{X}^{[1-k]^n} |
\mathbf{X}^{[1-2]^n}\!\!,\ldots,\mathbf{X}^{[1-(k-1)]^n}\!\!,\mathbf{X}^{[1-(k+1)]^n}\!\!,\ldots,\mathbf{X}^{[1-K]^n}\!\!,
\mathbf{X}^{[1-0]^n}). \label{eqn:c2}
\end{eqnarray}
Again, as $\mathbf{X}^{[1-k]^n}$ is totally determined by the
channels associate with RX $k$, they are linearly independent,
almost surely. Adding all the $K-1$ equations that we show above
together produces:
\begin{eqnarray}
(K\!-\!1)KnR-n\epsilon_n &\!\!\!\!\leq:\!\!\!\!& \sum^K_{i=2}\hbar(\mathbf{X}^{[1-i]^n}|\mathbf{X}^{[1-2]^n},\ldots,\mathbf{X}^{[1-(i-1)]^n},\mathbf{X}^{[1-(i+1)]^n},\ldots, \mathbf{X}^{[1-K]^n},\mathbf{X}^{[1-0]^n})\notag \\
&\!\!\!\!& +(K\!-\!1)nN\log(\rho)\\
&\!\!\!\!\leq:\!\!\!\!& (K\!-\!1)nN\log(\rho)+\hbar(\mathbf{X}^{[1 - 2]^n}, \ldots, \mathbf{X}^{[1 - K]^n} | \mathbf{X}^{[1 - 0]^n})\\
&\!\!\!\!\leq:\!\!\!\!& (K\!-\!1)nN\log(\rho)+\hbar(\mathbf{X}^{[1]^n})\\
&\!\!\!\!\leq:\!\!\!\!& (K\!-\!1)nN\log(\rho)+nR . \label{eqn:c3}
\end{eqnarray}
Rearranging the terms in (\ref{eqn:c3}) we obtain:
\begin{eqnarray}
(K^2\!-\!K\!-\!1)nR-n\epsilon_n\leq: (K\!-\!1)nN\log(\rho).
\end{eqnarray}
By letting $n\rightarrow \infty$ first and then $\rho\rightarrow
\infty$, we have the desired DoF outer bound
\begin{eqnarray}
d \leq \frac{(K-1)N}{K^2-K-1}.
\end{eqnarray}
Thus, we complete the information theoretical DoF outer bound proof
for the $M_T<M_R$ setting. \hfill\QED

\subsection{The Information Theoretical DoF Outer Bound for $M_T>M_R$}

In this section, we consider the reciprocal $M_T>M_R$ setting, which
means that $N=M_T,~M=M_R$. For the $K$ user $N\times M$ MIMO
interference channel, again if $\frac{M}{N}\leq \frac{K}{K^2-K-1}$,
the DoF outer bound can be directly obtained by the single-user DoF
bound and the cooperation DoF bound, as stated in Appendix
\ref{sec:appsub_linear}. What remains to be shown is for the case of
of $\frac{M}{N} \in [\frac{K}{K^2-K-1},\frac{K-1}{K(K-2)}]$. We will
use a two-stage approach as follows.

\begin{figure}[!ht]
\centering
\includegraphics[width=5.1in]{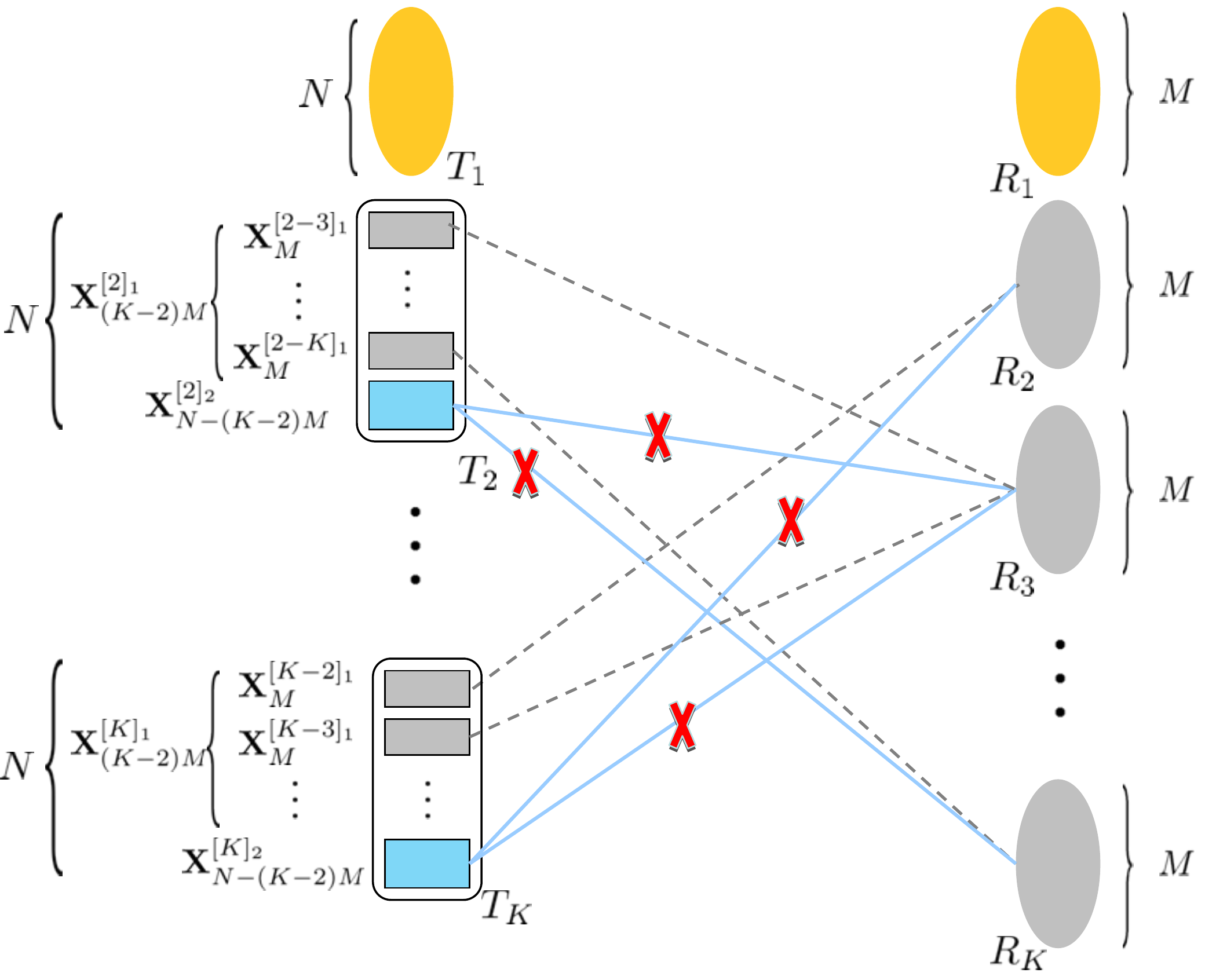}
\caption{The First Stage Linear Transformations for the $N \times M$
Case, $\frac{M}{N} \in [\frac{K}{K^2-K-1},\frac{K-1}{K(K-2)}]$ where
the red cross stands for nulling and the dashed lines stand for an
identity matrix. Note that we only show the transformed channels for
a clear presentation.} \label{fig:nmouter1}
\end{figure}

{\bf Stage 1:} In the first stage, we only deal with user pairs
$2,\ldots,K$. Let us first consider TX 2. Denote its first
$(K-2)M$ antennas as layer 1 symbol $\mathbf{X}^{[2]_1}_{(K-2)M}$
and last $N-(K-2)M$ antennas as layer 2 symbol
$\mathbf{X}^{[2]_2}_{N-(K-2)M}$. Now we take invertible linear
transforms on the symbols of these two layers. As shown in Figure
\ref{fig:nmouter1}, for layer 1 symbol
$\mathbf{X}^{[2]_1}_{(K-2)M}=[\mathbf{X}^{[2-3]_1}_{M};\ldots;
\mathbf{X}^{[2-K]_1}_{M}]$, invert its channel matrices to all the
RX $k$ where $k\in\{3,\ldots,K\}$ such that
$\mathbf{X}^{[2-k]_1}_{M}$ is only heard by RX $k$. For layer
2 symbol $\mathbf{X}^{[2]_2}_{N-(K-2)M}$, let it \emph{not} be heard
by all RXs $3,\ldots,K$ by nulling the channel matrices to
those RXs from TX 2. Once we complete the
transformation at TX 2, we can apply similar
transformations at TX $k \in \{3,\ldots,K\}$, i.e.,
dividing $\mathbf{X}^{[k]}$ to layer 1 symbols
$\mathbf{X}^{[k]_1}_{(K-2)M}$ associated with the first $(K-2)M$
antennas which sequentially have projections to RXs
$k'\in\mathcal{K}\setminus\{1,k\}$, and the remaining layer 2
symbols $\mathbf{X}^{[k]_2}_{N-(K-2)M}$ which is not heard by
RXs $k'\in\mathcal{K}\setminus\{1,k\}$. Now it can be seen
that if a genie provides $K-2$ messages
$W^{[1]},W^{[4]},\ldots,{W}^{[K]}$ to RX 3, it can remove the
interference signals carrying those messages and only hears
$\mathbf{X}^{[2 - 3]_1}_{M}$. Hence, if a genie further provides
$\mathbf{X}^{[2]}\setminus\mathbf{X}^{[2 - 3]_1}_{M}$ to RX 3,
it is then able to decode $W^{[2]}$ subject to the noise distortion.
That is, providing ${\bf
G}_3=\{W^{[1]},W^{[4]},\ldots,{W}^{[K]},\mathbf{X}^{[2]}\setminus\mathbf{X}^{[2
- 3]_1}_{M}\}$ to RX 3 allows it to decode all the messages
subject to the noise distortion. From the Fano's inequality, we have
\begin{eqnarray}
KnR-n\epsilon_n &\!\!\!\!\leq:\!\!\!\!& nM\log(\rho) +
\hbar(W^{[1]},W^{[4]},\ldots,{W}^{[K]},\mathbf{X}^{[2]^n}\setminus\mathbf{X}^{[2-3]^n_1}_{M}|{\bf S}^{[3]^n})\\
&\!\!\!\!\leq:\!\!\!\!& nM\log(\rho) + n(K\!-\!2)R + \hbar(\mathbf{X}^{[2]^n} \backslash\mathbf{X}^{[2 - 3]^n_1}_{M} | \mathbf{X}^{[2-3]^n_1}_{M})\\
&\!\!\!\!\leq:\!\!\!\!& nM\log(\rho) + n(K\!-\!2)R + \hbar(\mathbf{X}^{[2]^n}) - \hbar(\mathbf{X}^{[2 - 3]^n_1}_{M})\\
&\!\!\!\!\leq:\!\!\!\!& nM\log(\rho) + n(K\!-\!1)R - \hbar(\mathbf{X}^{[2 - 3]^n_1}_{M})\\
\Rightarrow nR-n\epsilon_n &\!\!\!\!\leq:\!\!\!\!&  nM\log(\rho) -
\hbar(\mathbf{X}^{[2- 3]^n_1}_{M}). \label{eqn:e1}
\end{eqnarray}
Following the same line, if a genie provides to the signals set
\begin{eqnarray*}
{\bf
G}_k=\{W^{[1]},W^{[3]},\ldots,{W}^{[k-1]},{W}^{[k+1]},\ldots,{W}^{[K]},\mathbf{X}^{[2]}\setminus\mathbf{X}^{[2
-k]_1}_{M}\}
\end{eqnarray*}
to RX $k\in\{4,\ldots,K\}$, RX $k$ can also decode all
messages subject to the noise distortion. Therefore, we have the sum
rate inequality as follows:
\begin{eqnarray}
nR-n\epsilon_n\leq:
nM\log(\rho)-\hbar(\mathbf{X}^{[2-k]^n_1}_{M})\label{eqn:e2}.
\end{eqnarray}
Adding all $K-2$ sum rate inequalities associated with RX
$k\in\{3,\cdots,K\}$ above, we have:
\begin{eqnarray}
(K-2)nR-n\epsilon_n &\!\!\!\!\leq:\!\!\!\!& (K\!-\!2)nM\log(\rho)-\sum_{k=3}^K \hbar(\mathbf{X}^{[2-k]^n_1}_{M})\\
&\!\!\!\!\leq:\!\!\!\!& (K\!-\!2)nM\log(\rho)-\hbar(\mathbf{X}^{[2-3]^n_1}_{M},\ldots,\mathbf{X}^{[2-K]^n_1}_{M})\\
&\!\!\!\!=:\!\!\!\!& (K\!-\!2)nM\log(\rho) -
\hbar(\mathbf{X}^{[2]^n_1}_{(K-2)M}) \label{eqn:e3}.
\end{eqnarray}
After obtaining the inequality above by considering layer 1 symbols
$\mathbf{X}^{[2]_1}_{(K-2)M}$ at TX 2, we proceed to
TX $k\in\{3,\ldots,K\}$ to make similar analysis, and we
obtain the following sum rate inequality:
\begin{eqnarray}
(K-2)nR-n\epsilon_n\leq:
(K-2)nM\log(\rho)-\hbar(\mathbf{X}^{[k]^n_1}_{(K-2)M}),~~~k\in\{3,\ldots,K\}.\label{eqn:e4}
\end{eqnarray}
Adding (\ref{eqn:e3}) and $K-2$ inequalities in (\ref{eqn:e4})
produces the inequality
\begin{eqnarray}
(K-1)(K-2)nR-n\epsilon_n\leq: (K-1)(K-2)nM\log(\rho)-\sum_{k=2}^{K} \hbar(\mathbf{X}^{[k]^n_1})\label{eqn:e5}
\end{eqnarray}
at the first stage where we omit the dimension denoting subscript for convenience.

\begin{figure}[h]
\centering
\includegraphics[width=5.1in]{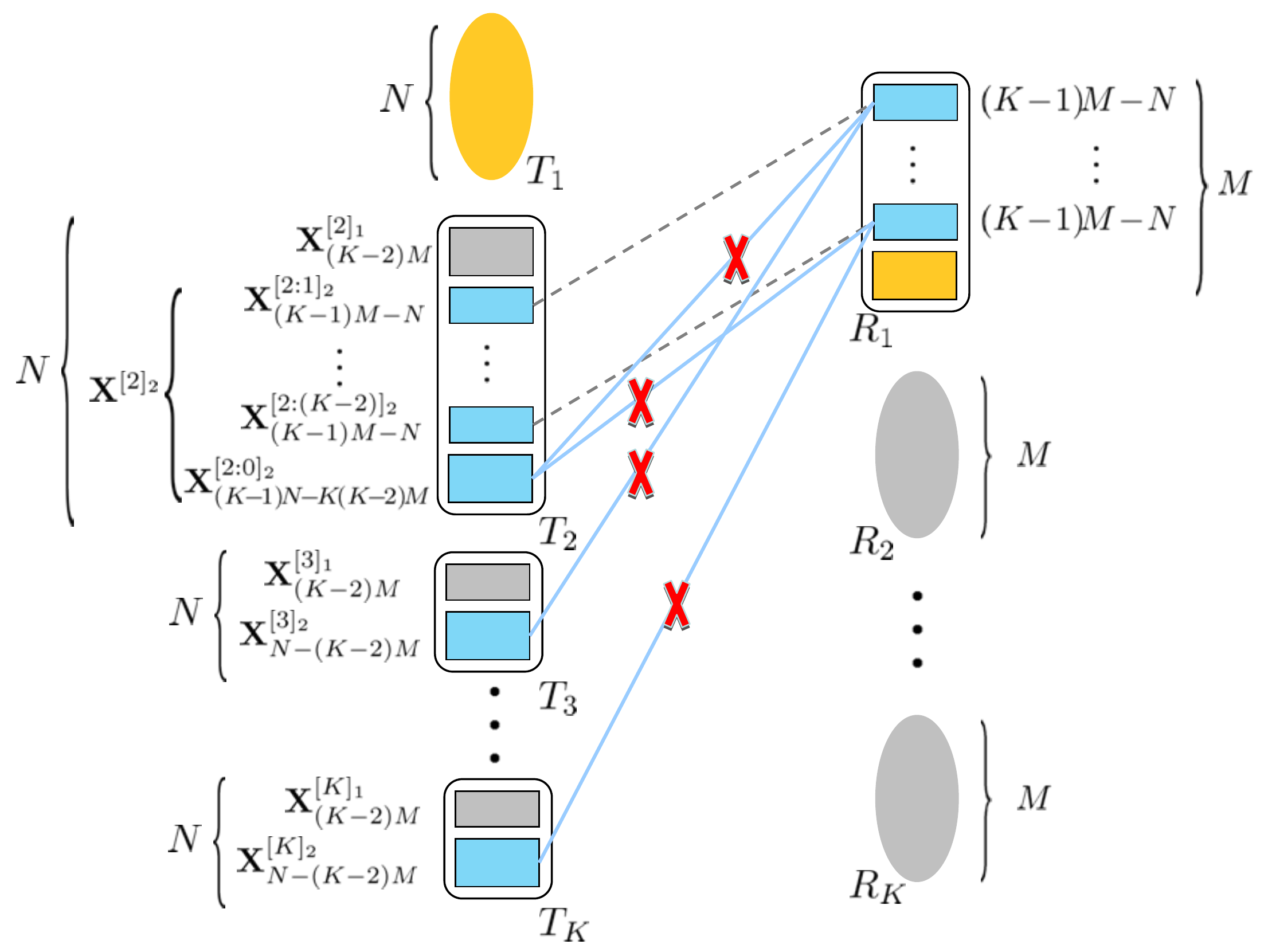}
\caption{The Second Stage Linear Transformations for the $N \times
M$ Case, $\frac{M}{N} \in [\frac{K}{K^2-K-1},\frac{K-1}{K(K-2)}]$,
where the red cross stands for nulling and the dashed lines stand
for an identity matrix. Note that we only show the transformed
channels for a clear presentation.} \label{fig:nmouter2}
\end{figure}

{\bf Stage 2:} Next we come to the second stage where we focus on
RX $1$. Note that all the linear transformations in the first
stage are all {\em not} associated with the channel matrix to
RX $1$ which guarantees that all the transmit symbols from
TX $2,\ldots,K$ are still generic for RX 1. Again, we
will take invertible linear transformations at both the TX
side and the RX side. We first describe the linear
transformation at RX 1. For $k\in\{3,\ldots,K\}$, as each
layer 2 symbols $\mathbf{X}^{[k]_2}_{N-(K-2)M}$ has $N-(K-2)M$
dimensions, denoted with the blue color at each TX in
Figure \ref{fig:nmouter2}, RX 1 can pick out
$M-(N-(K-2)M)=(K-1)M-N$ dimensions that do not hear these layer 2
symbols from one TX through zero forcing. Therefore, at
RX 1, let the first $(K-1)M-N$ antennas not hear
$\mathbf{X}^{[3]_2}_{N-(K-2)M}$, the next $(K-1)M-N$ antennas not
hear $\mathbf{X}^{[4]_2}_{N-(K-2)M}$ and so on. That is to say, the
$k^{th}$ $(K-1)M-N$ antennas at RX 1 in sequence do not hear
$\mathbf{X}^{[k]_2}_{N-(K-2)M}$ where $k\in\{3,\ldots,K\}$. So far,
we complete the linear transform at RX 1. Note that we
currently only deal with the first $(K-2)[(K-1)M-N]<M$ antennas due
to $\frac{M}{N} \leq \frac{K-1}{K(K-2)}$. Next we consider the layer
2 symbols $\mathbf{X}^{[2]_2}_{N-(K-2)M}$ at TX 2. We
invert the channel from its first $(K-2)[(K-1)M-N]$ antennas
(corresponding to the first $K-2$ blue boxes of TX 2 in
Figure \ref{fig:nmouter2}) to the first $(K-2)[(K-1)M-N]$ antennas
at RX 1 such that the channel between them becomes an identity
matrix. Moreover, at TX 2, the remaining
$N-(K-2)M-(K-2)[(K-1)M-N] = (K-1)N-K(K-2)M$ dimensions of
$\mathbf{X}^{[2]_2}_{N-(K-2)M}$ (the last blue box at TX 2
in Figure \ref{fig:nmouter2}) are chosen to be zero forced at the
first $(K-2)[(K-1)M-N]$ antennas at RX 1. Denote the symbols
of $\mathbf{X}^{[2]_2}_{N-(K-2)M}$ in sequence after linear
transforms as $\mathbf{X}^{[2]_2}_{N-(K-2)M} =
[\mathbf{X}^{[2:1]_2}_{(K-1)M-N};\ldots;\mathbf{X}^{[2:(K-2)]_2}_{(K-1)M-N};\mathbf{X}^{[2
: 0]_2}_{(K-1)N-K(K-2)M}]$. At the RX side, owing to our
linear transformations, the received signals from
$\mathbf{X}^{[2]_2}_{N-(K-2)M}$ seen by the first $(K-2)[(K-1)M-N]$
antennas of RX 1 are given by
$[\mathbf{X}^{[2:1]_2}_{(K-1)M-N};\ldots;\mathbf{X}^{[2:(K-2)]_2}_{(K-1)M-N}]$.
Now we finish the linear transforms at the second stage.

Now if a genie provides to RX 1 the $K-3$ messages
$W^{[4]},\ldots,W^{[K]}$ and layer 1 symbols
$\mathbf{X}^{[2]_1},\mathbf{X}^{[3]_1}$, RX 1 only hears
interference caused by layer two symbols
$\mathbf{X}^{[2]_2}_{N-(K-2)M},~\mathbf{X}^{[3]_2}_{N-(K-2)M}$ from
TX 2 and 3. Further, with these genie signals, RX 1
hears clean $\mathbf{X}^{[2:1]_2}_{(K-1)M-N}$, denoted as the first
$(K-1)M-N$ antennas at RX 1 where
$\mathbf{X}^{[3]_2}_{N-(K-2)M}$ are zero forced. So further giving
$(\mathbf{X}^{[2]_2}_{N-(K-2)M}\setminus\mathbf{X}^{[2:1]_2}_{(K-1)M-N})$
allows RX 1 to decode messages $W_{[2]}$ and $W_{[3]}$ subject
to the noise distortion. Therefore, providing ${\bf
G}_3=\{W^{[4]},\ldots,W^{[K]},\mathbf{X}^{[2]_1},\mathbf{X}^{[3]_1},\mathbf{X}^{[2]_2}_{N-(K-2)M}\setminus\mathbf{X}^{[2:1]_2}_{(K-1)M-N}\}$
to RX 3 allows it to decode all the messages subject to the noise
distortion. From the Fano's inequality, we have
\begin{eqnarray}
KnR-n\epsilon_n&\!\!\!\!\leq:\!\!\!\!& nM\log(\rho)\!+\!\hbar(W^{[4]},\ldots,W^{[K]},\mathbf{X}^{[2]_1},\mathbf{X}^{[3]_1},\mathbf{X}^{[2]_2}_{N-(K-2)M}\!\!\setminus\!\!\mathbf{X}^{[2:1]_2}_{(K-1)M-N}|{\bf S}^{[3]^n})\ \ \ \ \ \ \\
&\!\!\!\!\leq:\!\!\!\!& nM\log(\rho) + (K\!-\!3)nR + \hbar(\mathbf{X}^{[3]^n_1}) + \hbar(\mathbf{X}^{[2]^n_1},\mathbf{X}^{[2]^n_2} \backslash \mathbf{X}^{[2:1]^n_2} | \mathbf{X}^{[2:1]^n_2} )\label{eqn:ee}\\
&\!\!\!\!=:\!\!\!\!& nM\log(\rho) + (K\!-\!3)nR + \hbar(\mathbf{X}^{[3]^n_1}) + \hbar(\mathbf{X}^{[2]^n}) -  \hbar(\mathbf{X}^{[2:1]^n_2})\\
\Rightarrow 2nR-n\epsilon_n&\!\!\!\!\leq:\!\!\!\!&
nM\log(\rho)+\hbar(\mathbf{X}^{[3]^n_1})-\hbar(\mathbf{X}^{[2:1]^n_2})\label{eqn:e6}.
\end{eqnarray}
Similarly, for $k\in\{4,\ldots,K\}$, if a genie provides to RX
1 the signals set
\begin{eqnarray*}
{\bf
G}_k=\{W^{[3]},\ldots,{W}^{[k-1]},{W}^{[k+1]},\ldots,{W}^{[K]},\mathbf{X}^{[k]^n_1},\mathbf{X}^{[2]}\setminus\mathbf{X}^{[2:(k-2)]_2}_{(K-1)M-N}\},
\end{eqnarray*}
RX 1 can also decode all the messages subject to the noise
distortion. Therefore, we have the sum rate inequality as follows:
\begin{eqnarray}
2nR-n\epsilon_n\leq: nM\log(\rho) + \hbar(\mathbf{X}^{[k]^n_1})-
\hbar(\mathbf{X}^{[2:k]^n_2}),~~~k\in\{4,\ldots,K\}.\label{eqn:e7}
\end{eqnarray}
In order to make the addition operation shown later simple, we
rewrite the inequality for $k=K$, similar to (\ref{eqn:ee}), as
follows:
\begin{eqnarray}
3nR-n\epsilon_n &\!\!\!\!\leq:\!\!\!\!& nM\log(\rho)+\hbar(\mathbf{X}^{[K]^n_1})+\hbar(\mathbf{X}^{[2]^n_1},\mathbf{X}^{[2]^n_2}\!\!\setminus\!\!\mathbf{X}^{[2:(K-2)]^n_2}|\mathbf{X}^{[2:(K-2)]^n_2})\\
&\!\!\!\!\leq:\!\!\!\!& nM\log(\rho)+\hbar(\mathbf{X}^{[K]^n_1})+\hbar(\mathbf{X}^{[2]^n_1},\mathbf{X}^{[2:1]^n_2},\ldots,\mathbf{X}^{[2:(K-3)]^n_2},\mathbf{X}^{[2:0]^n_2})\ \ \ \ \\
&\!\!\!\!\leq:\!\!\!\!& nM\log(\rho) + \hbar(\mathbf{X}^{[K]^n_1})+
\hbar(\mathbf{X}^{[2]^n_1})+\sum_{k=1}^{K-3}\hbar(\mathbf{X}^{[2:k]^n_2})+
\hbar(\mathbf{X}^{[2:0]^n_2}). \label{eqn:e8}
\end{eqnarray}
Now adding (\ref{eqn:e6}), $K\!-\!4$ inequalities in (\ref{eqn:e7})
for $k\in\{4,\cdots,K\!-\!1\}$ and (\ref{eqn:e8}) produces the sum
rate inequality at the second stage:
\begin{eqnarray}
2(K-3)nR +3nR-n\epsilon_n\leq: (K-2)nM\log(\rho)+\sum_{k=2}^{K}
\hbar(\mathbf{X}^{[k]^n_1})+\hbar(\mathbf{X}^{[2:0]^n_2})\label{eqn:e9}.
\end{eqnarray}
Finally, adding up the two inequalities (\ref{eqn:e5})
and(\ref{eqn:e9}) that we obtain in the two stages, we have:
\begin{eqnarray}
(K^2-K-1)nR-n\epsilon_n &\!\!\!\!\leq:\!\!\!\!& K(K-2)nM\log(\rho)+\hbar(\mathbf{X}^{[2:0]^n_2})\\
&\!\!\!\!\leq:\!\!\!\!& K(K-2)nM\log(\rho)+[(K-1)N-K(K-2)M]n\log(\rho)\label{eqn:e10} \\
&\!\!\!\!=:\!\!\!\!& (K-1)Nn\log(\rho)
\end{eqnarray}
where (\ref{eqn:e10}) follows from the fact that
$\mathbf{X}^{[2:0]_2}$ has a total of $(K-1)N-K(K-2)M$ dimensions
and Property 1 in Lemma \ref{lemma:property}.

By letting $n\rightarrow \infty$ first and then $\rho\rightarrow
\infty$, we have the desired DoF outer bound
\begin{eqnarray}
d \leq \frac{(K-1)N}{K^2-K-1}.
\end{eqnarray}
Thus, we complete the information theoretical DoF outer bound proof
the $M_T>M_R$ setting.\hfill\QED

\subsection{The DoF Achievability}

We will show that schemes based on linear beamforming at the
TXs and zero-forcing at the RXs are sufficient to
achieve the DoF values in Theorem \ref{theorem_linear}. Due to the duality
of linear schemes, we only need to consider the
$M_T<M_R$ setting, i.e., $M=M_T,~N=M_R$.

We begin with the first two regions. First,
if $ \frac{M}{N}\in (0,\frac{1}{K}]$, i.e., $KM\leq N$, each
RX has enough antennas distinguish all the transmit signals
from all TXs. Thus, each user can achieve its interference
free DoF which is given by $d=\min(M,N)=M$. Second, if $\frac{M}{N}
\in [\frac{1}{K},\frac{1}{K-1}]$, the sum DoF $N$ are achievable
because each RX, after decoding its own message and
subtracting the signal carrying that message, still has enough
antennas to distinguish all the interference signals owing to
$(K\!-\!1)M\leq N$. Thus, the DoF value $d=N/K$ per user is
achievable in this region.

Next, we consider the remaining two cases $\frac{M}{N} \in
[\frac{1}{K-1},\frac{K}{K^2-K-1}]$ and $\frac{M}{N}
\in[\frac{K}{K^2-K-1},\frac{K-1}{K(K-2)}]$. Our aim is to show that
$d=\frac{(K-1)M}{K}$ and $d=\frac{(K-1)N}{K^2-K-1}$ are achievable
for these two regions, respectively. As a matter of fact, the DoF
value is linear piecewise on these two regions, depending on either
$M$ or $N$. This situation also appears in the $K=3$ user $M\times
N$ MIMO interference channel where we rely on the antenna
redundancies argument for the DoF achievability. Again, here we will
follow the similar approach as that we use in
\cite{Wang_Gou_Jafar_3userMxN} for the DoF achievability. That is,
we first prove $d=\frac{(K-1)M}{K}$ or equivalently
$d=\frac{(K-1)N}{K^2-K-1}$ are achievable at
$\frac{M}{N}=\frac{K}{K^2-K-1}$. Then by increasing the RX
antenna redundancies $N$ such that $M/N$ falls into the region
$[\frac{1}{K-1},\frac{K}{K^2-K-1}]$, the achievability of
$d=\frac{(K-1)M}{K}$ DoF should remain. Similarly, by increasing the
TX antenna redundancies $M$ such that $M/N$ falls into the
region $[\frac{K}{K^2-K-1},\frac{K-1}{K(K-2)}]$, the achievability
of $d=\frac{(K-1)N}{K^2-K-1}$ should not be affected as well.

We first investigate the case $\frac{M}{N}=\frac{K}{K^2-K-1}$ which
implies $(M,N) = (\beta K,\beta(K^2\!-\!K\!-\!1))$ where
$\beta\in\mathbb{Z}^+$. Our goal is to show $d =\frac{(K-1)M}{K}=
\frac{(K-1)N}{K^2-K-1}=\beta(K-1)$ are achievable. At each time
slot, TX $k\in\mathcal{K}$ sends $\beta(K\!-\!1)$
independent symbols using a $\beta K \times \beta(K-1)$ beamforming
matrix $\mathbf{V}^{[k]}=[
\mathbf{V}^{[k]}_1,\ldots,\mathbf{V}^{[k]}_{K-1}]$ where each block
$\mathbf{V}^{[k]}_i,~i\in\{1,\cdots,K-1\}$ is a $\beta K\times
\beta$ beamforming matrix. In the $\beta(K^2\!-\!K\!-\!1)$
dimensional vector space at each RX, the desired signal will
occupy $\beta(K-1)$ dimensions, thus leaving only a subspace with
$\beta(K^2-K-1)-\beta(K-1)=\beta((K-1)^2-1)$ dimensions to
accommodate a total of $\beta(K-1)^2$ dimensional interference. That
is to say, we need to align $\beta(K-1)^2-\beta((K-1)^2-1)=\beta$
dimensional interference at each RX. Specifically, we design
the beamforming vectors in the following way to satisfy these
alignment constraints. At RX 1, $\beta$ interference vectors
from TX 2 are aligned into the subspace spanned by
$(K-2)\beta$ interference vectors, $\beta$ from each of TX
$3,\ldots,K$, respectively. This operation gives us one alignment
equation:
\begin{eqnarray}
&&\mathbf{H}^{[12]} \mathbf{V}^{[2]}_1 = - (\mathbf{H}^{[13]} \mathbf{V}^{[3]}_1 + \cdots + \mathbf{H}^{[1K]} \mathbf{V}^{[K]}_1)\\
&&\Rightarrow
\underbrace{[\mathbf{H}^{[12]}~\mathbf{H}^{[13]}~\cdots~
\mathbf{H}^{[1K]}]_{\beta(K^2-K-1) \times \beta
K(K-1)}}_{\mathbf{\bar{H}}_1}
\underbrace { \left[ \begin{array}{c} \mathbf{V}^{[2]}_1 \\
\mathbf{V}^{[3]}_1 \\ \vdots \\ \mathbf{V}^{[K]}_1 \end{array}
\right]_{\beta K(K-1) \times \beta}}_{\mathbf{\bar{V}}_1} =
\mathbf{O}. \label{eqn:a2}
\end{eqnarray}
Since $\mathbf{\bar{H}}_1$ is a $\beta(K^2-K-1) \times \beta K(K-1)$
generic matrix which only consists of interference carrying channel
matrices from TX $2,\ldots,K$ to RX 1,
$\mathbf{\bar{V}}_1$ can be obtained as the basis in the null space
of $\mathbf{\bar{H}}_1$, and thus all
$\mathbf{V}^{[2]}_1,\ldots,\mathbf{V}^{[K]}_1$ can be derived. Note
that from the equation above associated with RX 1, we have
determined the directions of $\beta(K-1)$ symbols, $\beta$ from each
of TX $2,\ldots,K$, respectively. Similarly, at RX $k
\in \{2,\ldots,K\}$, by aligning only $\beta$ dimensional
interference, we obtain the following alignment equations:
\begin{small}
\begin{eqnarray}
&&\mathbf{H}^{[k1]} \mathbf{V}^{[1]}_{k-1} = - (\mathbf{H}^{[k2]} \mathbf{V}^{[2]}_{k-1} + \cdots + \mathbf{H}^{[k(k-1)]} \mathbf{V}^{[k-1]}_{k-1} + \mathbf{H}^{[k(k+1)]} \mathbf{V}^{[k+1]}_k + \cdots + \mathbf{H}^{[kK]} \mathbf{V}^{[K]}_k)\ \ \ \ \ \ \ \\
&&\Rightarrow
\underbrace{[\mathbf{H}^{[k1]}~\mathbf{H}^{[k2]}~\cdots~
\mathbf{H}^{[k(k-1)]}~\mathbf{H}^{[k(k+1)]}~\cdots~\mathbf{H}^{[iK]}
]_{\beta(K^2-K-1) \times \beta K(K-1)}}_{\mathbf{\bar{H}}_k}
\underbrace {
\left[\begin{array}{c} \mathbf{V}^{[1]}_{k-1} \\
\mathbf{V}^{[2]}_{k-1} \\ \vdots \\ \mathbf{V}^{[k-1]}_{k-1} \\
\mathbf{V}^{[k+1]}_k \\ \vdots \\ \mathbf{V}^{[K]}_k \end{array}
\right]_{\beta K(K-1) \times \beta}}_{\mathbf{\bar{V}}_k} =
\mathbf{O}. \label{eqn:a1}\ \ \ \ \ \ \ \
\end{eqnarray}
\end{small}
Again, $\mathbf{\bar{H}}_i$ is a $\beta(K^2-K-1) \times \beta
K(K-1)$ generic matrix and $\mathbf{\bar{V}}_k$ can be determined as
the basis of its null space. For the $k^{th}$ alignment equation we
show above where $k\in\mathcal{K}$, we determine the beamforming
directions of $\beta$ symbols per user. For each user $k$, with all
$K$ equations except for the $k^{th}$, we establish the beamforming
directions of all $\beta(K-1)$ symbols per user. After aligning
interference at each RX, we still need to ensure that the
desired signals do not overlap with the interference space. In fact,
this is guaranteed since the direct channels $\mathbf{H}^{[kk]}$ do
not appear in the alignment equations (\ref{eqn:a2}) and
(\ref{eqn:a1}), $k \in \mathcal{K}$. In addition, note that the
$k^{th}$ alignment equation only involves the interference carrying
links associated with RX $k$ and each channel matrices are
generic. Thus, the beamforming directions of all $\beta(K-1)$
symbols at each user are linearly independent, almost surely, and
can be separated from the interference at each RX. Therefore,
each user is able to achieve $d=\beta(K-1)$ DoF, almost surely.

After we establish the DoF achievability at $M/N=\frac{K}{K^2-K-1}$,
let us consider $\frac{M}{N}\in [\frac{1}{K-1},\frac{K}{K^2-K-1}]$,
i.e., $\frac{K^2-K-1}{K}M\leq N$. In this region, the DoF value only
depends on $M$, so we can reduce the number of RX antennas $N$
to $N'=\frac{K^2-K-1}{K}M$, without affecting the DoF, such that it
becomes the case $\frac{M}{N'}=\frac{K}{K^2-K-1}$ that we have
solved. Note that if the value of $N-N'$ is not an integer, then we
can scale the number of both TX and RX antennas by
the the same factor $\alpha$ such that $\alpha(N-N')$ is an integer,
i.e., we solve the DoF in the sense of spatial normalization
\cite{Wang_Gou_Jafar_3userMxN}. In the absence of spatial
extensions, we can also resort to time/freuqency extensions for the
DoF achievability. Specifically, using $K$ symbol extensions, the
value of $K(N-N')=KN-(K^2-K-1)M$ becomes an integer, and thus we
obtain an $KM\times (K^2-K-1)M$ MIMO interference channel over every
$K$ slots. For this new channel, notice that the channel matrices
now has a block diagonal structure, i.e., they are no longer
generic. For this structured channel, we can still use the same
achievable schemes by ignoring the special channel structure, except
for the cases of $M/N=\frac{K-1}{K(K-2)}$ where we can use
asymmetric complex signaling for the linear achievability, similar
to $M/N=p/(p+1)$ cases for the three-user MIMO interference channel
\cite{Wang_Gou_Jafar_3userMxNacs} where
$p\in\mathbb{Z}^+\setminus\{1\}$. Notice that for the three-user
MIMO interference channel, we have shown in
\cite{Wang_Gou_Jafar_3userMxN} that if the channel coefficients are
varying over channel extensions, then the linear achievable schemes
do work; and also if the channel coefficients are constant-valued
over channel extensions, then validity of the schemes \emph{in
general} rely on numerical tests. However, for $K\geq 4$ settings in
this paper, we do not need numerical tests because the coefficients
and variables comprising equations in (\ref{eqn:a2}) and
(\ref{eqn:a1}) are independent. Therefore, the value of
$d=\frac{(K-1)M}{K}$ DoF per user can be achievable by using ether
spatial/time/frequency extensions. Similarly, for the cases of
$M/N\in [\frac{K}{K^2-K-1},\frac{K-1}{K(K-2)}]$, i.e., $M \geq
\frac{K}{K^2-K-1}N$, the DoF value $d=\frac{(K-1)N}{K^2-K-1}$ only
depends on $N$ and we can reduce the number of transmit antennas
from $M$ to $M'=\frac{KN}{K^2-K-1}$, such that $M'/N$ becomes
$\frac{K}{K^2-K-1}$ again. Also, if $M-M'$ is not an integer, then
we can again use channel extensions over space/time/frequency to
establish the same results as we describe above. \hfill\QED

\section{The Linear Independence Proofs for the $\frac{M}{N}\geq \frac{K-1}{K(K-2)}$ Setting for the $K=4$ User $M\times N$ MIMO Interference Channel}

\subsection{$M/N\in[2/5,1/2)$ Cases (Algorithm 2)}\label{app:subsec_alg2}

\begin{figure}[!ht] \vspace{-0.1in}
\centering
\includegraphics[width=5.0in]{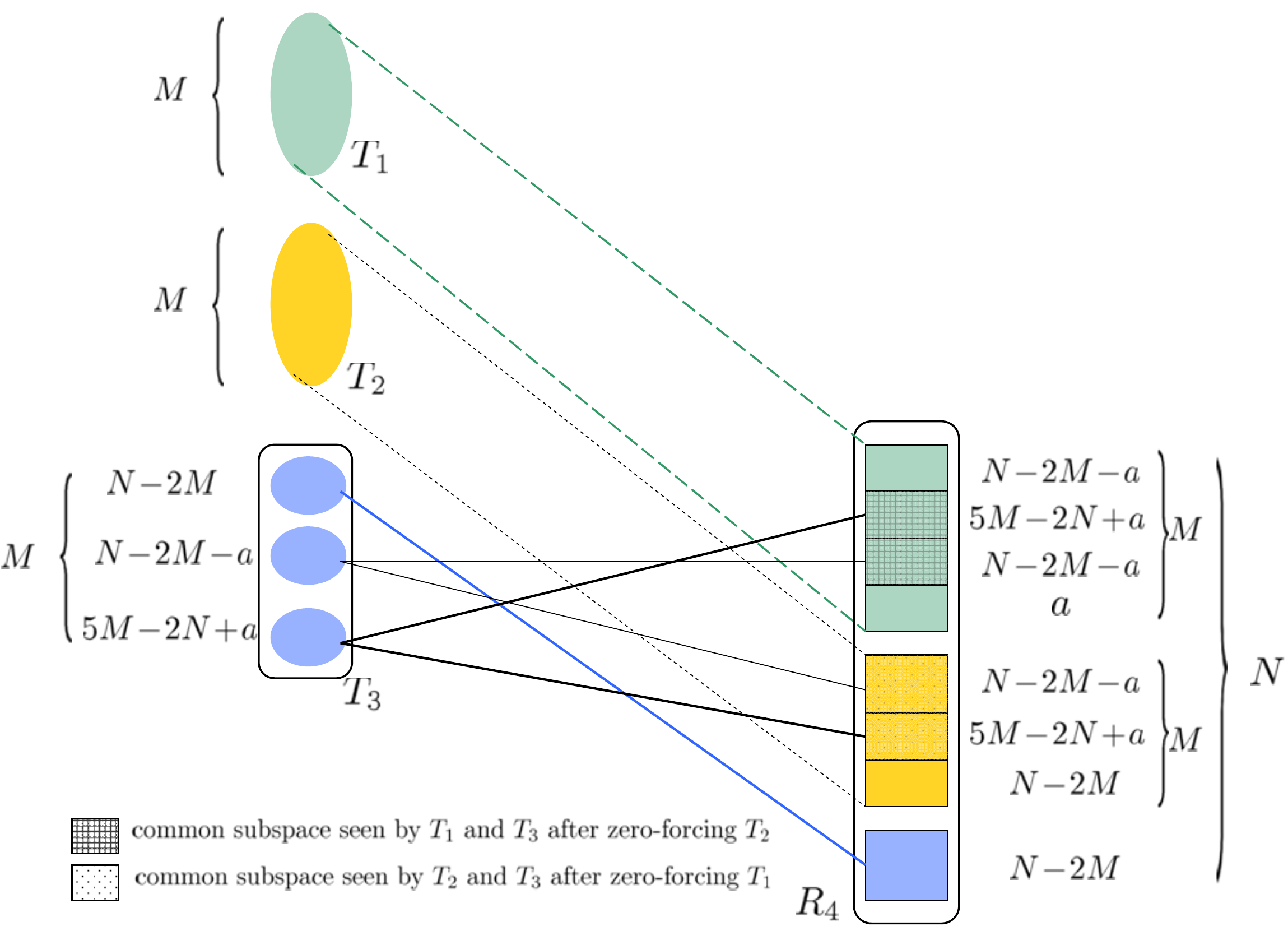}\vspace{-0.1in}
\caption{Linear Dimension Counting of Subspaces Participating in the
Interference Alignment for the $M\times N$ setting (the values
denote the dimensions of each corresponding subspace.)}
\vspace{-0.1in} \label{fig:special_channel_alg2}
\end{figure}

{\it Proof:} As shown in Algorithm 2, ${\bf G}$ may contain two
kinds of components, i.e., $\mathcal{O}$ and randomly generated
linear combinations of one interferer's symbols. Providing
$\mathcal{O}$ releases $|\mathcal{O}|$ dimensional observations of
the TX that we concern. Therefore, we need to show that the
$|{\bf G}|=(3M-N)$ dimensional observations of the TX that
we concern are linearly independent with $N-2M$ dimensional
observations which the RX has originally. In order to do this,
we need to show that the $M\times M$ square matrix whose entries are
the linear combination coefficients of the $M$ equations has full
rank, i.e., the determinant of this matrix, a polynomial function of
its entries, is zero almost surely. This polynomial is either a zero
polynomial or not equal to zero almost surely, since there are
finite number of solutions of the polynomial equation, which has
measure zero. Therefore, if the polynomial is not a zero polynomial,
the polynomial is not equal to zero almost surely for randomly
generated channel coefficients. Next, we will show the polynomial is
not a zero polynomial. To do that, we only need to find one specific
set of channel coefficients such that the polynomial is not equal to
zero. Next, we construct the channels for all interference carrying
links, i.e., ${\bf H}^{[ji]},~i,j\in\{1,2,3,4\},~j\neq i$.
\begin{eqnarray}\label{eqn:alg2_channel}
{\bf H}^{[k~k\!+\!1]}&\!\!\!\!=\!\!\!\!&\left[\begin{array}{c}{\bf
I}_M\\{\bf O}_M\\{\bf O}_{N\!-\!2M}\end{array}\right],~~~{\bf
H}^{[k~k\!+\!2]}=\left[\begin{array}{c}{\bf O}_M\\{\bf I}_M\\{\bf
O}_{N\!-\!2M}\end{array}\right],\notag\\
{\bf H}^{[k~k\!-\!1]}&\!\!\!\!=\!\!\!\!&\left[\begin{array}{ccc}{\bf
O}_{(N\!-\!2M\!-\!a)\times(N\!-\!2M)}&{\bf
O}_{(N\!-\!2M\!-\!a)\times(N\!-\!2M\!-\!a)}&{\bf
O}_{(N\!-\!2M\!-\!a)\times(5M\!-\!2N\!+\!a)}\\
{\bf O}_{(5M\!-\!2N\!+\!a)\times(N\!-\!2M)}&{\bf
O}_{(5M\!-\!2N\!+\!a)\times(N\!-\!2M\!-\!a)}&{\bf
I}_{(5M\!-\!2N\!+\!a)\times(5M\!-\!2N\!+\!a)}\\
{\bf O}_{(N\!-\!2M\!-\!a)\times(N\!-\!2M)}&{\bf
I}_{(N\!-\!2M\!-\!a)\times(N\!-\!2M\!-\!a)}&{\bf
O}_{(N\!-\!2M\!-\!a)\times(5M\!-\!2N\!+\!a)}\\
{\bf O}_{a\times(N\!-\!2M)}&{\bf O}_{a\times(N\!-\!2M\!-\!a)}&{\bf
O}_{a\times(5M\!-\!2N\!+\!a)}\\
{\bf O}_{(3M\!-\!N)\times(N\!-\!2M)}&{\bf
I}_{(3M\!-\!N)\times(N\!-\!2M\!-\!a)}&{\bf
O}_{(3M\!-\!N)\times(5M\!-\!2N\!+\!a)}\\
{\bf O}_{(3M\!-\!N)\times(N\!-\!2M)}&{\bf
O}_{(3M\!-\!N)\times(N\!-\!2M\!-\!a)}&{\bf
I}_{(3M\!-\!N)\times(5M\!-\!2N\!+\!a)}\\
{\bf O}_{(N\!-\!2M)\times(N\!-\!2M)}&{\bf
O}_{(N\!-\!2M)\times(N\!-\!2M\!-\!a)}&{\bf
O}_{(N\!-\!2M)\times(5M\!-\!2N\!+\!a)}\\
{\bf I}_{(N\!-\!2M)\times(N\!-\!2M)}&{\bf
O}_{(N\!-\!2M)\times(N\!-\!2M\!-\!a)}&{\bf
O}_{(N\!-\!2M)\times(5M\!-\!2N\!+\!a)}
\end{array}\right]
\end{eqnarray}
where $a=\gcd(M,N)$. While it is easy to check the specific matrices
above have full rank, we will show through Figure
\ref{fig:special_channel_alg2} using linear dimension counting
argument that these matrices keep the generic properties of linear
subspaces. For brevity, we only show the connectivity associated
with RX 4. The interference carrying links associated with
other RXs can be obtained by advancing user indices. Note that
the channel matrices of desired links are still generic.

From Figure \ref{fig:special_channel_alg2}, it can be easily seen
that in the $N$ dimensional space at RX 4, after zero-forcing
the two $M$-dimensional subspaces seen by two interferers, RX
4 has $N-2M$ dimensional clean observations of the remaining
interferer. For example, after zero-forcing the subspaces from $T_1$
and $T_2$, the remaining $N-2M$ dimensional observations of $T_3$
are with blue color. Similarly, the remaining $N-2M$ dimensional
observations of $T_1$ and $T_2$ after zero-forcing the rest of
interferers are with green and yellow colors, respectively.

Substituting the specific channel matrices in
(\ref{eqn:alg2_channel}) into Algorithm 2, we can easily check that
in each step the $M\times M$ square matrix that we obtain has full
rank. In our work, we use programming to check all cases for the
values of $M,N$ upto 100. Notice that our programming checking is
not numerical test, i.e., at each step we do not need to calculate
the determinant of the $M\times M$ square matrix. Instead, at each
step we produce a set that contains the indices of dimensions among
the $M$ dimensions (the dimensions here corresponds to $M$ antennas)
of the associated TX. Therefore, the programming checking
does not have numerical error sensitivity, unlike the numerical
tests taken in \cite{Wang_Gou_Jafar_3userMxN,
Annapureddy_vernu_CoMP, Feasibility} which are sensitive to
the numerical error especially if the values of $M,N$ are large.

\hfill\QED

\subsection{Special Cases of $M/N\in[3/8,2/5)$ (Algorithm 2)}\label{app:subsecalg2set}

\begin{figure}[!ht] \vspace{-0.1in}
\centering
\includegraphics[width=5.0in]{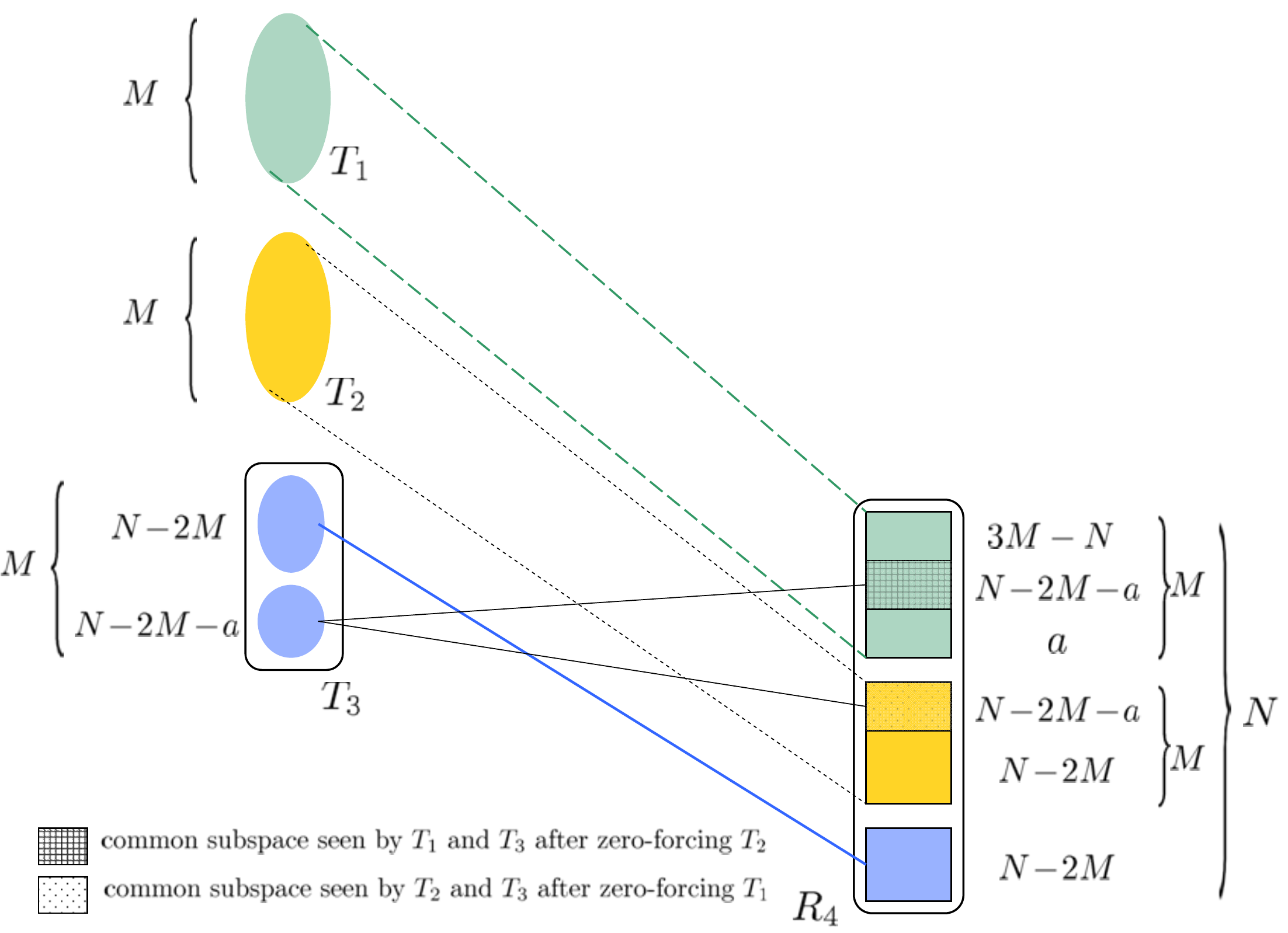}\vspace{-0.1in}
\caption{Linear Dimension Counting of Subspaces Participating in the
Interference Alignment for the $M\times N$ setting (the values
denote the dimensions of each corresponding subspace.)}
\vspace{-0.1in} \label{fig:special_channel_alg2alg3}
\end{figure}

{\it Proof:} Similar to the proof for
$M/N\in[2/5,1/2)$ case in Appendix \ref{app:subsec_alg2}, again, we
only need to find one specific set of channel coefficients such that
the polynomial is not equal to zero. We construct the channels for
all interference carrying links, i.e., ${\bf
H}^{[ji]},~i,j\in\{1,2,3,4\},~j\neq i$.
\begin{eqnarray}\label{eqn:alg2alg3_channel}
{\bf H}^{[k~k\!+\!1]}&\!\!\!\!=\!\!\!\!&\left[\begin{array}{c}{\bf
I}_M\\{\bf O}_M\\{\bf O}_{N\!-\!2M}\end{array}\right],~~~{\bf
H}^{[k~k\!+\!2]}=\left[\begin{array}{c}{\bf O}_M\\{\bf I}_M\\{\bf
O}_{N\!-\!2M}\end{array}\right],\notag\\
{\bf H}^{[k~k\!-\!1]}&\!\!\!\!=\!\!\!\!&\left[\begin{array}{cc}{\bf
O}_{(3M\!-\!N)\times(N\!-\!2M)}&{\bf
O}_{(3M\!-\!N)\times(N\!-\!2M\!-\!a)}\\
{\bf O}_{(N\!-\!2M\!-\!a)\times(N\!-\!2M)}&{\bf
I}_{(N\!-\!2M\!-\!a)\times(N\!-\!2M\!-\!a)}\\
{\bf O}_{a\times(N\!-\!2M)}&{\bf O}_{a\times(N\!-\!2M\!-\!a)}\\
{\bf O}_{(N\!-\!2M\!-\!a)\times(N\!-\!2M)}&{\bf
I}_{(N\!-\!2M\!-\!a)\times(N\!-\!2M\!-\!a)}\\
{\bf O}_{(N\!-\!2M)\times(N\!-\!2M)}&{\bf
O}_{(N\!-\!2M)\times(N\!-\!2M\!-\!a)}\\
{\bf I}_{(N\!-\!2M)\times(N\!-\!2M)}&{\bf
O}_{(N\!-\!2M)\times(N\!-\!2M\!-\!a)}
\end{array}\right]
\end{eqnarray}
where $a=\gcd(M,N)=2N-5M$. In addition, we will show through Figure
\ref{fig:special_channel_alg2alg3} using linear dimension counting
argument that these matrices keep the generic properties of linear
subspaces. For brevity, we only show the connectivity associated
with RX 4. The interference carrying links associated with
other RXs can be obtained by advancing user indices. Note that
the channel matrices of desired links are still generic.
Substituting the specific channel matrices in
(\ref{eqn:alg2alg3_channel}) into Algorithm 2, we can easily check
that in each step the $M\times M$ square matrix that we obtain has
full rank. In our work, we test all $(M,N)$ cases where
$M/N=(2c-1)/(5c-2),~c\in\mathbb{Z}^+\!\setminus\!\{1\},~M/N\geq
3/8,~M\leq 100,~N\leq 100$, and it produces our desired results.
\hfill\QED

\section{DoF of the Four-to-One MIMO Interference
Channel}\label{app:4to1}

As implied by Theorem \ref{theorem:4to1}, our goal is to show that
the DoF value per user is given by:
\begin{eqnarray}\label{eqn:4to1}
d=\left\{\begin{array}{ccl}M,&&M/N\leq 1/4,\\
N/4,&&1/4\leq M/N\leq 1/3,\\
3M/4,&&1/3\leq M/N\leq 4/9,\\
N/3,&&4/9\leq M/N\leq 1/2,\\
2M/3,&&1/2\leq M/N\leq 3/5,\\
2N/5,&&3/5\leq M/N\leq 2/3,\\
3M/5,&&2/3\leq M/N\leq 5/6,\\
N/2,&&5/6\leq M/N\leq 1.
\end{array}\right.
\end{eqnarray}

First, let us consider $M/N\leq 1/3$, which means $3M\leq N$.
Therefore, besides its own message $W_1$, RX 1 is able to
invert the channel matrix projected from three interferers, such
that RX 1 can reconstruct the three interfering signal vectors
subject to the noise distortion. Therefore, the DoF value of each
user is given by $d=\min(M,N/4)$, which produces the two cases in
(\ref{eqn:4to1}), i.e., $d=M$ for $M/N\leq 1/4$, and $d=N/4$ for
$1/4\leq M/N\leq 1/3$.

The remaining six cases are for the $M/N>1/3$ setting. Note that
$1/3\leq M/N$ or $3M\geq N$ implies that alignment of the signals
from three interferers becomes possible. In the following, we are
going to first investigate the information theoretic DoF outer
bound, and then the achievability.

\subsection{Information Theoretic DoF Outer Bound}

As have shown in this paper, we always apply genie aided argument to
produce the information theoretic DoF outer bound. For the $M\times
N$ Four-to-One MIMO interference channel, in order to resolve all
genie signals for RX 1 after decoding its own message and
subtracting its desired signals, a genie needs to provide $|{\bf
\bar{G}}|=3M-N$ dimensions to RX 1.

{\bf Case: $1/3\leq M/N\leq 4/9 \Rightarrow d\leq 3M/4$}

In this case, notice that $3M-N\leq N/3 \leq M$. Therefore, a genie
provides ${\bf \bar{G}}={\bf \bar{X}}^{[2]}_{(3M-N)}$ to RX 1. Since ${\bf
G}$ is linearly independent with ${\bf S}^{[1]}$, RX 1 is able
to decode all the messages subject to the noise distortion. Thus, we
obtain the following inequality:
\begin{eqnarray}
4nR-n\epsilon_n&\!\!\!\!\leq:\!\!\!\!& \hbar({\bf Y}^{[1]^n})+\hbar({\bf G}^n|{\bf S}^{[1]^n})\\
&\!\!\!\!\leq:\!\!\!\!& Nn\log\rho+\hbar(\mathbf{X}^{[2]^n}_{3M-N})\\
&\!\!\!\!\leq:\!\!\!\!& Nn\log\rho+(3M-N)n\log\rho=3Mn\log\rho.
\end{eqnarray}
By letting $n\rightarrow \infty$ first and then $\rho\rightarrow
\infty$, we have the desired DoF outer bound
\begin{eqnarray}
d \leq 3M/4.
\end{eqnarray}

{\bf Case: $4/9\leq M/N\leq 1/2 \Rightarrow d\leq N/3$}

In this case, we still have $3M-N\leq M$. Therefore, a genie
provides ${\bf \bar{G}}={\bf \bar{X}}^{[2]}$, no less than $3M-N$ dimensions, to
RX 1, such that after removing its desired signal and
interference from User 2, RX 1 is able to invert the channels
from TX 3 and TX 4 to recover the transmit signal
vectors from the remaining two users. Thus, we obtain the following
inequality:
\begin{eqnarray}
4nR-n\epsilon_n\leq: \hbar({\bf Y}^{[1]^n})+\hbar({\bf G}^n|{\bf
S}^{[1]^n})\leq: Nn\log\rho+\hbar(\mathbf{X}^{[2]^n})\leq:
Nn\log\rho+nR.
\end{eqnarray}
By letting $n\rightarrow \infty$ first and then $\rho\rightarrow
\infty$, we have the desired DoF outer bound
\begin{eqnarray}
d \leq N/3.
\end{eqnarray}

{\bf Case: $1/2\leq M/N\leq 3/5 \Rightarrow d\leq 2M/3$}

In this case, $3M-N\geq M$ implies that the genie signals contain
more than $M$ dimensions. Thus, we let a genie provides ${\bf
\bar{G}}=\{{\bf \bar{X}}^{[2]},{\bf \bar{X}}^{[3]}_{(2M-N)}\}$ to RX 1. After
removing its desired signal and interference from User 2, RX 1
is able to recover the signal vectors from User 3 and User 4 from
the $N$-dimensional observations of ${\bf X}^{[3]},{\bf X}^{[4]}$
and ${\bf G}$ subject to the noise distortion. Thus, we obtain the
following inequality:
\begin{eqnarray}
4nR-n\epsilon_n&\!\!\!\!\leq:\!\!\!\!& \hbar({\bf Y}^{[1]^n})+\hbar({\bf G}^n|{\bf S}^{[1]^n})\\
&\!\!\!\!\leq:\!\!\!\!& Nn\log\rho+\hbar({\bf X}^{[2]^n},{\bf X}^{[3]^n}_{(2M-N)})\\
&\!\!\!\!\leq:\!\!\!\!& Nn\log\rho+nR+(2M-N)n\log\rho.
\end{eqnarray}
By letting $n\rightarrow \infty$ first and then $\rho\rightarrow
\infty$, we have the desired DoF outer bound
\begin{eqnarray}
d \leq 2M/3.
\end{eqnarray}

{\bf Case: $3/5\leq M/N\leq 2/3 \Rightarrow d\leq 2N/5$}

In this case, we still have $3M-N\geq M$. Thus, we first let a genie
provides ${\bf \bar{G}}_1=\{{\bf \bar{X}}^{[2]},{\bf \bar{X}}^{[3]}_{(2M-N)}\}$ to
RX 1, such that it can decode all the messages subject to the
noise distortion. Therefore, we have the first inequality as
follows:
\begin{eqnarray}
4nR-n\epsilon_n&\!\!\!\!\leq:\!\!\!\!& \hbar({\bf Y}^{[1]^n})+\hbar({\bf G}_1^n|{\bf S}^{[1]^n})\\
&\!\!\!\!\leq:\!\!\!\!& Nn\log\rho+\hbar({\bf X}^{[2]^n},{\bf X}^{[3]^n}_{(2M-N)}|{\bf S}^{[1]^n})\\
&\!\!\!\!\leq:\!\!\!\!& Nn\log\rho+\hbar({\bf X}^{[2]^n})+\hbar({\bf X}^{[3]^n}_{(2M-N)}|{\bf S}^{[1]^n},{\bf X}^{[2]^n})\\
&\!\!\!\!\leq:\!\!\!\!& Nn\log\rho+nR+\hbar({\bf X}^{[3]^n}_{(2M-N)}|{\bf X}^{[3\sim1]^n}_{N-M})\label{eqn:4to1_ob_zf}\\
&\!\!\!\!\leq:\!\!\!\!& Nn\log\rho+2nR-\hbar({\bf
X}^{[3\sim1]^n}_{N-M})\label{eqn:4to1_2by3_ob1}
\end{eqnarray}
where (\ref{eqn:4to1_ob_zf}) is obtained by zero forcing the
interference from User 4 at the $N$-dimensional vector space. Then
subtracting the signal caused by ${\bf X}^{[2]}$ produces the $N-M$
dimensional observations from User 3.

Second, a genie provides ${\bf \bar{G}}_2=\{{\bf \bar{X}}^{[4]},{\bf
\bar{X}}^{[3\sim1]^n}_{N-M}\}$ to RX 1. Notice that $|{\bf
\bar{G}}_2|=M+(N-M)\geq 3M-N$. In addition, within the $M$-dimensional
subspace projection at RX 1 from TX 3, the directions
of ${\bf X}^{[3\sim1]^n}_{N-M}$ depends only on the channel from
User 4 (the zero-forcing step). Thus, providing ${\bf \bar{G}}_2$ to
RX 1 allows it to decode all the messages as well subject to the
noise distortion. Hence, the argument above produces the second
inequality as follows:
\begin{eqnarray}
4nR-n\epsilon_n&\!\!\!\!\leq:\!\!\!\!& \hbar({\bf Y}^{[1]^n})+\hbar({\bf G}_1^n|{\bf S}^{[1]^n})\\
&\!\!\!\!\leq:\!\!\!\!& Nn\log\rho+\hbar({\bf X}^{[4]^n},{\bf X}^{[3\sim1]^n}_{N-M})\\
&\!\!\!\!\leq:\!\!\!\!& Nn\log\rho+nR+\hbar({\bf
X}^{[3\sim1]^n}_{N-M}).\label{eqn:4to1_2by3_ob2}
\end{eqnarray}

Adding up the two inequalities (\ref{eqn:4to1_2by3_ob1}) and
(\ref{eqn:4to1_2by3_ob2}), we finally have the desired DoF outer
bound:
\begin{eqnarray}
8nR-n\epsilon_n \leq: 2Nn\log\rho+3nR\Rightarrow d \leq 2N/5.
\end{eqnarray}

{\it Remark:} The DoF value for the $3/5\leq M/N\leq 2/3$ setting of
four-to-one $M\times N$ MIMO interference channel is identical to
that of three-user $M\times N$ MIMO interference channel with the
same $M/N$ value. While there is only one interfering RX in
the four-to-one setting, it is interesting to observe that we have
enough users in the four-to-one network, such that zero-forcing the
observations from User 2 and User 4 produces generic two subspaces
projection from User 3. In the three-user $M\times N$ MIMO
interference channel, however, we can only zero force one
interferer, leaving the observation of the other. Thus, producing
generic subspace (observations) will have to rely on the other
undesired RX.

{\bf Case: $2/3\leq M/N\leq 5/6 \Rightarrow d\leq 3M/5$}

The proof for this case is similar to that for the $3/5\leq M/N\leq
2/3$ setting by reproducing the second sum rate inequality.
Specifically, a genie provides ${\bf \bar{G}}_2=\{{\bf \bar{X}}^{[4]},{\bf
\bar{X}}^{[3\sim1]}_{N-M},{\bf \bar{X}}^{[3]}_{3M-2N}\}$ to RX 1. Again,
it can be easily seen that RX 1 is able to decode all the messages
subject to the noise distortion. Therefore, we have the second
inequality:
\begin{eqnarray}
4nR-n\epsilon_n&\!\!\!\!\leq:\!\!\!\!& \hbar({\bf Y}^{[1]^n})+\hbar({\bf G}_2^n|{\bf S}^{[1]^n})\\
&\!\!\!\!\leq:\!\!\!\!& Nn\log\rho+\hbar({\bf X}^{[4]^n},{\bf X}^{[3\sim1]^n}_{N-M},{\bf X}^{[3]^n}_{3M-2N})\\
&\!\!\!\!\leq:\!\!\!\!& Nn\log\rho+\hbar({\bf X}^{[4]^n})+\hbar({\bf X}^{[3\sim1]^n}_{N-M})+\hbar({\bf X}^{[3]^n}_{3M-2N})\\
&\!\!\!\!\leq:\!\!\!\!& Nn\log\rho+nR+\hbar({\bf
X}^{[3\sim1]^n}_{N-M})+(3M-2N)n\log\rho.\label{eqn:4to1_2by3_ob3}
\end{eqnarray}
Adding up (\ref{eqn:4to1_2by3_ob1}) and (\ref{eqn:4to1_2by3_ob3}) we
have the desired DoF outer bound:
\begin{eqnarray}
8nR-n\epsilon_n\leq: 2Nn\log\rho+3nR+(3M-2N)n\log\rho\Rightarrow
d\leq 3M/5.
\end{eqnarray}

{\bf Case: $5/6\leq M/N\leq 1 \Rightarrow d\leq N/2$}

In this case, a genie provides ${\bf \bar{G}}=\{{\bf \bar{X}}^{[2]},{\bf
\bar{X}}^{[3]}\}$ to RX 1. After removing its desired signal and
interference from User 2 and User 3, RX 1 is able to invert
the channels from TX 4 to recover ${\bf X}^{[4]}$ and thus
decode $W_4$ subject to the noise distortion. Thus, we obtain the
following inequality:
\begin{eqnarray}
4nR-n\epsilon_n\leq: \hbar({\bf Y}^{[1]^n})+\hbar({\bf G}^n|{\bf
S}^{[1]^n})\leq:
Nn\log\rho+\hbar(\mathbf{X}^{[2]^n},\mathbf{X}^{[3]^n})\leq:
Nn\log\rho+2nR.
\end{eqnarray}
By letting $n\rightarrow \infty$ first and then $\rho\rightarrow
\infty$, we have the desired DoF outer bound
\begin{eqnarray}
d \leq N/2.
\end{eqnarray}

So far, we finish all the DoF outer bound proofs. \hfill\QED

\subsection{DoF Achievability}

For the $M/N>1/3$ setting, as similar to the three user MIMO
interference channel \cite{Wang_Gou_Jafar_3userMxN}, it suffices to
show the achievability at $M/N=4/9,~3/5,~5/6$, and all the other
cases directly follow from the intuition that increasing antenna
redundancies does not hurt the DoF of the networks. As only RX
1 suffers from interference, we merely need to design the precoding
matrices $\mathbf{V}^{[i]},~i\in\mathcal{K}$ at each TX so
that the interference at RX 1 spans a vector subspace as small
dimensions as possible.

{\bf Case: $M/N=4/9 \Rightarrow d=3M/4=N/3$}

We suppose $(M,N) = (4\beta,9\beta)$ where $\beta\in\mathbb{Z}^+$.
In this case, Theorem \ref{theorem:4to1} implies that the $d =
3\beta$ DoF per user are achievable. Consider the
$9\beta$-dimensional vector subspace at the Receive 1, the desired
signal occupies $3\beta$ dimensions, leaving the rest $6\beta$
dimensions for interference. Since there are a total of $3d =
9\beta$ dimensional interference from TX 2 to TX
4, we need to align $9\beta-6\beta=3\beta$ interference symbols.
Note that TX 2 and TX 3 together project at
RX 1 an $8\beta$-dimensional subspace, which has a
$8\beta+4\beta-9\beta=3\beta$-dimensional intersection with the
$4\beta$-dimensional subspace seen from TX 4 at RX 1.
Therefore, we can align the $3\beta$ symbols from TX 4
\emph{totally} into the subspace spanned by the interference from
TX 2 and TX 3. Mathematically, we have
\begin{eqnarray}
&&\mathbf{H}^{[14]}\mathbf{V}^{[4]}=-(\mathbf{H}^{[12]}\mathbf{V}^{[2]}+\mathbf{H}^{[13]}\mathbf{V}^{[3]})\notag\\
&&\Rightarrow \left[ \mathbf{H}^{[12]}~~\mathbf{H}^{[13]}
~~\mathbf{H}^{[14]} \right]_{9\beta \times 12\beta} \left[
\begin{array}{c} \mathbf{V}^{[2]} \\ \mathbf{V}^{[3]} \\
\mathbf{V}^{[4]}
\end{array} \right]_{12\beta \times 3\beta}
= \mathbf{0},
\end{eqnarray}
and we can solve
$\mathbf{V}^{[2]},\mathbf{V}^{[3]},\mathbf{V}^{[4]}$ by truncating
the $3\beta$ basis of the null space of the matrix consisting the
matrices associated with RX 1. Finally, we randomly generate
$\mathbf{V}^{[1]}$ to ensure the linear independence between the
desired signals and interference at RX 1, such that RX 1
is able to decode its $3\beta$ symbols. Since RX 2 to RX
4 hears no interference, they are able to decode their own messages
as well.

{\bf Case: $M/N=3/5 \Rightarrow d=2M/3=2N/5$}

In this case, we suppose $(M,N) = (3\beta,5\beta)$ where
$\beta\in\mathbb{Z}^+$, and Theorem \ref{theorem:4to1} implies that
$d = 2\beta$ DoF per user are achievable. In the
$5\beta$-dimensional vector space at RX 1, the desired signal
occupies $2\beta$ dimensions, thus leaving the rest
$3\beta$-dimensional subspace for interference. As a result, a total
of $3d = 6\beta$ dimensional interference from TX 2 to 4
should be aligned into a $3\beta$-dimensional subspace. Note that
each TX projects a $3\beta$-dimensional subspace at
RX 1, and any two of them have a $3\beta+3\beta-5\beta=\beta$
dimensional intersection. Therefore, for any two users among the
three interferers, we align $\beta$ symbols and thus saving $3\beta$
interference dimensions. Suppose the beamforming matrix of
TX $i$ is given by $\mathbf{V}^{[i]}=
[\mathbf{V}^{[i]}_1,\mathbf{V}^{[i]}_2]$ which consists of $2\beta$
columns. Thus, we produce the following alignment equations:
\begin{eqnarray}
&&\mathbf{H}^{[12]}\mathbf{V}^{[2]}_1 = - \mathbf{H}^{[13]}\mathbf{V}^{[3]}_1\Rightarrow \left[ \mathbf{H}^{[12]} \mathbf{H}^{[13]} \right]_{5\beta \times 6\beta} \left[  \begin{array}{c} \mathbf{V}^{[2]}_1 \\ \mathbf{V}^{[3]}_1 \end{array} \right]_{6\beta \times \beta} = \mathbf{0}, \\
&&\mathbf{H}^{[12]}\mathbf{V}^{[2]}_2 = - \mathbf{H}^{[14]}\mathbf{V}^{[4]}_1\Rightarrow \left[ \mathbf{H}^{[12]} \mathbf{H}^{[14]}\right]_{5\beta \times6\beta} \left[  \begin{array}{c} \mathbf{V}^{[2]}_2 \\\mathbf{V}^{[4]}_1\end{array} \right]_{6\beta \times \beta}= \mathbf{0}, \\
&&\mathbf{H}^{[13]}\mathbf{V}^{[3]}_2 = -
\mathbf{H}^{[14]}\mathbf{V}^{[4]}_2\Rightarrow \left[
\mathbf{H}^{[13]} \mathbf{H}^{[14]} \right]_{5\beta \times 6\beta}
\left[
\begin{array}{c}\mathbf{V}^{[3]}_2 \\\mathbf{V}^{[4]}_2\end{array}
\right]_{6\beta \times \beta} = \mathbf{0},
\end{eqnarray}
and thus we can solve the equations by truncating the basis of each
null space of corresponding matrices. The linear independence among
$2\beta$ columns of each ${\bf V}^{[i]},~i=2,3,4$, as shown in
\cite{Wang_Gou_Jafar_3userMxN}, can be established by choosing a set
of special matrices and showing it has full rank. Finally, we choose
$\mathbf{V}^{[1]}$ randomly to ensure the linear independence of the
desired signal and interference at RX 1. So far, each user
achieves $2\beta$ DoF.

{\bf Case: $M/N=5/6 \Rightarrow d=3M/5=N/2$}

In this case, suppose $(M,N) = (5\beta,6\beta)$ where
$\beta\in\mathbb{Z}^+$, and Theorem \ref{theorem:4to1} implies that
$d=3\beta$ DoF per user are achievable. In the $6\beta$-dimensional
vector space at RX 1, the desired signal occupies $3\beta$
dimensions, leaving the rest $3\beta$-dimensional subspace for
interference. Because we have a total of $3d=9\beta$ symbols from
three interferers, we need to align the interference from
TX 3 and TX 4 into the same subspace projected
from TX 2. Therefore, we have the following alignment
equations:
\begin{eqnarray}
&&\mathbf{H}^{[12]}\mathbf{V}^{[2]} = - \mathbf{H}^{[13]}\mathbf{V}^{[3]} =  - \mathbf{H}^{[14]}\mathbf{V}^{[4]}\\
&&\Rightarrow \left[ \begin{array}{ccc} \mathbf{H}^{[12]} &
\mathbf{H}^{[13]} & {\bf O} \\ \mathbf{H}^{[12]} & {\bf O} & \mathbf{H}^{[14]}
\end{array}
 \right]_{12\beta \times 15\beta}
\left[  \begin{array}{c} \mathbf{V}^{[2]} \\ \mathbf{V}^{[3]} \\
\mathbf{V}^{[4]}
\end{array} \right]_{15\beta \times 3\beta}
= \mathbf{0},
\end{eqnarray}
and again each $\mathbf{V}^{[i]}, i=2,3,4$ can be obtained
correspondingly. Finally, we choose $\mathbf{V}^{[1]}$ randomly to
ensure the linear independence of the desired signal and
interference at RX 1. Therefore, each user achieves $3\beta$
DoF.

\hfill\QED

\section{DoF Counting Bound of the Many-to-One MIMO Interference Channel}\label{app:counting}

In the $K$-user many-to-one $M\times N$ MIMO Gaussian interference
channel where each TX has $M$ antennas and each RX
has $N$ antennas, only RX 1 hears interference from
TXs $2$ to $K$ while all the other RXs only hear
their desired signals. Consider the channels have no structure,
i.e., without symbol extensions, and we only use the linear
transmission scheme, i.e., through beamforming and zero-forcing
steps. We are only interested in the symmetric DoF value per user
where each user requires $d$ DoF. Basically, this problem is within
the linear feasibility framework. While the achievable scheme is
limited to linear scheme, the DoF outer bound is based on the
intuition of counting free variables and alignment bilinear
equations associated with feasibility conditions. In this work, we
only consider the DoF counting bound, following the rules
established by Yetis et. al. in \cite{Cenk_Gou_Jafar} in the fully
connected MIMO interference channel, and then applied in the $X$
channel setting, as recently shown by Sun et. al. in
\cite{Sun_Geng_Gou_Jafar}.

Using the linear transmission scheme, TX $i$ intends to
send $d$ independent streams using a $M\times d$ precoding matrix
$\mathbf{V}^{[i]}$ and RX $i$ extracts its desired signal by
using an $N \times d$ interference filtering matrix
$\mathbf{U}^{[i]}$ where $i\in\mathcal{K}$. Then the feasibility of
linear interference alignment problem is equivalent to solving the
following algebraic equations:
\begin{eqnarray}
\mathbf{U}^{[1]^{\dagger}} \mathbf{H}^{[1i]} \mathbf{V}^{[i]} = 0,
~~~~~~i \in \{2,3,\ldots,K\} \label{eqn:e1mto}
\end{eqnarray}
which essentially means that only RX 1 needs to zero force the
interference. Note that we omit the full rank condition of the
desired signals for User 1 because the channel matrix of the desired
link is generic and also we are primarily interested in the counting
(outer) bound. Now we can proceed to count the number of equations
$N_e$ and variables $N_v$ in (\ref{eqn:e1mto}) which are given by:
\begin{eqnarray}\label{eqn:count_nenv}
N_e &\!\!\!\!=\!\!\!\!& (K-1)d^2, \\
N_v &\!\!\!\!=\!\!\!\!& (K-1)(M-d)d + (N-d)d.
\end{eqnarray}
A system is defined as \emph{proper} if $N_e \leq N_v$. Otherwise,
the system is \emph{improper}. Thus, substituting
(\ref{eqn:count_nenv}) into (\ref{eqn:e1mto}), we have the proper
system condition:
\begin{eqnarray}
d \leq \frac{(K-1)M+N}{2K-1}.
\end{eqnarray}
For example, $d\leq \frac{3M+N}{7}$ for $K=4$, and it can be
verified to be always larger than the decomposition bound
$d=\frac{MN}{M+N}$.

%\newpage

%\appendix
%\section{Appendix}
%
%
%Suppose both $h({\bf L}^{[1]}|{\bf L}^{[2]})>d/2$ and $h({\bf
%L}^{[2]}|{\bf L}^{[1]})>d/2$. Then we have:
%
%\begin{eqnarray*}
%d &<& h({\bf L}^{[1]}|{\bf L}^{[2]})+ h({\bf L}^{[2]}|{\bf L}^{[1]})\\
%&\leq& h({\bf L}^{[1]}|{\bf L}^{[2]})+ h({\bf L}^{[2]})\\
%&=& h({\bf L}^{[1]},{\bf L}^{[2]})\\
%&=& h(X)\\
%&=& d
%\end{eqnarray*}
%which is a contradiction.
%
% \hfill\QED

\begin{thebibliography}{1}

\bibitem{Wang_Gou_Jafar_3userMxN}
C.~Wang, T.~Gou, S.~Jafar, ``Subspace Alignment Chains and the
Degrees of Freedom of the Three-User MIMO Interference Channel",
{\em e-print arXiv:1109.4350}, Sept. 2011.

\bibitem{Bresler_Cartwright_Tse}
G. Bresler, D. Cartwright, D. Tse, ``Geometry of the 3-user MIMO interference channel",
{\em 2011 Allerton Conference}, Monticello, IL, September 2011.

\bibitem{Cenk_Gou_Jafar}
C. Yetis, T. Gou, S. Jafar, A. Kayran, ``On Feasibility of
Interference Alignment in MIMO Interference Networks," {\em IEEE
Trans. Signal Processing}, Sep. 2010, Vol. 58, Issue: 9, Pages:
4771-4782.

\bibitem{Bresler_Cavendish_Tse}
G. Bresler, D. Cartwright, D. Tse, ``Settling the
feasibility of interference alignment for the MIMO interference
channel: the symmetric square case," {\em arXiv:1104.0888}, April
2011.

\bibitem{Razaviyayn_MIMO}
M. Razaviyayn, G. Lyubeznik, Z. Luo, ``On the Degrees
of Freedom Achievable Through Interference Alignment in a MIMO
Interference Channel", {\em arXiv:1104.0992}, April 2011.

\bibitem{Jafar_ITA}
S. Jafar, ``Understanding spatial signal dimensions: recent results and some conjectures on the DoF of MIMO interference networks," {\em 2012 Information Theory and Applications Workshop}, January 2012.
%
%\bibitem{Gou_Jafar_MIMO}
%T. Gou and S. A. Jafar, ``Degrees of Freedom of the $K$ User
%$M\times N$ MIMO Interference Channel", {\em IEEE Transactions on
%Information Theory}, Dec. 2010, Vol. 56, Issue: 12, pp.~6040--6057.

\bibitem{Ghasemi_Motahari_Khandani_MIMO}
A. Ghasemi, A. Motahari, A. Khandani, ``Interference Alignment for
the $K$ User MIMO Interference Channel," {\em arXiv:0909.4604}, Sep.
2009.

\bibitem{Cadambe_Jafar_int}
V.~Cadambe and S.~Jafar, ``Interference alignment and the degrees of
freedom of the $K$ user interference channel'', {\em IEEE Trans. on
Information Theory}, vol.~54, pp.~3425--3441, Aug. 2008.
%
%\bibitem{Motahari_Gharan_Maddah-Ali_Khandani}
%A. Motahari, S. Gharan, M. A. Maddah-Ali, A, Khandani, ``Real
%Interference Alignment: Exploiting the Potential of Single Antenna
%Systems", {\em arxiv.org/pdf/0908.2282}, August 2009.
%
%\bibitem{Jafar_Shamai}
%S.~Jafar, S.~Shamai, ``Degrees of Freedom Region for the MIMO $X$
%Channel'', {\em IEEE Trans. on Information Theory}, vol.~54, No.~1,
%pp.~151-170, Jan. 2008.
%

%\bibitem{feasibility_spain}
%O. Gonzalez, C. Beltran, I. Santamaria, ``On the feasibility of
%interference alignment for the K-user MIMO channel with constant
%coefficients", {\em http://arxiv.org/abs/1202.0186}, Feburary 2012.
%
%\bibitem{Avestimehr_Diggavi_Tse}
%S. Avestimehr, S. Diggavi, and D. Tse, ``A Deterministic Approach to
%Wireless Relay Networks", {\em Allerton Conf. Commun., Control,
%Comput.}, Monticello, IL, Sep. 2007.
%
%
%\bibitem{Etkin_Ordentlich}
%R. Etkin and E. Ordentlich, ``On the Degrees-of-Freedom of the
%$K$-User Gaussian Interference Channel'', {\em
%http://arxiv.org/abs/0901.1695}, Jan. 2009.
%
%\bibitem{Real_IA}
%A.S. Motahari, S. O. Gharan and A. K. Khandani, ``Real Interference
%Alignment with Real Numbers,'' {\em http://arxiv.org/abs/0908.1208},
%Aug. 2009.


%\bibitem{Gou_Jafar_Wang}
%T.~Gou, S.~Jafar and C.~Wang, ``On the Degrees of Freedom of Finite
%State Compound Wireless Networks - Settling a Conjecture by
%Weingarten et. al", {\em arXiv:0909.4177}, 2009.
%
%\bibitem{Maddah_Ali}
%M.A. Maddah-Ali, ``On the Degrees of Freedom of the Compound MIMO
%Broadcast Channels with Finite States'', {\em arXiv:0909.5006v3},
%Oct. 2009.
%
%\bibitem{Lee_Jafar}
%N. Lee and S. Jafar, ``Aligned Interference Neutralization and the
%Degrees of Freedom of the 2 User Interference Channel with an
%Instantaneous Relay", {\em http://arxiv.org/abs/1102.3833}, Feb.
%2011.

%\bibitem{Wang_Jafar}
%C. Wang and S. Jafar, ``Aligned Interference Neutralization and the
%Degrees of Freedom of the 2 User X Channel with an Instantaneous
%Relay", {\em CPCC Technical Report
%http://escholarship.org/uc/item/0112p0hf}, June 2011.
%
%\bibitem{Gou_2x2x2}
%Tiangao Gou, Syed A. Jafar, Sang-Woon Jeon, Sae-Young Chung,
%``Aligned Interference Neutralization and the Degrees of Freedom of
%the 2x2x2 Interference Channel", {\em e-print arXiv:1012.2350},
%December 2010.
%
%\bibitem{MMK}
%M.A. Maddah-Ali, A.S. Motahari, and A.K. Khandani, "Communication
%Over MIMO $X$ Channels: Interference Alignment, Decomposition, and
%Performance Analysis," IEEE Transaction on Information Theory, Vol.
%54, No. 8, pp. 3457-3470, Aug. 2008.

\bibitem{Wang_Sun_Jafar_ISIT}
C. Wang, H. Sun, and S. Jafar, ``Genie chains and the Degrees of Freedom of the K-user MIMO Interference Channel,'' {\em Proceedings of 2012 IEEE International Symposium on Information Theory (ISIT)}, 1-6 July 2012, MIT, USA.

\bibitem{Sridharan_Yu_2}
G. Sridharan and W. Yu, ``Degrees of Freedom of MIMO Cellular Networks
with Two Cells and Two Users Per Cell'', in {\em ISIT}, 2013.

\bibitem{Sridharan_Yu_3}
G. Sridharan and W. Yu, ``Degrees of Freedom of MIMO Cellular Networks:
Two-Cell Three-User-Per-Cell Case'', in {\em GLOBECOM}, 2013.

\bibitem{Jafar_Fakhereddin}
S.~Jafar, M.~Fakhereddin, ``Degrees of Freedom for the MIMO
Interference Channel," {\em IEEE Transactions on Information
Theory}, July 2007, Vol. 53, No. 7, Pages: 2637-2642.

\bibitem{Feasibility}
O. Gonz‡lez, C. Beltr‡n, and I. Santamar'a, ``A Feasibility Test for Linear Interference Alignment
in MIMO Channels with Constant CoefÞcients'', {\em IEEE Trans. on Information theory}, vol.~60, no.~3, pp.~1840-1856, March 2014.

\bibitem{Wang_Gou_Jafar_3userMxNacs}
C.~Wang, T.~Gou, S.~Jafar, ``On Optimality of Linear Interference
Alignment for the Three-User MIMO Interference Channel with Constant
Channel Coefficients", {\em
http://escholarship.org/uc/item/6t14c361}, October 2011.

\bibitem{Annapureddy_vernu_CoMP}
V. Annapureddy, A. Gamal, V. Veeravalli,
``Degrees of Freedom of Interference Channels with CoMP Transmission
and Reception'', {\em IEEE Transactions on Information Theory}, vol. 58, no. 9, pp. 5740-5760, Sep. 2012.

\bibitem{Sun_Geng_Gou_Jafar}
H. Sun, C. Geng, T. Gou, S. Jafar, ``Degrees of Freedom of MIMO X Networks: Spatial Scale Invariance, One-Sided Decomposability and Linear Feasibility,'' {\em Proceedings of 2012 IEEE International Symposium on Information Theory (ISIT),} 1-6 July 2012, MIT, USA.

%
%\bibitem{Jafar_SCBC}
%S.~Jafar, ``Exploiting Channel Correlations - Simple Interference
%Alignment Schemes with no CSIT'', {\em arXiv:0910.0555}, 2009.
%
%
%\bibitem{Nazer_ergodic}
%B. Nazer, M. Gastpar, S. A. Jafar, and S. Vishwanath, ``Ergodic
%Interference Alignment,'' Proceedings of the IEEE International
%Symposium on Information Theory (ISIT 2009), Seoul, Korea, June
%2009.

%\bibitem{Gesbert_Hanly_Huang_Shitz_Simeone_Yu}
%D. Gesbert, S. Hanly, H. Huang, S. Shitz, O. Simeone, and W. Yu,
%``Multi-cell MIMO cooperative networks: A new look at interference",
%{\em IEEE Journal on Selected Areas in Communications}, vol. 28, no.
%9, pp. 1380--1408, Dec. 2010.

%\bibitem{Kramer_Maric_Yates_cooperative}
%G. Kramer, I. Mari\'{c}, and R. Yates, ``Cooperative
%communications", {\em IEEE Foundations and Trends \textregistered in
%Networking}, vol. 1, no. 3, pp. 271--425, 2006.

%\bibitem{Cadambe_Jafar_relay_feedback}
%V. R. Cadambe and S. A. Jafar, ``Degrees of freedom of wireless
%networks with relays, feedback, cooperation and full duplex
%operation", {\em IEEE Transactions on Information Theory}, vol. 55,
%pp. 2334--2344, May 2009.

%\bibitem{Host-Madsen_Nosratinia_ISIT05}
%A. Host-Madsen and A. Nosratinia, ``The multiplexing gain of
%wireless networks", {\em in Proc. of ISIT}, 2005.
%

%\bibitem{Bresler_Tse_Alleton09}
%G.~Bresler and D.~Tse, ``Degrees-of-freedom for the 3-user Gaussian
%interference channel as a function of channel diversity", {\em
%Allerton Conference on Communication, Control, and Computing},
%Monticello, IL, September 2009.

%\bibitem{Levy_Shamai_ITA2009}
%N. Levy and S. Shamai (Shitz), ``Clustered local decoding for
%Wynertype cellular models," {\em IEEE Transactions on Information Theory}, Vol. 55, No. 11, Pages:  4967--4985, 2009.
%
%\bibitem{Lapidoth_Shamai_ISIT2009}
%A. Lapidoth, N. Levy  S. Shamai (Shitz) and  M. A. Wigger, ``A
%Cognitive Network with Clustered Decoding", {\em Proceedings of IEEE
%International Symposium on Information Theory (ISIT)}, Seoul, Korea,
%June 28 -- July 3, 2009.
%
%\bibitem{Shamai_ITA2010}
%D. Gesbert,  S. Hanly, H. Huang, S. Shamai, O.Simeone and Wei Yu,
%"Multi-cell MIMO cooperative networks: A new look at interference,"
%Jornal Selected Areas in Communications (JSAC),
%vol. 28, no. 9, pp. 1380-1408, December 2010.



\end{thebibliography}
\end{document}